\documentclass[12pt]{article}
\usepackage{amsmath, amssymb}
\usepackage{jheppub}
\usepackage{booktabs}
\usepackage{siunitx}
\usepackage{float}
\usepackage{graphicx}
\usepackage[normalem]{ulem}
\input epsf.tex
\usepackage{graphicx}
\usepackage{color}
\usepackage[dvipsnames]{xcolor}

\definecolor{darkblue}{rgb}{0.0, 0.0, 0.55}
\definecolor{grey}{rgb}{0.57, 0.64, 0.69}
\definecolor{lightbrown}{rgb}{0.71, 0.4, 0.11}
\newcommand{\tcb}{\textcolor{blue}}
\newcommand{\tcc}{\textcolor{cyan}}

\newcommand{\tcfg}{\textcolor{ForestGreen}}

\newcommand{\tcr}{\textcolor{red}}
\newcommand{\tcg}{\textcolor{green}}

\newcommand{\tco}{\textcolor{orange}}
\newcommand{\tcy}{\textcolor{yellow}}
\newcommand{\tcp}{\textcolor{purple}}
\newcommand{\tcgr}{\textcolor{grey}}
\newcommand{\tcbr}{\textcolor{brown}}
\newcommand{\tclb}{\textcolor{lightbrown}}

\newcommand{\nn}{\nonumber}

\newcommand{\beq}{\begin{equation}}
\newcommand{\eeq}{\end{equation}}
\newcommand{\beqa}{\begin{eqnarray}}
\newcommand{\eeqa}{\end{eqnarray}}
\newcommand{\beqar}{\begin{eqnarray*}}
\newcommand{\eeqar}{\end{eqnarray*}}

\newcommand{\bea}{\begin{eqnarray}}
\newcommand{\eea}{\end{eqnarray}}

\newcommand{\ie}{{\it i.e.,}\ }
\newcommand{\reef}[1]{(\ref{#1})}

\newcommand\cS{{\cal S}}

\newcommand\cL{{\cal L}}
\newcommand\cG{{\cal G}}
\newcommand\cI{{\cal I}}

\newcommand\cO{{\cal O}}

\title{On generalized quasi-topological cubic-quartic gravity:  thermodynamics and holography}

\author[a,b]{Mozhgan Mir}
\author[a,c]{and Robert B. Mann}

\affiliation[a]{Department of Physics and Astronomy, University of Waterloo,
Waterloo, Ontario, Canada, N2L 3G1}
\affiliation[b]{Department of Physics, Faculty of Science, Ferdowsi University of Mashhad\\
P.O. Box 1436, Mashhad, Iran}
\affiliation[c]{Perimeter Institute, 31 Caroline Street North, Waterloo,
ON, N2L 2Y5, Canada}

\emailAdd{mozhganmir@um.ac.ir}
\emailAdd{rbmann@uwaterloo.ca}

\abstract{
 We  investigate the thermodynamic behaviour of asymptotically anti de Sitter black holes in generalized quasi-topological gravity  containing terms both cubic and quartic  in the curvature.   We investigate the general
 conditions required for  physical phase transitions and critical behaviour in any dimension and then consider
 in detail specific properties in spacetime dimensions 4, 5, and 6.  We find for spherical black holes
  that there are respectively at most two and three physical critical points in five and six dimensions.  For hyperbolic black holes we find  the occurrence of  Van der Waals phase transitions in four dimensions
  and reverse Van der Waals phase transitions in dimensions greater than 4
  if both  cubic and quartic curvature terms are present.  We also observe the occurrence of
 phase transitions in for fixed chemical potential. We consider some applications of our work in the dual CFT, investigating how the ratio of viscosity to entropy is modified by inclusion of these higher curvature terms. We conclude that the presence of the quartic curvature term results in a violation of  the KSS bound in five dimensions, but not in other dimensions.
 }


\begin{document}
\maketitle

\section{Introduction } \label{intro}

In recent years there has been an increased appreciation of the role that higher-curvature theories of gravity play in our understanding of several areas of physics, including supergravity and string theory,  holography, cosmology,
and black holes.  These studies are motivated both by a desire to understand the ultraviolet behaviour of gravity and by a realization that terms non-linear in the curvature generically appear in perturbative calculations, particularly in    string theory \cite{Gross:1986mw}.    Furthermore, an analysis of this class of theories provides us with new insights into general relativity (or Einstein gravity) and may even provide new empirical tests of gravitational physics \cite{Amendola:2008vd,Navarro:2006mw,sotiriou2010f, AliHaimoud:2011fw,clifton2012modified,Yagi:2012gp,Delsate:2014hba,Hennigar:2018hza,Poshteh:2018wqy,mir:2017m}.

Originally proposed nearly a century ago \cite{weyl1923allgemeine, carmichael1925eddington}, a revival of interest in these theories in the theoretical physics community occurred once significant effort began to be expended on constructing a quantized theory of gravity.   Adding terms quadratic in the curvature to the
 Einstein-Hilbert action were found to yield a power-counting renormalizable theory~\cite{Stelle:1977ry}, and later a Gauss-Bonnet term was found to appear in the low energy effective action of string theory  \cite{Zwiebach:1985uq}.   More recently it has been shown from a variety of perspectives
\cite{Myers:2010ru,Hofman:2009ug,Sinamuli:2017rhp} that a proper investigation of dual theories beyond large $N$  in the  context of the AdS/CFT correspondence conjecture \cite{Maldacena:1998,mir:1307} entails  inclusion of higher curvature terms.

A key challenge presented by a generic higher-curvature theory of gravity is that the equations of motion are
greater than second order in the derivatives, leading to a number of  pathological properties such as the appearance of ghost degrees of freedom and other instabilities.  However there exist a  few classes of theories in which such pathologies are ameliorated and in certain cases are  absent.  The best known example is the   Lovelock class of gravitational theories~\cite{Lovelock:1971yv}. This class yields second order equations of motion in arbitrary dimensions, with  the Einstein-Hilbert term being one of several terms that constitute Lovelock theory in a given dimension.  In this sense Einstein gravity can be regarded as a special case of Lovelock gravity.  Since this class of  theories is ghost-free~\cite{Zwiebach:1985uq} they are viable candidates for generalizations of Einstein gravity in higher dimensions $d \geq 2k+1$ for a Lovelock theory that is $k$th order in the curvature.  For dimensions $d < 2k+1$ such terms play no role in the equations of motion.  Hence only $k=1$ Einstein gravity
has non-trivial field equations and so one must look beyond Lovelock gravity to obtain interesting higher curvature theories with implications in $(3+1)$ dimensions.

In the past several years considerable progress has been made along these lines. A  broader class of \textit{quasi-topological gravity} theories~\cite{Myers:2010ru, Oliva:2010eb} have been constructed that retain many of the nice properties of Lovelock gravity under certain symmetry restrictions. They
are non-trivial in any dimension $d \ge 5$ regardless of the order in the curvature. Cubic-curvature quasi-topological gravity, for example exists in $d \ge 5$ whereas cubic Lovelock gravity exists in $d \ge 7$.  Furthermore, their field equations, while  generally greater than second-order,   become second order under the imposition of spherical symmetry.  Moreover, the linearized of the equations of motion of quasi-topological gravity coincide  with those of Einstein gravity on maximally symmetric spacetime backgrounds up to an overall prefactor
\cite{Myers:2010tj},  ensuring  that  negative energy excitations do not propagate to asymptotic regions of constant curvature.

Even more recently a more general class of higher-curvature gravity theories have been discovered that are of
considerable interest both holographically and phenomenologically.  This is because they are free of ghosts on constant curvature backgrounds,  solutions of their field equations yield a metric that depends on a single metric function in the spherically symmetric case, and they are dynamically non-trivial even in four dimensions.  First obtained in $(3+1)$ dimensions for cubic curvature~\cite{Bueno:2016xff} (a theory known as    \textit{Einsteinian cubic gravity}  or ECG),  they were found to have generalizations to any dimension
\cite{Hennigar:2017ego}  and to quartic powers in the curvature  \cite{Ahmed:2017jod}.  Generalizations to any power in the curvature exist \cite{Bueno:2017qce} but have not been explicitly constructed.

This class of theories  -- referred to as Generalized Quasitopological Gravity (GQG) -- has several remarkable features.  The constraint that  spherically symmetric solutions depend on only a single metric function is found to also eliminate ghosts upon linearization of the theory on a constant curvature background \cite{Hennigar:2017ego}.
Very recently it has been shown that they have a well-posed  initial value problem for cosmological solutions  and the potential to provide a late-time cosmology arbitrarily close to  the $\Lambda$-Cold dark matter scenario whilst having  a purely geometrical inflationary period in the early universe with a graceful exit \cite{Arciniega:2018fxj,Cisterna:2018tgx,Arciniega:2018tnn}.   While the field equations can be solved exactly in certain special cases~\cite{Feng:2017tev}, it is possible to analytically investigate the thermodynamics of black holes even in the generic case where analytic solutions are not available~\cite{Bueno:2016lrh,Hennigar:2016gkm}.
Charged black branes have an interesting phase structure that is absent for both their Lovelock and quasi-topological black brane counterparts~\cite{Hennigar:2017umz}. The Kovtun-Son-Starinets bound on the ratio of entropy density to shear viscosity always holds~\cite{Bueno:2018xqc}, and  small asymptotically flat black hole solutions  were found to be stable~\cite{Bueno:2017qce}, which may have implications for the information loss problem.  Further holographic applications of this class of theories have been carried out,
with discussions of  the a-theorem, the  universal stress-tensor two-point function and a universal relation for central charges \cite{Li:2017ncu,Li:2017txk,Li:2018drw}; more recently
an extensive investigation of  the holographic properties of the  cubic case without massive modes was carried out, including holographic central charges, energy flux, Renyi entropies, and shear viscosity to entropy ratio \cite{Li:2019auk}.  Other
recent work has shown that the shadows of GQG black holes have potentially interesting phenomenological signatures~\cite{Hennigar:2018hza,Poshteh:2018wqy}.

Thermodynamics of GQG black holes has yet to be fully explored.
Previous studies for asymptotically flat solutions and AdS black holes have appeared in restricted contexts~\cite{Bueno:2017sui, Bueno:2017qce, Hennigar:2017ego, Hennigar:2017umz}, but a full study combining both cubic \cite{Hennigar:2017ego} and quartic GQG \cite{Ahmed:2017jod} has yet to be carried out.  The purpose of this paper is to conduct such a study for both spherical and hyperbolic charged black holes.

Our  investigation will be carried out in the context of  black hole chemistry, in which the cosmological constant is
taken to be a thermodynamic variable \cite{Henneaux:1985tv, Creighton:1995au} that is interpreted as pressure in the first law of black hole mechanics \cite{Kastor:2010gq, Kastor:2011qp}.   An extensive amount of work over the past six years has been carried out in this subject \cite{Kubiznak:2016qmn} and has indicated that black holes can exhibit a broad range of phase behaviour that has been observed in other areas of physics.  Examples include
triple points \cite{Altamirano:2013uqa}, re-entrant phase transitions \cite{Altamirano:2013ane}, polymer-like behaviour \cite{Dolan:2014vba}, and even superfluid-like phase transitions \cite{Hennigar:2016xwd,Hennigar:2016ekz,Dykaar:2017mba}, as well as a deep analogy between charged anti-de Sitter black holes and  Van der Waals fluids \cite{Kubiznak:2012wp}.  Higher-curvature gravity theories have, using this approach, likewise been seen to have a very rich thermodynamic structure~\cite{Wei:2012ui, Cai:2013qga, Xu:2013zea, Mo:2014qsa, Wei:2014hba, Mo:2014mba, Zou:2013owa, Belhaj:2014tga, Xu:2014kwa, Frassino:2014pha, Dolan:2014vba, Sherkatghanad:2014hda, Hendi:2015cka, Hendi:2015oqa, Hennigar:2015esa, Hendi:2015psa, Nie:2015zia, Hendi:2015pda, Hendi:2015soe, Zeng:2016aly, Hennigar:2016gkm, Hennigar:2016ekz, Hennigar:2016xwd, Cvetic:2010jb, Hennigar:2014cfa, Johnson:2014yja, Karch:2015rpa, Caceres:2015vsa, Dolan:2016jjc}, with even more results surveyed in a recent review \cite{Kubiznak:2016qmn}.

Our paper is organized as follows.  In section \ref{sec:bhsolution} we present charged static, spherically symmetric AdS black holes in the {cubic-quartic GQG theory}.
This includes an asymptotic solution, a near horizon solution and then their match in the form of a numerical solution. In section \ref{sec:thermo} we investigate the thermodynamic properties of charged black holes in {cubic-quartic} generalized quasi-topological gravity by applying the black hole chemistry formalism. In section \ref{sec: thermoce} we classify the phase structure and critical points for these black holes by considering the perspective of black hole chemistry, working in the fixed charge ensemble. In section \ref{sec: thermofpe} we consider the thermodynamics for fixed potential ensemble.  We will analyze the four dimensional case in detail, and then present relevant results in higher dimensions. In section \ref{sec: holog} we present some results of holographic hydrodynamics to understand these theories in the context of AdS/CFT correspondence. We summarize our work in section \ref{discuss} and present some directions for further research.

\section{Charged black hole solutions in cubic-quartic GQG} \label{sec:bhsolution}

We begin by setting up charged static, spherically symmetric AdS black holes obtained
from the equations of motion that follow from a  combination of cubic and quartic terms of generalized quasi-topological gravity (GQG).

\subsection{Construction of equations of motion}

Consider the class of static radially symmetric metrics with radial coordinate $r$ and time coordinate $t$. Choosing coordinates so that the radius of a $(d-2)$ sphere behaves as $r^{d-2}$,
higher-curvature gravity theories up to quartic order are obtained by applying  the condition $g_{tt}g_{rr} = -1$ in order to get a single metric function.  They result in both Lovelock and quasi-topological curvature terms as well as GQG curvature terms.

Concentrating only on the properties of GQG theory, we put aside the Lovelock and quasi-topological terms (for a discussion see e.g.~\cite{Frassino:2014pha, Hennigar:2015esa}) and consider Einstein gravity accompanied by cubic and quartic generalized quasi-topological terms, with minimal coupling to an Abelian gauge field. The
action\footnote{Here our convention for the cubic coupling differs by a minus sign from that used in \cite{Hennigar:2017ego}.} in $d$ dimensional spacetime reads \cite{Ahmed:2017jod}
\beqa
\cI&=&\frac{1}{16\pi G}\int d^dx \sqrt{-g}\left[\frac{(d-1)(d-2)}{\ell^2}+R+\hat{\mu}\cS_{3,d}
+\hat{\lambda}\cS_{4,d}
-\frac{1}{4}F_{a b}F^{a b}
\right],\label{action0}
\eeqa
where the cosmological constant $\Lambda = -\frac{(d-1)(d-2)}{2\ell^2}$,
\beqa\label{Sd}
\mathcal{S}_{3, d} &=&
14 R_{a}{}^{e}{}_{c}{}^{f} R^{abcd} R_{bedf}+ 2 R^{ab} R_{a}{}^{cde} R_{bcde}- \frac{4 (66 - 35 d + 2 d^2) }{3 (d-2) (2 d-1)} R_{a}{}^{c} R^{ab} R_{bc}
\nonumber\\
&&-  \frac{2 (-30 + 9 d + 4 d^2) }{(d-2) (2 d-1)} R^{ab} R^{cd} R_{acbd}
-  \frac{(38 - 29 d + 4 d^2)}{4 (d -2) (2 d  - 1)} R R_{abcd} R^{abcd}
\nonumber\\
&&+ \frac{(34 - 21 d + 4 d^2) }{(d-2) ( 2 d - 1)} R_{ab} R^{ab} R -  \frac{(30 - 13 d + 4 d^2)}{12 (d-2) (2 d - 1)}  R^3,
\eeqa
and the quartic generalized quasi-topological term \cite{Ahmed:2017jod} is given at appendix \reef{app:lag_dens}.

The rescaled cubic coupling $\hat{\mu}$, and quartic coupling $\hat{\lambda}$, are given by
\begin{align}
\hat{\mu} &=  \frac{12(2d-1)(d-2)\; \mu}{(d-3)}  \, ,
\nonumber\\
\hat{\lambda} &= -{\frac { d\left( 3{d}^{3}-27{d}^{2}+73\,d-57 \right)  \lambda }{16
 \left( {d}^{5}-14{d}^{4}+79{d}^{3}-224{d}^{2}+316\,d-170
 \right)   }},
 \label{rescall}
\end{align}
where $\mu$ and $\lambda$ are arbitrary coupling constants, and rescaling is done to simplify the field equations.

As per our requirements for radially symmetric metrics, we employ the following ansatz
\beqa
\label{eqn:metricAnsatz}
ds^2&=& -N(r)^2f(r)dt^2+\frac{dr^2}{f(r)}+r^2d\Sigma^2_{(d-2),k},
\eeqa
and we find that the field equations of GQG yield $N(r)=constant$
\cite{Hennigar:2017ego}; we shall set $N(r)=1$ for simplicity\footnote{We note that the choice $N=1/\sqrt{f_{\infty}}$ has been used  \cite{Myers:2010ru} to normalize the speed of light on the boundary or to get $c=1$ in the dual CFT.  Here we shall set $N=1$, and note that by time reparametrization of the metric we can obtain $c=1$ on the boundary if desired.}. Here $d\Sigma^2_{(d-2),k}$ describes the $(d-2)$-dimensional line element of the transverse space, where $k=+1,0,-1$ stand for spherical, flat and hyperbolic geometries of a surface of constant scalar curvature. As an investigation of the $k=0$ case has previously been carried out \cite{Hennigar:2017umz}, we shall in the sequel consider only non-planar black holes.

For a maximally symmetric space, the metric ~\eqref{eqn:metricAnsatz} becomes,
\beqa\label{finf}
f_{\rm AdS}(r) = k+f_{\infty} \frac{r^2}{ \ell^2} \, ,
\eeqa
where $\ell$ is the AdS radius that is related to the cosmological constant.  The quantity $f_\infty = \lim_{r\to\infty} f(r) \ell^2/r^2$ and is obtained by solving following polynomial equation
\beqa\label{asympf}
1 - f_\infty +\frac{\mu}{\ell^4} (d-6)(4d^4 - 49 d^3 + 291 d^2 - 514 d + 184) f_\infty^3 + \frac{\lambda}{\ell^6} \frac{(d-8)}{3} f_\infty^4  = 0,
\eeqa
which is independent of the choice of  $k$ in the transverse section. While at least one coupling is non-zero, the higher curvature terms drive away $f_\infty$ from unity. Since we require the same asymptotics as AdS space we only pick positive real solutions of the above polynomial. The effective radius of the AdS space is given by $\ell_{\rm eff} = \ell/\sqrt{f_\infty}$.

In fact it turns out the negative of the derivative of eq.~\eqref{asympf} with respect to $f_\infty$ yields the prefactor of the linearized equations of motion~\cite{Hennigar:2017ego},
\beqa\label{PF}
P(f_\infty) =  1 -  3\frac{\mu}{\ell^4} (d-6)(4d^4 - 49 d^3 + 291 d^2 - 514 d + 184) f_\infty^2 - \frac{\lambda}{\ell^6} \frac{4(d-8)}{3} f_\infty^3
\eeqa
and to prevent the appearance of ghosts in the particle spectrum
 we require $P(f_\infty) > 0$.

For charged black holes, we include a Maxwell field  $F_{a b}=\partial_a A_b-\partial_b A_a$, with electromagnetic one form defined as
\beqa
A &=& q E(r) dt,
\eeqa
where, by inserting the above expression into the Maxwell equation, we get
\beqa
E(r) &=& \sqrt{\frac{2(d-2)}{(d-3)}}\frac{1}{r^{d-3}},
\eeqa
for the electric field. The specific choice of prefactor makes for greater simplification later on in the field equations;   we choose the constant term in the potential to be zero.

The field equation for the action \eqref{action0} yields the relation
\beqa
F&=&r^{d-3}\left(k-f(r)+\frac{r^2}{\ell^2}\right)+\mu F_{\cS_{3,d}}+\lambda F_{\cS_{4,d}}+r^{3-d}q^2 = m,
\label{Feq0}
\eeqa
where $m$ is a constant of integration and where \cite{Hennigar:2017ego}
\beqa
F_{\cS_{3,d}}&=&12  \Bigl[ (d^2+5d-15)\Bigl(
 \frac{4}{3}  r^{d-4} f'^3- 8 r^{d-5} f f'' \bigl(\frac{r f'}{2} + k - f \bigr) \label{eqn:fullEFE}\\
&& - 2 r^{d-5} ((d-4)f -2k) f'^2
 + 8(d-5) r^{d-6} ff'( f - k) \Bigr)
 -\frac{1}{3} (d-4) r^{d-7}(k-f)^2 \nonumber\\
 &&\times \Bigl( \bigl(-d^4 + \frac{57}{4} d^3 - \frac{261}{4} d^2 + 312 d - 489  \bigr)f + k\bigl( 129 - 192 d + \frac{357}{4} d^2 - \frac{57}{4} d^3 + d^4 \bigr) \Bigr)  \Bigr] \nonumber
\eeqa

and~\cite{Ahmed:2017jod}
\beqa
\label{eqn:GQT_eom}
F_{\cS_{4,d}} &=&   \left( k-f \right) \left[   \left( d-4
 \right) f \left( k-f \right)  f''+{f'}^{2} \left(  \left({d}^{2}-\frac{23}{2} d + 32
 \right) f- \frac{1}{2}\,k \left( d-4 \right)  \right)  \right] {r}^{d-7}
\nonumber\\
&&\left.+
  2\,f f' f'' \left(  \left( k-f \right)  \left( d-5 \right) {r}^{d-6}+
\frac{f'}{8} \left( 3d- 16 \right) {r}^{d-5} \right)
\right.\nonumber\\
&&\left. +f f'
 \left( k-f \right) ^{2} \left( d-4 \right)  \left( d-7 \right) {r}^{d-8}+\frac{f'^3}{12} \bigg[  \left(  \left( 3d- 16 \right) f-8 k
 \right)  \left( d-5 \right) {r}^{d-6}
 \right.\nonumber\\
&&\left.
- 3 \frac{f'}{4}\left( 3d-16
 \right) {r}^{d-5} \bigg] \, ,\right.
\eeqa
are respectively generated by the cubic  and quartic generalized quasi-topological terms.
The parameter $m$ has scaling dimension $[{\rm length}]^{d-3}$ and we will see that appears in the formula for the mass of black hole.

While exact solutions to the field equation seem hard to find (except in special cases~\cite{Feng:2017tev}), studying the far and near horizon behaviour of the metric perturbatively is still feasible, and permits us to analytically  obtain the thermodynamic quantities associated with black hole solutions. Specifically, we shall utilize   information from the near horizon expansion to describe the thermodynamics of the black holes.

\subsection{Degenerate Vacuum Solutions}

Before discussing solutions with nonzero mass and charge, we first consider solutions with
degenerate vacua.  These are analogous to the Lovelock-Unique-Vacuum (LUV) solutions
\cite{Wheeler:1985nh,Wheeler:1985qd,Kastor:2006vw,Arenas-Henriquez:2019rph}, and occur when the field equation \eqref{Feq0} has solutions of the form \eqref{finf} with multiple degenerate solutions for
$L = \ell/\sqrt{f_\infty}$ for $q=m=0$, or alternatively when \eqref{asympf} has multiple degenerate solutions.

It is straightforward to show that there is a one-parameter family of doubly-degenerate solutions to
\eqref{Feq0} for $q=m=0$, given by
\begin{equation}\label{dbleroot}
\lambda = \frac{3(3L^2-2\ell^2)L^6}{(d-8)\ell^2} \qquad   \mu = -\frac{(4L^2-3\ell^2)L^4}{\ell^2(d-6)(4d^4-49d^3+291d^2-514d+184)}
\end{equation}
valid for any $f_\infty$ and $\ell$.  There is also a triply-degenerate solution
\begin{equation}\label{trpleroot}
\lambda = - \frac{3 \ell^6}{16(d-8)} \qquad \mu = \frac{\ell^4}{4(d-6)(4d^4-49d^3+291d^2-514d+184)}
\end{equation}
with $f_\infty = 2$.   Note that for $d=8$ and $d=6$ these solutions do not exist.

There are no fully degenerate solutions  of the LUV type to \eqref{Feq0}.  This situation could presumably be altered if we were to add the quadratic Lovelock term (the Gauss-Bonnet term) to
\eqref{action0}, but we shall not pursue this here.

\subsection{Far region Solution}

In the asymptotic limit, the form of the metric function is
\beqa
f(r)_{asymp}=k+f_{\infty}\frac{r^2}{\ell^2} +\sum_{n=1}^{\infty}\frac{b_n}{r^n},\label{fexp}
\eeqa
Inserting the above expansion into eq. \reef{Feq0} and requiring that it be satisfied at each order in a $1/r$ expansion yields
\beqa
f(r)_{asymp}&=&f_{\infty}\frac{r^2}{\ell^2}+k
-\frac{m}{P(f_{\infty})r^{d-3}}
+\frac{q^2}{P(f_{\infty})r^{2d-6}}\label{asympsol}\\
&&\left.
+\frac{f_{\infty} m^2 }{\ell^4 [P(f_{\infty})]^3 r^{2 d-4}} \Big[\left(36 d^5-147 d^4+1179 d^3-5940 d^2+9444 d-3312\right) \right.\nonumber\\
&&\left.\times\ell^2 \mu+\left(-\frac{d^4}{2}+4 d^3-\frac{13 d^2}{2}+5 d-16\right) f_{\infty} \lambda\Big]+\frac{k m^2 }{\ell^2 [P(f_{\infty})]^3 r^{ 2 d-2}}\right.\nonumber\\
&&\left.\times \Big[24 (d-2) (d-1)^2 \left(d^2+5 d-15\right) \ell^2 \mu+\big(-\frac{d^4}{2}+5 d^3-\frac{29 d^2}{2}+16 d-6\big)\right.\nonumber\\
&&\left.\times f_{\infty} \lambda\Big]+\frac{f_{\infty} m q^2 }{\ell^4 [P(f_{\infty})]^3 r^{ 3 d-7}}\Big[\big(-216 d^5+342 d^4+2442 d^3-5064 d^2+1992 d\right.\nonumber\\
&&\left.-2016\big) \ell^2 \mu+\big(4 d^4-42 d^3+134 d^2-172 d+104\big) f_{\infty} \lambda\Big]+\frac{k m q^2  }{\ell^2 [P(f_{\infty})]^3 r^{ 3 d-5}} \right.\nonumber\\
&&\left.\times\Big[-96 (d-2) (d-1) (2 d-5) \left(d^2+5 d-15\right) \ell^2 \mu\right.\nonumber\\
&&\left.+\left(4 d^4-46 d^3+170 d^2-248 d+120\right) f_{\infty} \lambda\Big]\right.\nonumber\\
&&\left.+\cO\left(\frac{g_1(\mu,d )m^3}{[P(f_\infty)]^4 r^{3 d-5}},\frac{g_2(\lambda,d )f_{\infty}m^3}{\ell^2 [P(f_\infty)]^4 r^{3 d-5}},\frac{g_3(\mu, d) f_{\infty}q^4}{\ell^2 [P(f_\infty)]^3 r^{3 d-5}},\frac{g_4(\lambda, d) f_{\infty}^2q^4}{\ell^4 [P(f_\infty)]^3 r^{3 d-5}}\right),
\right.\nonumber
\eeqa
where $P(f_{\infty})$ was introduced in \reef{PF}. We have presented the six leading terms, and displayed the schematic structure of the next order corrections to $f(r)_{asymp}$. It is obvious that for $\mu\to 0$ and $\lambda\to 0$ that $P(f_\infty) \to 1$ and so
 \beqa
f^{Ein}(r)=k+f_{\infty}\frac{r^2}{\ell^2}-\frac{m}{r^{d-3}}
+\frac{q^2}{r^{2d-6}},
\eeqa
as expected from the solution in Einstein gravity.

 Degenerate (multiple) solutions of  \eqref{asympf}  will yield weaker falloff behaviour
for nonzero $m$ and $q$, as is the case with LUV-type solutions \cite{Arenas-Henriquez:2019rph}.
Setting $m=q=0$, $f = f_{ADS}$ turns out to be an exact  solution to the cubic theory
\cite{Feng:2017tev} and is also an exact solution to the field equations of the cubic-quartic theory \eqref{action0} provided \eqref{asympf} holds.   The thermodynamic behaviour
of this class of solutions is uninteresting (essentially it is the same as that of the BTZ black hole
\cite{Frassino:2015oca}). We shall briefly consider this situation in section \ref{vacua}, where we shall see it is connected with a maximal pressure.

 To find the asymptotic behaviour in the degenerate case, and for simplicity,
we consider the cubic and quartic parts of the action separately.
In both cases, only for doubly degenerate vacuum solutions (referred to as the critical state) in which
$P(f_{\infty})$ in \reef{PF} vanishes do we get nonvanishing coupling. For
higher orders of degeneracy  (where
 all derivatives of eq. \reef{asympf} are zero up to a specific order) we find vanishing coupling.

 For the cubic case, where  the associated critical coupling is given in \reef{muc0} and $f_{\infty}=3/2$, higher derivatives of $P(f_\infty)$ at this point are nonzero up to the third derivative of \reef{asympf}. Studying the asymptotic behaviour for degenerate vacua can be carried out by inserting
\beqa
f(r)=k+f_{\infty}\frac{r^2}{\ell^2}+\epsilon(r)
\eeqa
into the equation of motion \reef{Feq0} with vanishing quartic coupling for now. For the parameters at the critical state we get a differential equation for $\epsilon(r)$. To find an asymtotic solution to this equation we write
\beqa
\epsilon(r)=\frac{A}{r^x} +\cdots, \quad\quad   x\geq 0, \quad\quad  A\neq 0
\eeqa
where $\epsilon(r)$ is the correction to the  asymptotic AdS metric, and
$A$ and $x$ are determined upon insertion into \reef{Feq0}. Setting $q=0$ (for simplicity),
we find
\beqa
x=\frac{d-5}{2}, \quad \quad A^2=\frac{3 (d-6) \left(4 d^4-49 d^3+291 d^2-514 d+184\right) m}{2 \left(4 d^5-71 d^4+591 d^3-2308 d^2+3338 d-1134\right)   \ell^2}
\eeqa
showing (as before) that in six dimensions there is no  non-vanishing solution. For other dimensions there are two values for $A$ allowed (both signs of  the square root). Our result for $x$ is compatible with the result given in \cite{Arenas-Henriquez:2017xnr} in the case of double degeneracy.
Finding subleading terms requires including more terms in the initial expansion of $\epsilon(r)$.

For the quartic case   the critical state only exists at the quartic coupling given in \reef{muclac} and $f_{\infty}=4/3$, which yields two degenerate vacuum states while $P(f_{\infty})$ vanishes and up to forth derivative of eq. \reef{asympf} are nonzero at this critical value. Repeating a similar procedure for the cubic part, at leading order for large $r$ and $A\neq0$ we obtain
\beqa
x=\frac{d-5}{2}, \quad \quad A^2= -\frac{128 (d-8) m}{9 \left(d^3-8 d^2-3 d+122\right)   \ell^2}
\eeqa
where for $d=8$ there is no nonzero solution. For $d<8$   there are two solutions for $A$; for
$d>8$  there is no real value for $A$ unless $m<0$.  We obtain  the same falloff for the radial coordinate as in the cubic case, since it governs   the value of the degeneracy.

A more detailed study of `excitations' of the degenerate cases \eqref{dbleroot} and \eqref{trpleroot} with nonzero $m$ and $q$ are somewhat more complicated than the Lovelock case since \eqref{Feq0} is no longer a simple polynomial; we leave this for future investigation.

 The homogeneous solution in the far region is found by inserting $f(r)=f(r)_{asymp}+\epsilon f_h(r)$ in eq. \reef{Feq0}, where $\epsilon$ parameterizes the strength of these corrections. Substituting this expression in the field equation, we get an inhomogeneous second order differential equation for the function $f_h(r)$.
At leading order in $\epsilon$, assuming that $\mu\neq0$ and $\lambda\neq0$  the homogenous part of the equation at large $r$ becomes\footnote{Since the coefficients of $f_h''$ and $f_h'$ appearing in the differential equation are zero for vanishing $\mu$ and $\lambda$,  we recover in this limit the AdS black hole solution of Einstein gravity.}
\beqa
f_h''-\frac{4}{r}f_h'-\gamma^2 r^{d-3}f_h=0
\label{homog}
\eeqa
where
\beqa
\gamma^2&=& -\frac{\ell^4 [P(f_{\infty})]^2}{(d-1) f_{\infty} m \left(48 \left(d^2+5 d-15\right) \ell^2 \mu-(d-6) f_{\infty} \lambda\right)} \label{gamma2}
\eeqa
and we note that it is independent of the value of $k$; for vanishing $\mu$ and $\lambda$ it yields the well known AdS Reissner-Nordstrom ($RN$) solution.

For $d=6$ there is an ambiguity in the above expression for $\mu=0, \lambda\neq0$. For this particular case the applicable equation becomes
\beq
f_h''-\frac{9}{r}f_h'-\frac{2    \ell^2 [P(f_{\infty})]^3 }{25  \lambda f_{\infty} m^2}  r^{8}f_h=0,
\label{deq6d}
\eeq
with$P(f_{\infty})=1+\frac{8 f_{\infty}^3  \lambda}{3 \ell^6} $
and the explicit form of $\gamma^2$ can be read from \eqref{deq6d}.

The solution  of  \reef{homog}  in the case of $\gamma^2>0$ is
\beqa
f_{h+} = A r^{5/2} I_{\frac{5}{d-1}}\left(\frac{2 \gamma r^{\frac{d-1}{2}}}{d-1}\right)+B r^{5/2} K_{\frac{5}{d-1}}\left(\frac{2 \gamma r^{\frac{d-1}{2}}}{d-1}\right),
\label{bessel}
\eeqa
where $I$ and $K$ are the modified Bessel functions of the first and second kinds, and $A$ and $B$ are constants. In the limit of large $r$
\beq
f_{h+} \sim  A r^{5/2} \exp\left(\frac{2 \gamma r^{\frac{d-1}{2}}}{d-1}\right)+B r^{5/2}\exp \left(-\frac{2 \gamma r^{\frac{d-1}{2}}}{d-1}\right),\label{exp}
\eeq
and so we must set $A = 0$ to ensure the AdS boundary conditions are satisfied.  As a result no  ghost excitations can propagate to infinity.  We shall see shortly that   the contribution of the second term can be  dismissed.
The solution for the particular case  \eqref{deq6d} can be obtained in a similar fashion using its corresponding  value for $\gamma^2$.

Notice that $k$ does not appear in the asymptotic solution and the numerator of $\gamma^2$
in conjunction with the  positivity condition  \reef{PF} ensures freedom from ghosts \cite{Hennigar:2017ego,Ahmed:2017jod}.
Indeed, the positivity of the numerator relation \reef{gamma2} gives the same no-ghost condition as in \reef{PF}, and one only needs to check whether the denominator is positive as well.

To have the correct asymptotics we must choose  ($\mu,~ \lambda$) and the mass parameter in such a way that $\gamma^2 > 0$. To see this, note that if  $\gamma^2<0$ the homogenous solution asymptotically takes following form
\beqa
f_{h-} = C_1 r^{5/2} J_{\frac{5}{d-1}}\left(\frac{2 |\gamma| r^{\frac{d}{2}-\frac{1}{2}}}{d-1}\right)+C_2 r^{5/2} Y_{\frac{5}{d-1}}\left(\frac{2 |\gamma| r^{\frac{d}{2}-\frac{1}{2}}}{d-1}\right),
\eeqa
where $J$ and $Y$ are  respectively   Bessel functions of the first and second kind. In this situation, in any dimension the solution oscillates rapidly and its amplitude becomes larger than $\frac{r^2}{\ell^2}$ at large $r$. It therefore does not approach AdS asymptotically, and so we set $C_1 = C_2 = 0$ to get rid of this homogenous part of the solution.
For the rest of our considerations, to avoid any oscillating behaviour near infinity we restrict the solutions to the constraint $\gamma^2>0$.
Finally we note that the particular solution \reef{asympsol} polynomially decreases with $1/r$, and is the dominant part of  the total solution $f(r) = f_{h+} + f(r)_{asymp}$
for sufficiently large $r$; we therefore neglect the term $f_{h+}$ in eq. \reef{bessel} in the sequel.

\subsection{Near horizon solution}\label{nearsol}

To  construct the solution near the horizon  we consider the following expansion
\beqa\label{eqn:nh_ansatz}
f(r)=4\pi T (r-r_+)+\sum_{n=2}a_n (r-r_+)^n,
\eeqa
for the metric function, where $T$ is the Hawking temperature of the black hole. It is found by imposing the regularity condition for the Euclidean sector of the complex manifold (under $t\to i\tau$)
and reads as
\beqa
T=\frac{f'}{4 \pi} \, .
\eeqa
Substituting the near horizon expansion of the metric function into the field equation and imposing that it holds  at each order of $(r-r_+)$ we obtain for the zeroth and first order terms
\beqa
m &=&k r_+^{d-3}+\frac{r_+^{d-1}}{\ell^2}+\frac{q^2 }{ r_+^{d-3}}
-\frac{12(2 d-1)}{(d-3)}\mu\Bigl[   - \frac{ (d-3)}{2d-1}\label{MT01}\\
&&\left.\times \Bigl(-\frac{{k} ( d-4 ) (
129-192d+{\frac {357}{4}}\,{d}^{2}-{\frac {57}{4}}\,{d}^{3}+{d}^{4}
 ) r_+^{d-7}}{3}
\right.\nonumber\\
&&  \left.+ (d^2 + 5d - 15) \bigl(64 k \pi^2 r_+^{d-5}  + \frac{256}{3} \pi^3 T r_+^{d-4}\bigr) T^2 \Bigr)\Bigr]\right.\nonumber\\
&&\left.
+\lambda\Big[8 \pi ^2 (4-d)  k^2  T^2 r_+^{d-7}+\frac{128}{3} \pi ^3 (5-d)   k  T^3 r_+^{d-6}+16 \pi ^4 (16-3 d)    T^4 r_+^{d-5}\Big],\right.
\nonumber
\eeqa
\beqa
0&=&(d-3) k r_+^{d-4}+(d-1)\frac{ r_+^{d-2}}{\ell^2}-(d-3) \frac{q^2}{ r_+^{d-2}}-4 \pi  T r_+^{d-3}
+\frac{12\mu}{(d-3)} \nonumber\\
&&\left.\times\Bigl[
-\frac{{k}}{12}   \left( d-3 \right) \left( d-4 \right) \left( d-7 \right)
 \left( 516-768d+357{d}^{2}-57{d}^{3}+4{d}^{4} \right) r_+^{d-8}\right.\nonumber\\
&& \left.  -{\frac {128}{3}}{\pi }^{3} \left( d-4 \right)
 \left( d-3 \right)  \left( {d}^{2}+5\,d-15 \right) r_+^{d-5} T^3\right.\nonumber\\
&& \left.
 -64{\pi }^{2}  (d-3)(d-5) \left( {d}^{2}+5\,d-15 \right) kr_+^{d-6} T^2\right.\nonumber\\
&& \left.
  +  \left( d-3 \right)  \left( d-4 \right)\left( d-6 \right)  \pi \left( 4{d}^{3}- 33 {d}^{2}+127 d- 166 \right) {k}^{2}
  r_+^{d-7} T  \Bigr]\right.\nonumber\\
 &&\left.
+\lambda\Big[8 \pi ^2 (d-4)(d-7)  k^2  T^2 r_+^{d-8}+\frac{64}{3} \pi ^3 (d-5)(d-6)   k  T^3 r_+^{d-7}\right.
\nonumber\\
 &&\left.
+\frac{16}{3} \pi ^4 (d-5) (3 d-16)  T^4 r_+^{d-6}\Big],\right.
\label{MT02}
\eeqa
which specify the formula for the mass parameter $m$ and temperature $T$ in terms of the horizon radius and coupling constants. We shall use these equations for our thermodynamic investigation later on. Continuing to higher order terms one is able to find all other series coefficients in terms of $a_2$, which is a free parameter and its value is determined by using the boundary condition at infinity.

We pause to comment that  in a general non-linear
theory of gravity a spherically symmetric metric depends on two functions, and the mass and the temperature are determined by two parameters.  However
it is a special feature of the generalized quasi-topological theories (as well as the Lovelock
gravity theories) that both the mass   and the temperature are  determined in terms of
one parameter: the horizon length, as \eqref{MT01} and \eqref{MT02} indicate. This is an important feature of these theories, since it allows us to analyze thermodynamic behaviour without full knowledge of the solution \cite{Ahmed:2017jod,Bueno:2016lrh,Hennigar:2016gkm,Bueno:2018xqc,Bueno:2017sui}.

Now that we constructed the asymptotic and near horizon solutions, we next find a numerical solution that interpolates between them. For this purpose we define the rescaled metric function
\beqa
g(r)=\frac{\ell^2}{f_{\infty}r^2}f(r),\label{gf}
\eeqa
where $g(r)\rightarrow 1$ as $r\rightarrow \infty$. Here $f_{\infty}$ is a positive real root of eq. \reef{asympf}. Choosing some specific values for the coupling constants, electric charge and horizon radius, while $\ell=1$, we find the associated values of mass parameter and temperature referring to \eqref{MT01} and \eqref{MT02}.

To numerically solve the second order differential equation \eqref{Feq0} we need to identify initial values for the metric function $f$ and its first derivative. We use the value of $f$  close to the horizon to set up the seed solution for $g$. We then fix the value of $a_2$ to desired order, using the shooting method such that the numerical solution for $g$ approaches unity asymptotically.

As there are several branches of solutions, we select the one that tends to the Reissner-Nordstrom ($RN$) solution in the $\mu\to 0, ~\lambda\to 0$ limit;  otherwise we get other solutions that are not physically interesting.
Furthermore, because the differential equation is stiff, the solution can be obtained only to a certain precision. For our choice of $a_2$ the asymptotic solution up to $\cO(r^{-12})$ is precise to one part in 1,000 or better.

\begin{figure*}[htp]
\centering
\begin{tabular}{cc}
\includegraphics[scale=.3]{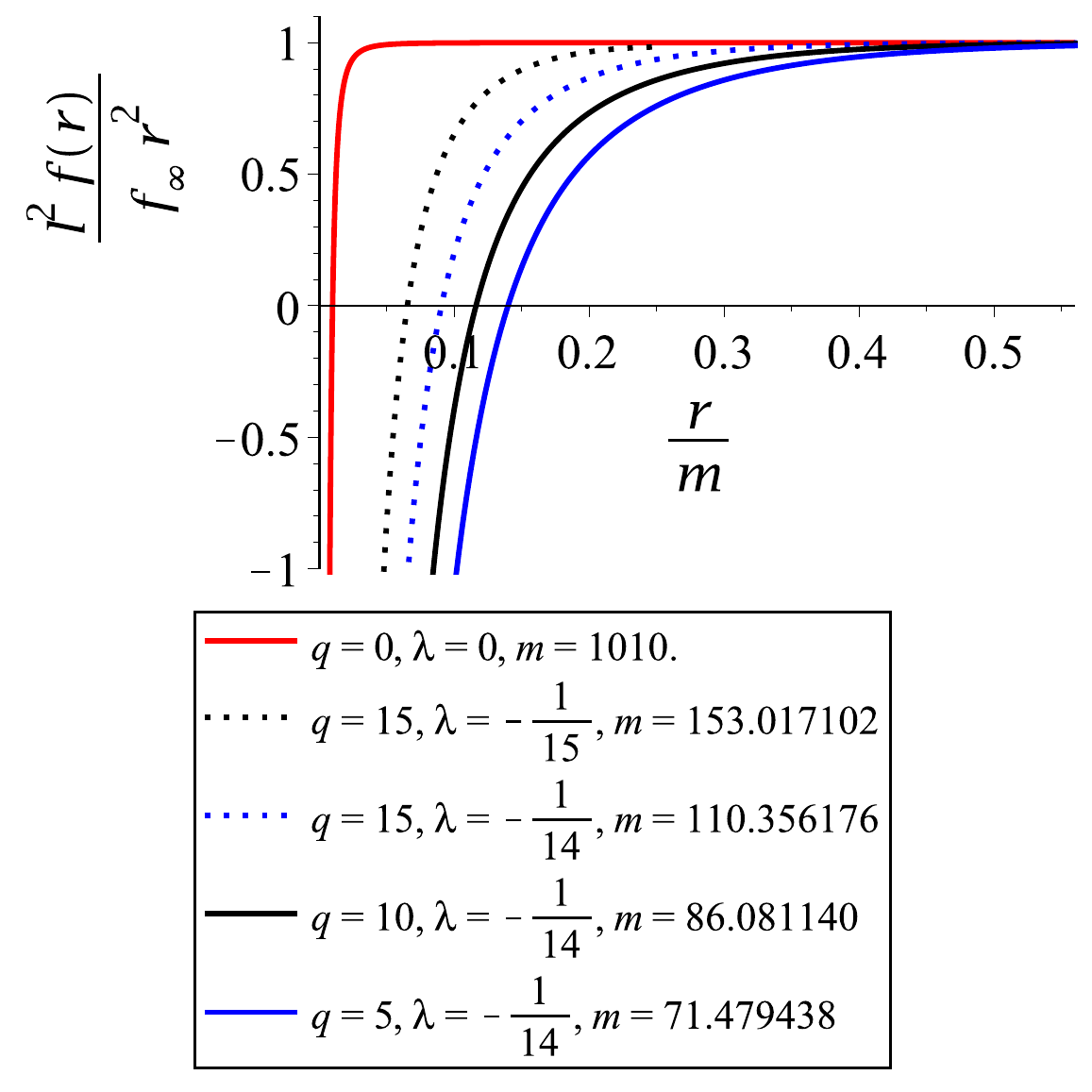}&\quad\quad\quad\quad
\includegraphics[scale=.3]{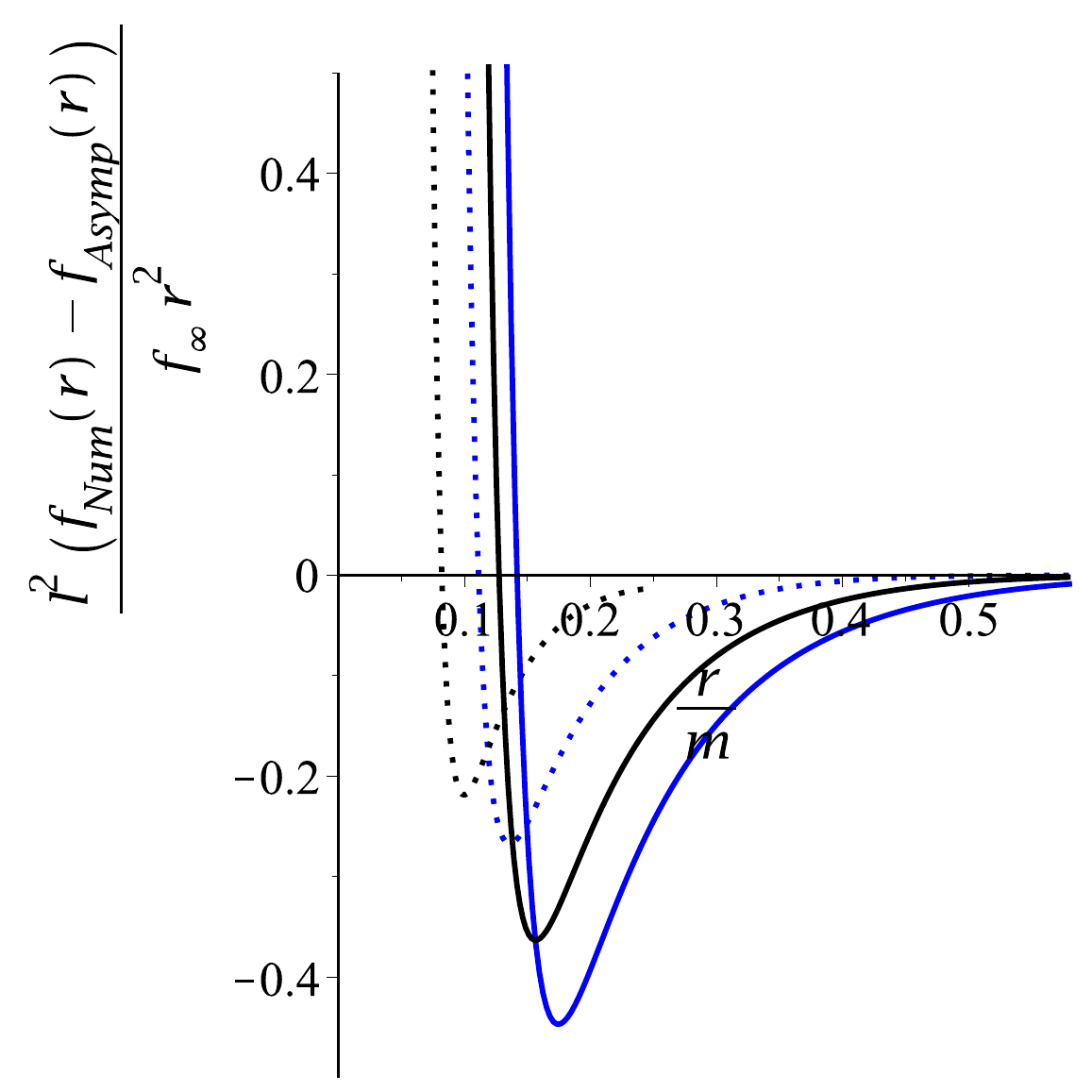}
\\
\includegraphics[scale=.3]{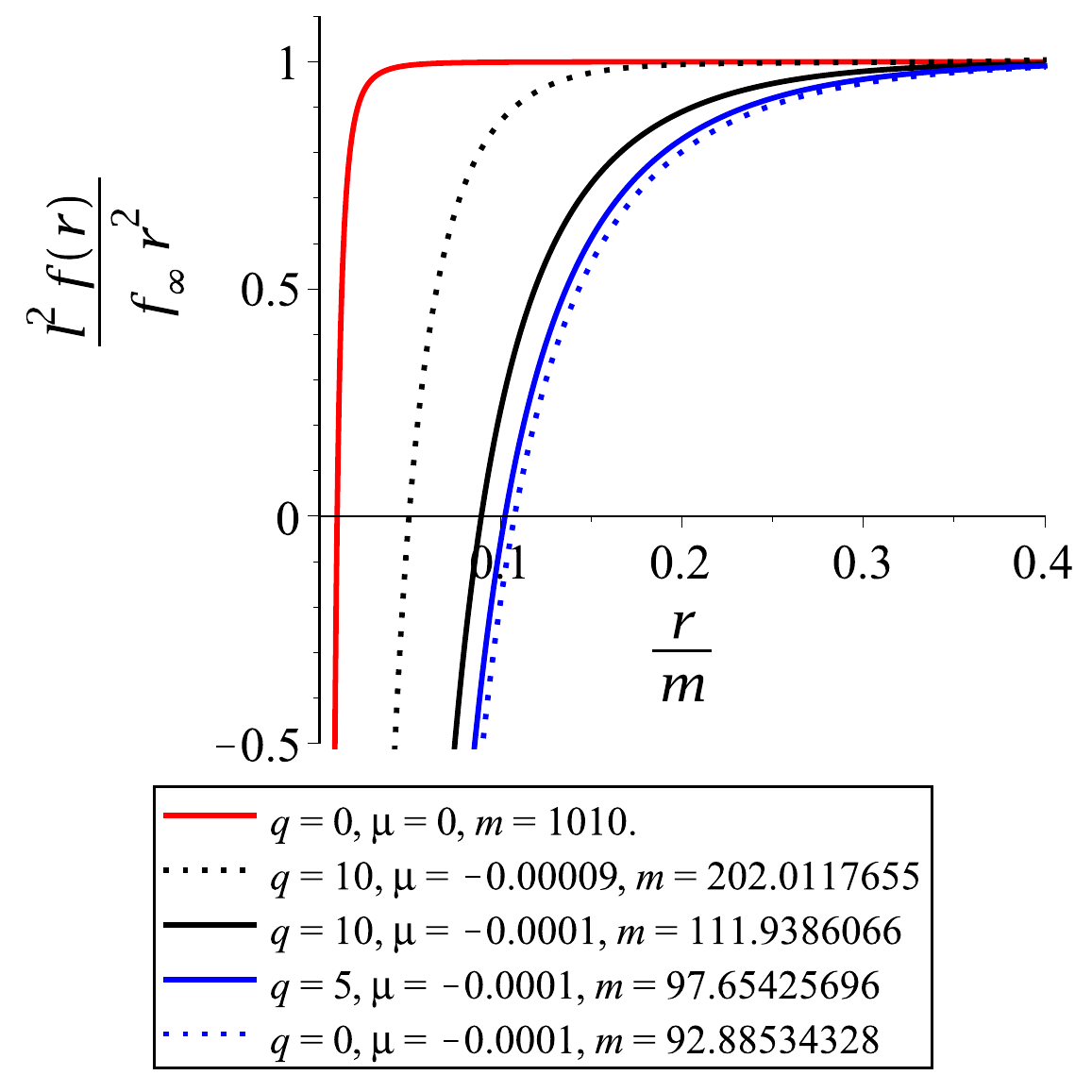}&\quad\quad\quad\quad
\includegraphics[scale=.3]{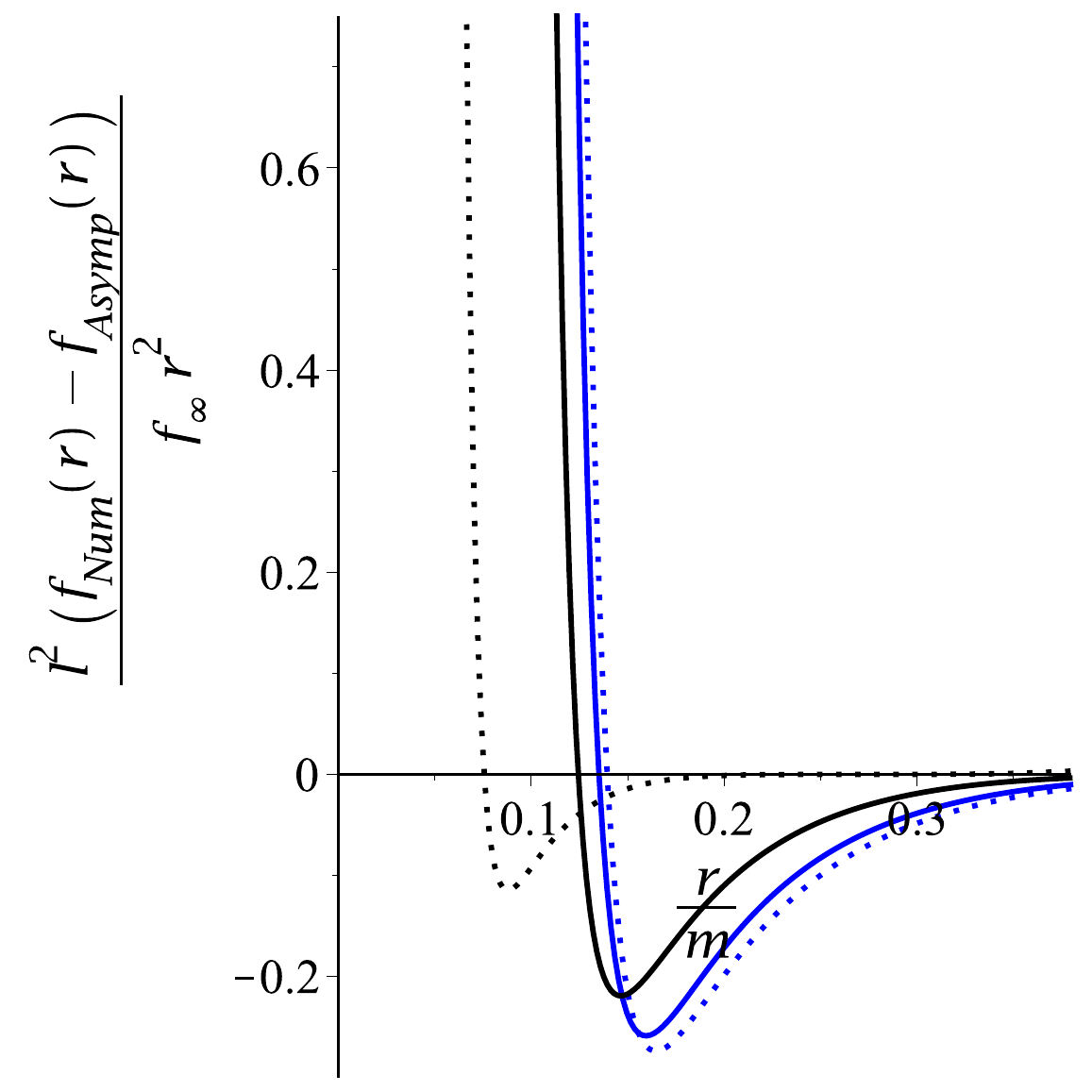}
\\
\end{tabular}
\caption{\textbf{Numerical solutions for cubic and quartic gravity} (color online).
\textit{Top left}: The plot presents the rescaled metric function \eqref{gf} in four dimensions for quartic gravity
(with $\mu=0$). The red line shows the solution for Einstein gravity (for which the quartic coupling $\lambda=0$) with zero  electric charge. For the other curves we choose different values for these parameters. \textit{Top right}: This panel depicts the difference between the numerical solution and its corresponding large-$r$ analytic solution in quartic gravity -- we see that convergence holds asymptotically.
\textit{Bottom left}: The plot presents the rescaled metric function \eqref{gf} in four dimensions for cubic gravity
(with $\lambda=0$); the red line is again the solution for Einstein gravity with zero charge.
 \textit{Bottom right}:  The difference between the numerical solution and its corresponding large-$r$ analytic solution
 in cubic gravity; we see again the asymptotic convergence.
 In all cases, we set $\ell=1$ and $r_+=10$. The mass parameter $m$ is defined in~\reef{MT01}.
}
\label{numeric0}
\end{figure*}

In order  to exhibit the behaviour of the solution while varying the cubic and quartic couplings individually,
we performed the computation for the cubic and quartic parts of the equation separately to see the impact of each of these terms individually.
Figure \ref{numeric0} illustrates  the numerical solution  in four dimensions and shows that at fixed quartic coupling, increasing charge drives the horizon inward, whereas at fixed electric charge,  larger values of $|\lambda|$ displace the event horizon outward\footnote{To better highlight the distinctions between the various cases we have plotted $f$ as a function of $r/m$ instead of $r/\ell$.}. The right panel demonstrates the difference between the numerical result for the metric function and  its corresponding asymptotic behaviour from \eqref{asympsol}; we see that these converge at enough large $r$. We also plotted the corresponding graphs for cubic gravity and find that similar behaviour takes place as charge and/or
the cubic coupling are varied.

To see the behaviour of the metric function in four dimensions as $r\rightarrow 0$, we expand the field equations in powers of $r$. In general we have the following expansion
\beqa
f(r)=r^s(a_0+r a_1+r^2 a_2+\cdots).
\eeqa

 Considering only the quartic curvature part of \eqref{Feq0}, for small $r$ the first term in the above expansion is the dominant  contribution to the metric function. To find $s$, we use the same numerical procedure described before, and depict the behaviour of $r f'(r)/f(r)$ near $r=0$. We find that in four dimensions $s$ vanishes. However in higher dimensions depending on the choice of parameters $s$ gains an non-integer negative value, except in six dimensions where it becomes positive for the choice of physical parameters as prescribed in the next section.
A similar argument was  given in \cite{mir:2018mmm} for the cubic part.  Therefore in four dimensions  the metric is regular at the origin.  However, the Kretschmann scalar $R_{a b c d}R^{a b c d}\sim r^{-4}$; the spacetime is still singular,
but its singularity is softer than its counterpart in Einstein gravity, in which $R_{a b c d}R^{a b c d}\sim r^{-6}$.

We also find that the associated plots for the five-dimensional metric function \eqref{gf} are similar, provided  the parameters are chosen to satisfy a physical constraint that we will discuss in the next section.

\section{Thermodynamic properties \label{sec:thermo}} \label{prop}

Our aim is to study the effects of including cubic and quartic generalized quasi-topological terms on the known behaviour Einstein black hole thermodynamics in four and higher dimensions.
We begin by investigating the first law and Smarr relation, where we apply the black hole chemistry formalism \cite{Kubiznak:2016qmn} by taking the cosmological constant, $\Lambda$ and the couplings $\mu, ~ \lambda$ as thermodynamic variables. We then study the physical constraints and find out whether at different domains in terms of the couplings and the charge they satisfy these constraints. We also elucidate the critical behaviour for these black holes.

\subsection{First law and Smarr relation}

As discussed in section \ref{nearsol}, equations \eqref{MT01} and \eqref{MT02} provide the relations for  obtaining the mass and temperature of the black holes without requiring knowledge of an exact solution. Since the explicit form for the temperature is complicated,  we apply the second equation implicitly to verify the first law of thermodynamics is satisfied.

We utilize the Iyer-Wald formalism~\cite{Wald:1993nt, Iyer:1994ys} to compute the entropy
\beqa
S = -2\pi \oint d^{d-2}x \sqrt{\gamma}E^{a b c d}\hat{\varepsilon}_{a b}\hat{\varepsilon}_{c d},
\eeqa
where
\beqa
E^{a b c d}=\frac{\partial \cL}{\partial R_{a b c d}},
\eeqa
and $\hat{\varepsilon}_{a b}$ is the binormal to the horizon, which is normalized as $\hat{\varepsilon}_{a b}\hat{\varepsilon}^{a b}=-2$. The induced metric on the horizon is $\gamma_{a b}$ and $\gamma=\textrm{det} \gamma_{a b}$.  From the action \reef{action0}  we find
\beqa
S&=&\frac{\Sigma_{(d-2),k}}{4} r_+^{d-2} \Big[1+\frac{48\mu}{r_+^4} (d-2)  \Big(8 \pi  \left(d^2+5 d-15\right) r_+ T (k+\pi  r_+ T)-\frac{1}{16} (d-4) \nonumber\\
&&\left.\times\left(4 d^3-33 d^2+127 d-166\right) k^2\Big)
-\frac{4 \pi   \lambda T}{r_+^5}(d-2) \Big((d-4) k^2+4 \pi  (d-5) k r_+ T\right.\nonumber\\
&&\left.
\qquad +\frac{4}{3} \pi ^2 (3 d-16) r_+^2 T^2\Big)\Big],\right.
\label{ENT}
\eeqa
for the entropy,  where $\Sigma_{(d-2),k}$ is the volume of the submanifold with line element $d\Sigma_{(d-2),k}$. For $k=1$ this is the volume of the $(d-2)$-dimensional sphere and finite, although when $k=0$ and $k=-1$ one needs to perform some kind of identification to define finite volume. Identifying the pressure  as
\cite{Kastor:2009wy,Kubiznak:2016qmn}
\beqa\label{press}
P =-\frac{\Lambda}{8\pi} = \frac{(d-1)(d-2)}{16 \pi \ell^2},
\eeqa
the other thermodynamic quantities are
\begin{align}
V &=\frac{\Sigma_{(d-2),k} r_+^{d-1}}{(d-1)} \, , \quad Q=\Sigma_{(d-2),k} \frac{\sqrt{2(d-2)(d-3)}}{16\pi}q \, ,\quad
\Phi =\sqrt{\frac{2(d-2)}{d-3}}\frac{q}{r_+^{d-3}} ,
\nonumber\\
\Psi_{\mu} =& -32(d-2)(d^2 + 5d - 15) \Sigma_{(d-2),k} \left(\pi^2 r_+^{d-4} T^3 + \frac{3}{2} \pi k T^2 r_+^{d-5} \right)
\nn\\
&+ \frac{(d-2)(d-4)\Sigma_{(d-2),k}}{4} \bigg[3\left(4d^3 - 33 d^2 + 127 d - 166 \right) k^2 T r_+^{d-6}
\nn\\
&\qquad\qquad - \left( 129 - 192 d + \frac{357}{4} d^2 - \frac{57}{4} d^3 + d^4 \right) \frac{k^3 r_+^{d-7}}{\pi} \bigg],
\nonumber\\
\Psi_{\lambda} &=\frac{\pi  (d-2)r_+^{d-7}\Sigma_{(d-2),k}}{6}    \left[3 (d-4) k^2 T^2+8 \pi  (d-5) k r_+ T^3+2 \pi ^2 (3 d-16) r_+^2 T^4\right]
\end{align}
and~\cite{Deser:2002jk} the mass is
\beqa
\label{eqn:adm_mass}
M =\frac{(d-2)\Sigma_{(d-2),k} m }{ 16 \pi }.  \eeqa

It is straightforward to show these quantities satisfy the extended first law of black hole thermodynamics,
\beqa
d M =T dS+V dP+\Phi dQ+\Psi_{\mu}d\mu+\Psi_{\lambda}d\lambda,
\eeqa
where $V$ is the thermodynamic volume conjugate to the pressure, and $\Psi_\mu,~\Psi_\lambda$ are the respective thermodynamic conjugates to the couplings $\mu,~\lambda$.  Furthermore,
a scaling argument \cite{Kastor:2011qp} applied to the various thermodynamic quantities above
yields the Smarr formula
\beqa
(d-3)M = (d-2)T S-2 P V+(d-3)\Phi Q+4 \mu \Psi_{\mu}+6 \lambda \Psi_{\lambda},
\eeqa
which can be shown to hold for these quantities.

To investigate the critical behaviour of these black holes, an equation of state is required.  This is obtained by substitution of $\ell^2$ in  \eqref{MT02} in terms of pressure.  Hence
\beqa\label{eos0}
P &=&\frac{T}{v}-\frac{(d-3)}{\pi  (d-2)} \frac{k}{v^2}+\frac{e^2}{v^{2 d-4}}
+(d-7) (d-4)\frac{\beta_0}{v^6}
-(d-6) (d-4)\beta_1\frac{  T}{v^5} \\
&&\left.
+\left((d-5)\frac{\beta_2}{v^4}
-(d-4) \frac{\alpha_2}{v^6}\right)T^2
+\left((d-4)\frac{\beta_3}{v^3}
-(d-5)\frac{\alpha_3}{v^5}\right)T^3
- (d-5) \alpha_4 \frac{T^4}{v^4},\right.\nonumber
\eeqa
where    the different parameters are
\beqa
v&=&\frac{4 r_+}{(d-2)}, \quad \quad
e^2=\frac{16^{d-3}}{\pi } (d-3) (d-2)^{5-2 d} q^2\nonumber\\
\alpha_2&=&\frac{2^{11} \pi  (d-7) k^2 }{(d-2)^5}\lambda, \quad
\alpha_3=\frac{2^{12} \pi ^2 (d-6) k }{3 (d-2)^4}\lambda, \quad
\alpha_4=\frac{2^8 \pi ^3 (3 d-16) }{3 (d-2)^3}\lambda,
\nonumber\\
\beta_0&=&\frac{2^8 \left(4 d^4-57 d^3+357 d^2-768 d+516\right) k }{\pi  (d-2)^5}\mu, \quad \beta_2=\frac{3\times 2^{12} \pi  \left(d^2+5 d-15\right) k }{(d-2)^3}\mu,\nonumber\\
\beta_1&=&\frac{3\times 2^8 \left(4 d^3-33 d^2+127 d-166\right) k^2 }{(d-2)^4}\mu,
 \quad
\beta_3=\frac{2^{11} \pi ^2 \left(d^2+5 d-15\right) }{(d-2)^2}\mu,
\label{rescaled}
\eeqa
where $v$ is the specific volume \cite{GunasekaranEtal:2012}.  As in previous studies  \cite{Hennigar:2016gkm, Hennigar:2017umz}, we see from \eqref{eos0} that there is a non-linear dependence of the equation of state on the temperature.
For future reference we choose the free parameters to be $e$, $\beta_3$ and $\alpha_4$; these are  independent of the choice of $k$.

The explicit form of the Gibbs free energy as $G=M-T S$ is
\beqa
\cG &=&\left[\frac{4}{d-2}\right]^{d-1}\frac{G}{\Sigma_{(d-2),k} } =\frac{ v^{d-1}P}{d-1} +\frac{ v^{d-3}k}{\pi  (d-2)}+\frac{e^2 }{(d-3)v^{d-3}} -\beta_0 (d-4) v^{d-7}\nonumber\\
&&\left.+ \left(-\frac{v^{d-2}}{d-2}+\beta_1 (d-4) v^{d-6}\right)T+ \left(\alpha_3\frac{ (d-5) v^{d-6}}{d-6}-\beta_3 v^{d-4}\right)T^3+\alpha_4  v^{d-5}T^4\right.\nonumber\\
&&\left.+ \left(\alpha_2\frac{ (d-4) v^{d-7}}{d-7}-\beta_0\frac{48 \pi ^2  (d-2)^2 \left(d^2+5 d-15\right) v^{d-5}}{4 d^4-57 d^3+357 d^2-768 d+516}\right) T^2,\right.
\eeqa
where we pulled out an overall positive factor  to simplify the expression; the explicit form of the other parameters is given in eq. \reef{rescaled}.   The equilibrium state is the one that minimizes the Gibbs free energy $\cG$ for fixed temperature and pressure.

\subsection{Physical constraints}\label{constraints}

We now explicate  the constraints on the cubic and quartic couplings required for physical solutions.

Generalized quasi-topological theories have the property that only the massless graviton propagates on
constant curvature backgrounds  provided the parameters are appropriately constrained.  To ensure this, the effective Newton constant of gravity must have the same sign as that in Einstein gravity. This implies that the pre-factor in the linearized equations of motion about the AdS solution is positive \cite{Hennigar:2017ego}, $\ie$ $P(f_{\infty})>0$ with $P(f_{\infty})$ defined in eq. \eqref{PF} and the value of $f_{\infty}$ is given by solution of eq. \eqref{asympf} which is positive in order to get an asymptotic AdS solution. The same relation occurs if we require $\gamma^2>0$ (see the discussion after \eqref{exp}).

 In terms of the rescaled parameters in eq. \reef{rescaled} and the pressure given
 in \eqref{press},  the no-ghost constraint  \eqref{PF}  becomes
 \beqa
1-\frac{3  (d-6) \left(4 d^4-49 d^3+291 d^2-514 d+184\right) f_{\infty}^2  P^2 \beta_3}{8 (d-1)^2 \left(d^2+5 d-15\right)} -\frac{64 (d-8) f_{\infty}^3    P^3 \alpha_4}{(d-1)^3 (3 d-16)} >0
\qquad
\label{nghost}
\eeqa
and we note that in the limit $\beta_3\rightarrow 0,~\alpha_4\rightarrow 0$ (or $\mu\rightarrow 0,~\lambda\rightarrow 0$) that  we reach the Einstein branch of the theory.

Disregarding solutions with $\gamma^2<0$ (since they are not asymptotically AdS)  we have
from \eqref{gamma2}
\beqa
  \frac{8 \pi  (d-1) (3 d-16) [P(f_{\infty})]^2}{3 (d-2) f_{\infty} m P \left(8 \alpha_4 (d-6) f_{\infty} P-\beta_3 \left(3 d^2-19 d+16\right)\right)} >0,\label{gamma21}
\eeqa
upon using eq. \reef{rescaled},  where $P(f_{\infty})$ is given in \reef{nghost}.
It is well-known that in higher curvature gravity  black hole entropy can be negative in some regions of  parameter space, perhaps indicative of an occurrence of instability~\cite{Cvetic:2001bk, Nojiri:2001pm}.
Imposing the requirement for positive black hole entropy  yields
\beqa
S>0 \Rightarrow && 1+(d-2)\Big[-\frac{ (d-4) {\beta_1}}{v^4} +\left(\frac{2  {\beta_2}}{v^3}-\frac{2  (d-4) {\alpha_2}}{(d-7) v^5}\right)T \nonumber\\
&&\qquad\qquad\qquad \left.+ \left(\frac{3  {\beta_3}}{v^2}-\frac{3 (d-5) {\alpha_3}}{(d-6) v^4}\right)T^2-\frac{4   {\alpha_4}T^3}{v^3}\Big] >0\right. \, .\label{sratio}
\eeqa
When temperature and specific volume are positive, in each dimension the values of couplings must be chosen to satisfy the above inequality.

We search for the domains in parameter space where these conditions are valid for various   charge and coupling constants  in the next section.

\section{Thermodynamics in the canonical ensemble} \label{sec: thermoce}

Equipped with the field equations and  relevant thermodynamic relations, we  consider  first the fixed charge ensemble. We aim to investigate the phase structure and critical points for these black holes.

The equation of state  in terms of the rescaled parameters $e,~\beta_3$ and $\alpha_4$ introduced in eq. \reef{rescaled} is
\beqa
P &=&\frac{T}{v}-\frac{(d-3) k}{\pi  (d-2) v^2}+\frac{e^2}{ v^{2 d-4}}+\frac{(d-7) (d-4) \left(4 d^4-57 d^3+357 d^2-768 d+516\right) k {\beta_3} }{8 \pi ^3 (d-2)^3 \left(d^2+5 d-15\right) v^6}\nonumber\\
&&\qquad\left.+ \left(-\frac{3  (d-6) (d-4) \left(4 d^3-33 d^2+127 d-166\right) k^2 {\beta_3}}{8 \pi ^2 (d-2)^2 \left(d^2+5 d-15\right) v^5}\right)T\right.\nonumber\\
&&\qquad\qquad\left.+ \Big(\frac{6  (d-5) k {\beta_3}}{\pi  (d-2) v^4}-\frac{24  (d-7) (d-4) k^2 {\alpha_4}}{\pi ^2 (d-2)^2 (3 d-16) v^6}\Big)T^2\right.\nonumber\\
&&\qquad\qquad\qquad\left.+ \left(\frac{ (d-4){\beta_3}}{v^3}-\frac{16  (d-6) (d-5) k {\alpha_4}}{\pi  (d-2) (3 d-16) v^5}\right)T^3-\frac{ (d-5){\alpha_4} }{v^4}T^4,\right.
\eeqa
for any $d \ge 4$; we note that only $e^2$ appears everywhere, so our results are valid for both positive and negative charge.

Phase transitions occur if the equation of state demonstrates some oscillatory behaviour, with $P(v)$  having at least one minimum and one maximum. This in turn depends on the signs of the coefficients of different powers of $v$, as these determine how many roots exist in the equations
for critical volume and temperature. To get a critical point, the following equations must hold
\beqa\label{criteqs}
\frac{\partial P}{\partial v}=\frac{\partial^2 P}{\partial v^2}=0\label{dPd2p} \, .
\eeqa

We also find that when $\mu=0$  in four and five dimensions $\lambda$ must be negative, whereas  in higher  dimensions $\lambda$ must be positive in order to get physical points that satisfy all physical constraints mentioned in the previous section.  To get the critical volume and temperature in terms of charge and couplings, we solve equation \label{crit2} in various dimensions. As the explicit form is lengthy, we do not present results explicitly; in practice it is easier to solve the equations parametrically for $T$ and $v_c$ in terms of the other parameters  in  certain dimensions.

\subsection{ $AdS_4$ vacua and maximum pressure}\label{vacua}

 We consider here the structure of the $AdS_4$ vacua of \reef{action0} in four dimensional spacetime, with curvature scale $1/\ell_{\textrm{eff}}^2=f_{\infty}/\ell^2$.  Setting the action length scale to $\ell=1$ (implying a fixed pressure of $3/(8\pi)$), we analyze solutions to \reef{asympf}, considering the cubic and quartic couplings in
 distinction for simplicity.

\begin{figure*}[htp]
\centering
\begin{tabular}{cc}
\includegraphics[scale=.35]{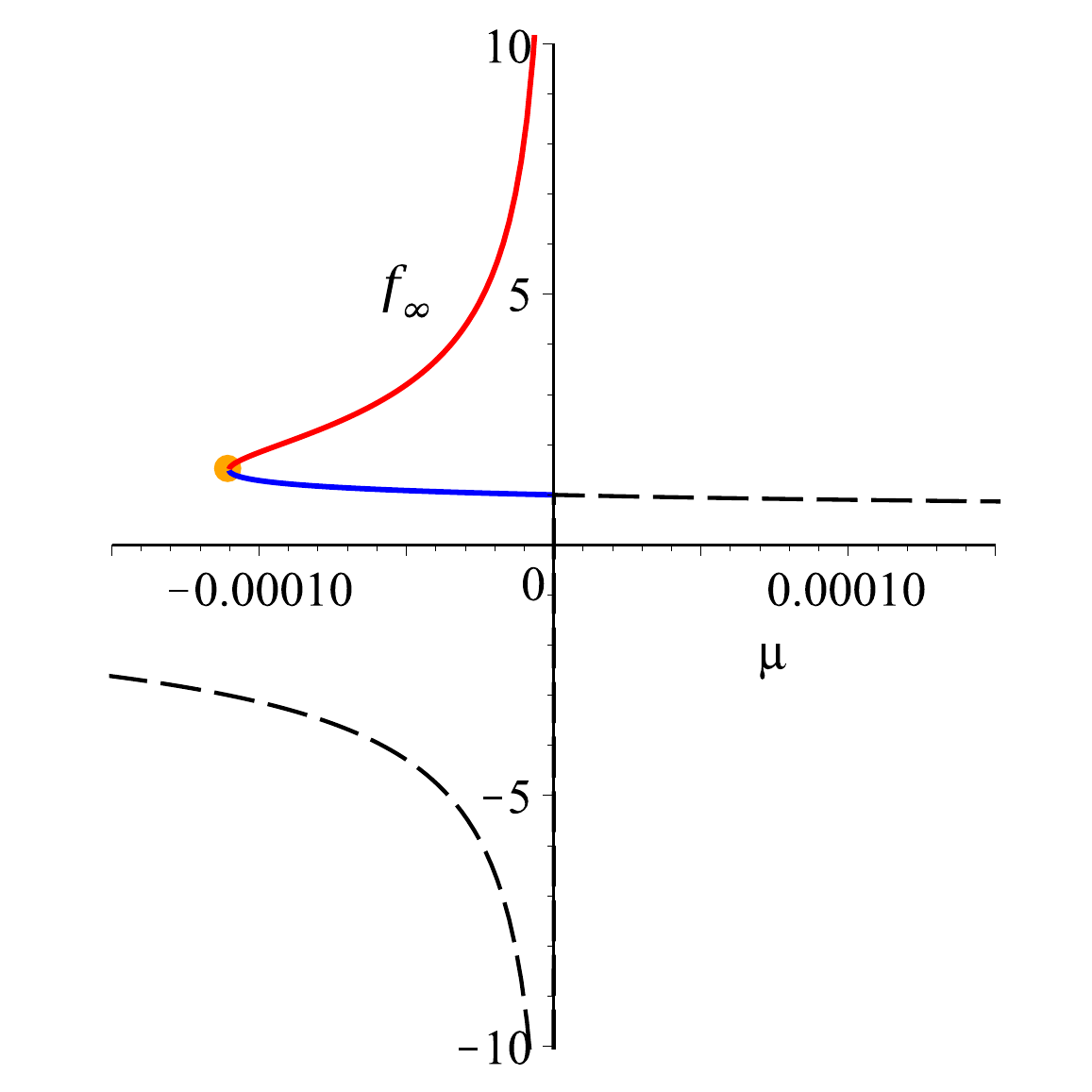}&\quad\quad
\includegraphics[scale=.35]{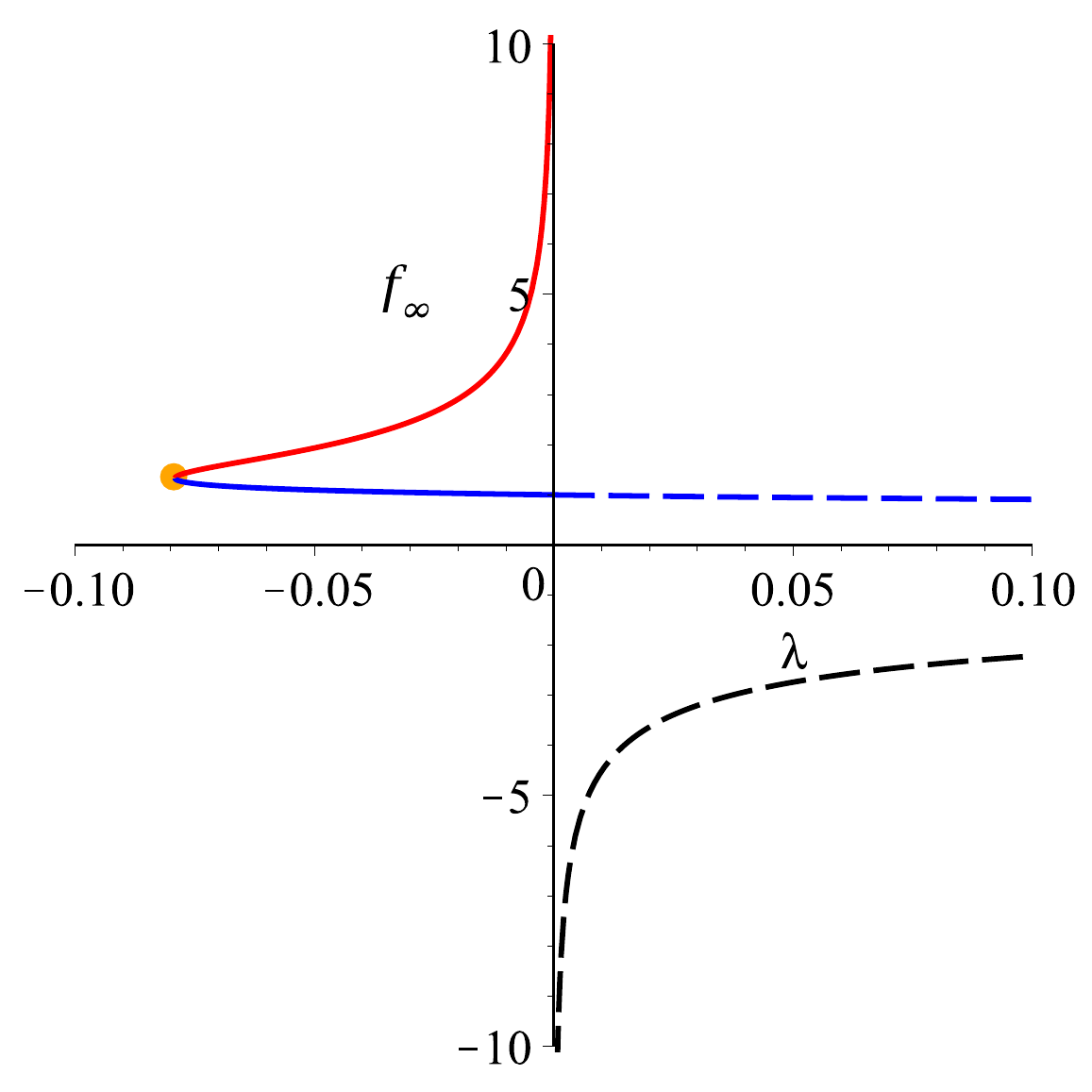}
\\
\end{tabular}
\caption{\textbf{Real possible values for $f_{\infty}$ in terms of coupling in four dimensions} (color online).
 \textit{Left}: The plot presents the real solutions to the equation \reef{eqfinf4d} for different values of $\mu$. The lower dashed black line corresponds to $f_{\infty}<0$, upper dashed black one depicts solutions with $\gamma^2<0$, and so is unstable.
 The red line, relates to existence of ghosts. Only blue line is associated with a stable branch of solutions. The orange dot corresponds to the critical case $\mu_c=-\ell^4/9072$.
 \textit{Right}: The plot denotes the real solutions to the equation \reef{eqfinf4dq} versus $\lambda$. The lower dashed black curve corresponds to $f_{\infty}<0$, upper dashed blue line depicts unstable solutions with $\gamma^2<0$.
 The red line shows appearance of a ghost solution. Blue solid line corresponds to stable branch of solutions. The orange dot denotes the critical case $\lambda_c=-81\ell^6/1024$.
 In both plots we have set $\ell=1$.
}
\label{finfplot}
\end{figure*}

 Starting with only a non-zero cubic coupling $\mu$, we have
\beqa
1-f_{\infty}-\frac{1344  \mu}{\ell^4}f_{\infty}^3 =0,\label{eqfinf4d}
\eeqa
and illustrate in the left graph of  figure \ref{finfplot}  three possible branches of real solutions to this equation, depicted in different colours.
An analysis of the discriminant of \reef{eqfinf4d} indicates that  there is a critical value  $\mu_c=-\ell^4/9072$ of the coupling where the discriminant $\Delta$ vanishes. For $\mu_c<\mu<0$,   $\Delta>0$ and there are two positive real solutions for $f_{\infty}$;  however only the smaller branch is free of ghosts  $\ie$ $P(f_{\infty})>0$ in \reef{PF}. Conversely $\Delta<0$ for both   $\mu<\mu_c$ and $\mu>0$; the first of these yields a negative real valued solution,  and the second implies $\gamma^2<0$ in   \reef{gamma2}. Both regions are unphysical and we exclude them from further analysis.  Note that from
the linearized equations of Einstein gravity,
the following relation
\beqa
G_{\textrm{eff}}=\frac{G}{1+\frac{4032}{\ell^4}\mu f_{\infty}^2},
\eeqa
holds between the effective Newton constant $G_{\textrm{eff}}$  and $G$.
We see that for $f_{\infty}^2=-\frac{\ell^4}{4032\mu}$, $G_{\textrm{eff}}\rightarrow \infty$,
and inserting this value inside eq.~\reef{eqfinf4d} yields the critical coupling of $\mu_c$
(noted previously  \cite{Feng:2017tev}).  At this  point  the discriminant of the cubic changes  sign and there are distinct branches of solutions for values of $\mu < \mu_c$ and $\mu > \mu_c$.

Repeating the same approach for higher dimensions, the overall behaviour of $f_{\infty}$ given in figure \ref{finfplot} is similar to  that in four dimensions. The critical limit is given by
\beqa
\mu_c&=&\frac{4 \ell^4}{27 \left(4 d^5-73 d^4+585 d^3-2260 d^2+3268 d-1104\right)}\quad \quad \textrm{if}\ \lambda=0.\label{muc0}
\eeqa

More generally we  can reconsider the above discussion for arbitrary values of the pressure $P$.  In four and five dimensions there is a maximum value for the pressure that results from the condition that the discriminant $\Delta>0$, which is a constraint on parameter space for physically acceptable solutions. In general we have
\beqa
P_{\textrm{max}}&=&\frac{4}{3} \sqrt{\frac{2}{3}} \sqrt{\frac{(d-1)^2 (d^2+5 d-15)}{  ( 4 d^4-49 d^3+291 d^2-514 d+184)(6-d)\beta_3}} \quad \quad \textrm{if}\ \alpha_4=0,
\eeqa
where in four and five dimensions $\beta_3>0$ and is given in \reef{rescaled}.  For $d=6$  the pressure is unbounded and for $d\geq 7$ a similar procedure does not yield an upper bound for the pressure.

Turning now to the  quartic case,  equation \reef{asympf} becomes
\beqa
1-f_{\infty}-\frac{4\lambda}{3\ell^6} f_{\infty}^4=0,
\label{eqfinf4dq}
\eeqa
in $d=4$ with $\ell=1$.  The  right graph in figure \ref{finfplot} indicates three possible real solutions to the above equation. The discriminant of   \reef{eqfinf4dq}   vanishes for $\lambda = \lambda_c=-81\ell^6/1024$. Again, we note that $G_{\textrm{eff}}$ becomes infinity or equivalently $P(f_{\infty})$ vanishes for $f_{\infty}^3=-3\ell^6/(16 \lambda)$ and that \eqref{eqfinf4dq} in turn implies $\lambda=\lambda_c$.  Similar to the previous case, for
 $\lambda_c<\lambda<0$,  we have $\Delta<0$ and there are two positive real solutions for $f_{\infty}$. Only the smaller of these  has positive $P(f_{\infty})$ and $\gamma^2$.  For  $\lambda<\lambda_c$,  $\Delta>0$ and there are no real solutions for $f_{\infty}$; for $\lambda>0$, although there is one real positive solution to \reef{eqfinf4dq}, it implies $\gamma^2<0$ and is therefore physically inadmissible.

Requiring  $\lambda_c<\lambda<0$ we conclude that there is a maximum value for the pressure given by
\beqa
P_{\textrm{max}}&=&\frac{3}{16} \sqrt[3]{\frac{(d-1)^3 (3 d-16)}{ (d-8)\alpha_4}}\quad \quad \textrm{if}\ \beta_3=0,\label{PMAX}
\eeqa
where $\alpha_4$ is given in \reef{rescaled}  and
\beqa
\lambda_c&=&\frac{81}{256}\frac{\ell^6}{d-8}\quad \quad \textrm{if}\ \mu=0,\label{muclac}
\eeqa
 Only for $d=4,5$ is $\alpha_4>0$ and $P_{\textrm{max}} > 0$, yielding an upper bound on the pressure; in higher dimensions there is no bound on the pressure.

In what follows, we concentrate on several specific dimensions and investigate the thermodynamic behaviour in some detail. In the figures \ref{domain}, \ref{domain5d}, \ref{domain6d} given in the following sections
we will adhere to the colour coding explained in table \ref{table:nonlin}.
\subsection{Critical behaviour in four dimensions}

Our next task is to determine how many of these possible critical points are actually physical, and study their critical behaviour.  We proceed by examining each value of $d$ in succession.

Preceding studies have shown that critical points exist for four dimensional charged black holes in Einstein gravity ($\beta_3=0,~\alpha_4=0$) \cite{Kubiznak:2012wp} and in ECG \cite{Hennigar:2016gkm}.  A recent study was carried out for cubic GQG (for which $\alpha_4=0$) in $d$ dimensions \cite{mir:2018mmm}.

Here we consider the effects of both cubic and quartic GQG in $d=4$.
The equation of state~\eqref{eos0} becomes
\beqa
P &=&\frac{T}{v}-\frac{k}{2 \pi  v^2}+\frac{e^2}{v^4}-\frac{3 \beta_3 k T^2}{\pi  v^4}+\frac{\alpha_4 T^4}{v^4}+\frac{4 \alpha_4 k T^3}{\pi  v^5}.
\eeqa
It is obvious that for small $v$ (i.e. for small black holes) that the term cubic in $T$   (coming from quartic GQG)  dominates.
 By taking different linear combinations of \eqref{criteqs} it is possible to obtain an equation linear in $T$;   the resultant critical temperature  and volume are then easily
seen to satisfy the equations
\beqa
T_c&=&\frac{1}{18}\Big(8000 \pi ^2 \alpha_4 e^4 k^2-12000 \pi  \alpha_4 e^2 k v_c^2+ \left(4500 \alpha_4 k^2-3600 \pi^2 \beta_3 e^2 k^2\right)v_c^4\nonumber\\
&&\left.+540 \pi  \beta_3 k v_c^6+180 \pi^4 e^2 v_c^8-27 \pi^3 k v_c^{10}\Big)
\Big/\Big( v_c^3 \big(-400 \pi^2 \alpha_4 e^2 k^2+300 \pi  \alpha_4 k v_c^2\right.\nonumber\\
&&\left.-400 \beta_3^2 k^2+40 \pi^2 \beta_3 k^2 v_c^4-\pi^4 v_c^8\big) \Big),
\right.
\label{Tc4}
\eeqa
and
\beqa
 \left(3 \pi ^2 v_c^4-60 \beta_3 k^2\right)T_c^2+18 \pi  k  v_c^3T_c+20 \pi  e^2 k-15 k^2 v_c^2  = 0,
 \label{Vc4}
\eeqa
which can be solved numerically for any choice of parameters for $T_c,~v_c$.

For simplicity, consider  the behaviour of the critical temperature and volume, with only the quartic coupling active. 
Equations \eqref {Tc4} and  \eqref {Vc4} become
\beqa
t_c &=& \frac{8000 \pi^2 \alpha_4 e^4 k^2-12000 \pi  \alpha_4 e^2 k  v_c^2+4500 \alpha_4 k^2 v_c^4+180 \pi^4 e^2 v_c^8-27 \pi^3 k  v_c^{10}}{18 \pi  v_c^3 \left(-400 \pi  \alpha_4 e^2 k^2+300 \alpha_4 k v_c^2-\pi^3 v_c^8\right)  } \quad
 \label{Tc4Q}
 \eeqa
with $t_c \equiv {T_c}|_{\beta_3\rightarrow 0}$, and the critical volume $ v_c$ satisfies
\beqa
20 \pi  e^2 k-15 k^2 v_c^2+18 \pi  k t_c  v_c^3+3 \pi^2 t_c^2 v_c^4 &=&0.
 \label{Vc4Q}
\eeqa
We see that critical temperature is singular if the critical volume is such that the polynomial quartic in
$v_c^2$ in the denominator vanishes.  An exception to this is if $\alpha_4= 6400/729 \pi^6 e^6 $:  the numerator in \eqref{Tc4Q} also vanishes and $t_c$ remains finite.  Note that is only occurs if  $k\neq 0$, in accord with earlier work  on  black branes in GQG \cite{Hennigar:2017umz}. However the corresponding $T_c$ and $v_c$ become imaginary in the case of $k=-1$, so  for hyperbolic black holes this singularity is absent.

 For $k=1$, if $\beta_3=0$ and $\alpha_4= 6400/729 \pi^6 e^6 $, the resulting values of $T_c,~v_c$
 correspond to a   critical point with standard critical  exponents   (see \reef{exponents} and the discussion following), and the  phase transition in the vicinity of this point is a standard first order VdW transition  similar to what is depicted in figure \ref{PT4d}.

In studying the behaviour at this point, we note that the equations of state need to be solved for these specific  parameter values,  instead of using  \reef{Tc4Q} and \reef{Vc4Q}, since the latter becomes invalid if the denominator of $t_c$ vanishes.
If  both couplings are non-zero, again the denominator of $T_c$ in \eqref{Tc4} is quartic in $v^2$, and a similar procedure can be employed to write a formula for $\alpha_4$ in terms of $\beta_3$ and $e$.  Doing so, we find   that for any values of the  parameters the only solution is $\alpha_4=0$, however in cubic gravity critical temperature does not have a singularity in four dimensions \cite{mir:2018mmm}.  In other words, this particular occurrence of  this apparent thermodynamic singularity is obtained only in
Einstein-quartic GQG (for which the  cubic coupling vanishes).

\begin{table}[ht]
\centering 
\begin{tabular}{c c c c c c} 
\hline 
Color & Number of Critical Points & $\gamma^2$ & Entropy & $f_{\infty}$ \\ [0.5ex] 
\hline 
\tcg{Green} & 1 & + & + & + \\ 
\tcfg{Dark Green} & 2 & + & + & + \\
\tco{Orange} & 3 & + & + & + \\
\tcc{Blue} & 1 & $-$ & + & + \\
\tcb{Dark Blue} & 2 & $-$ & + & + \\
\tcp{Purple} & 3 & $-$ & + & +\\
\tclb{Brown} & 1 & + & $-$ & +\\
\tcr{Red} & 2 & + & $-$ & +\\
Black & 1 & $-$ & $-$ & +\\
\tcy{Yellow} & 2 & $-$ & $-$ & +\\
\tcbr{Light Brown} & 1 & $\times$ & $+/-$ & $\times$\\
\tcgr{Grey} & 0 & $\times$ & $\times$ & $\times$  \\
[1ex] 
\hline 
\end{tabular}
\caption{ \textbf{Color Coding for Phase Space of Constraints}: This table illustrates the code
for figures \ref{domain}, \ref{domain5d} and \ref{domain6d} that illustrate how many critical points are present at each point in
the parameter space ($\beta_3$,$\alpha_4$). For completeness we consider   both the existence and signs of $\gamma^2$, the entropy, and  $f_{\infty}$. It is only when the signs of all three are positive that we get physical critical points.  The `$\times$' for $f_{\infty}$ means that Eq. \reef{asympf} does not have any positive real solution; in other cases it means the corresponding critical quantities are either negative or not real-valued.
  }
\label{table:nonlin}
\end{table}

\begin{figure*}[htp]
\centering
\begin{tabular}{cc}
\includegraphics[scale=.3]{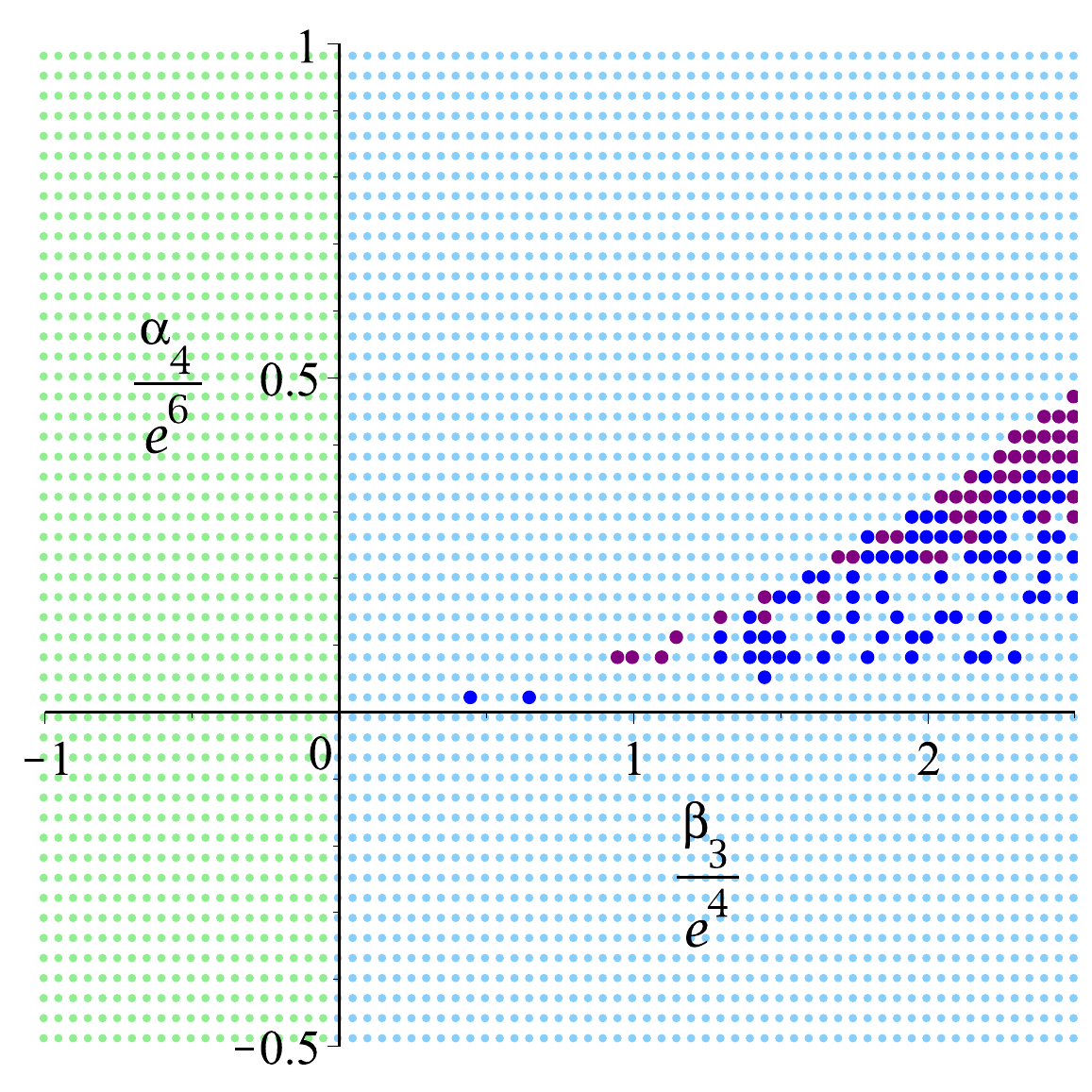}&\quad\quad\quad\quad
\includegraphics[scale=.3]{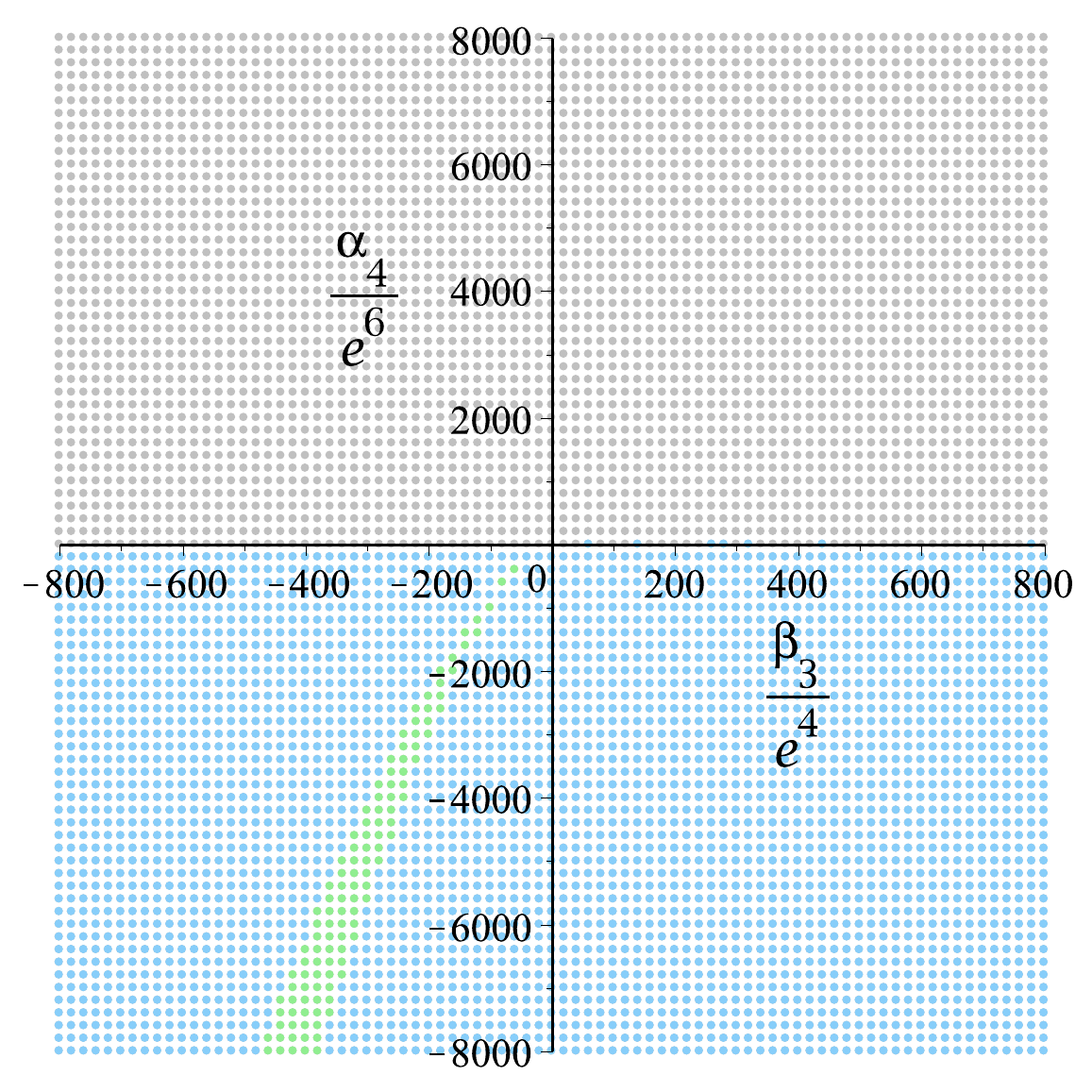}
\\
\includegraphics[scale=.3]{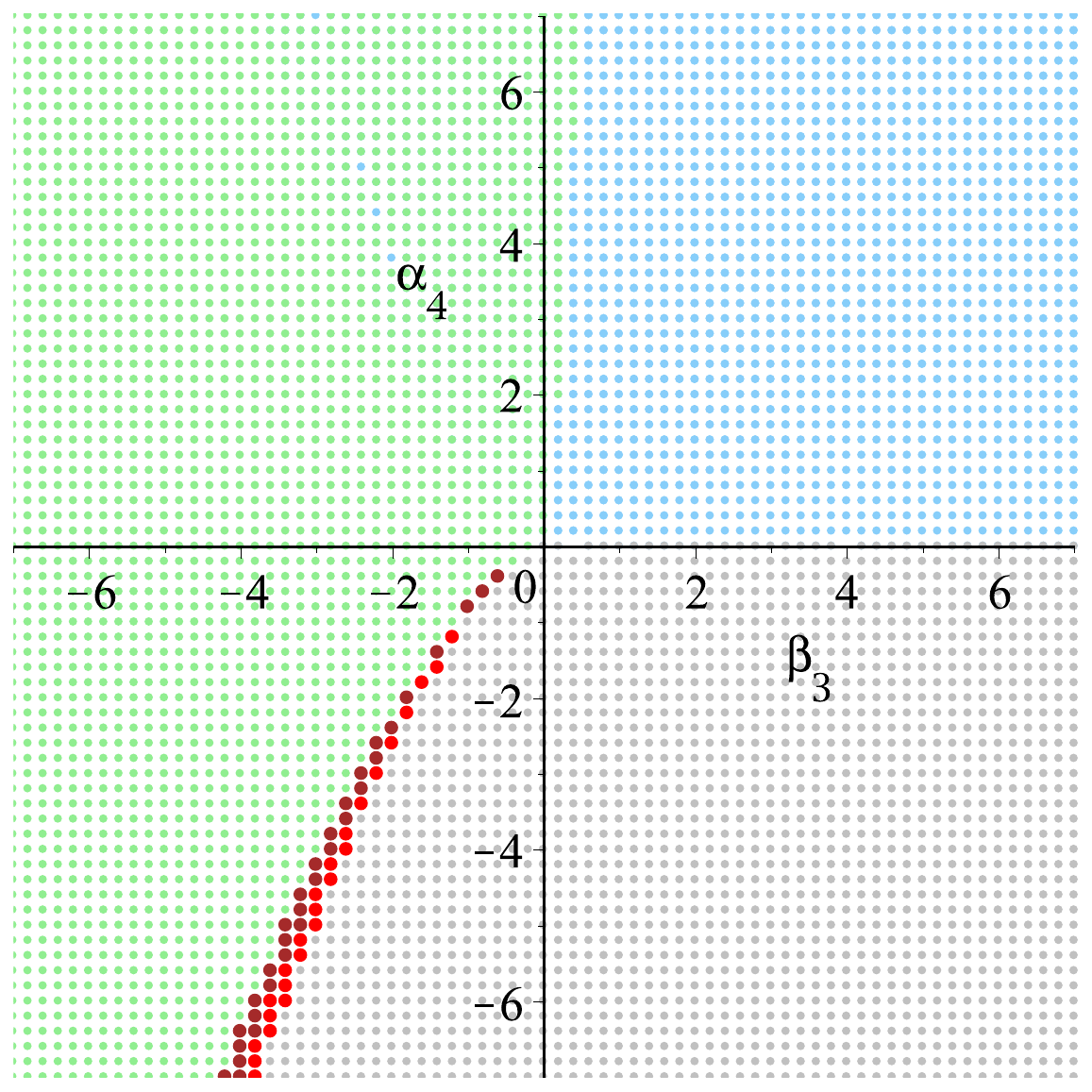}&\quad\quad\quad\quad
\includegraphics[scale=.3]{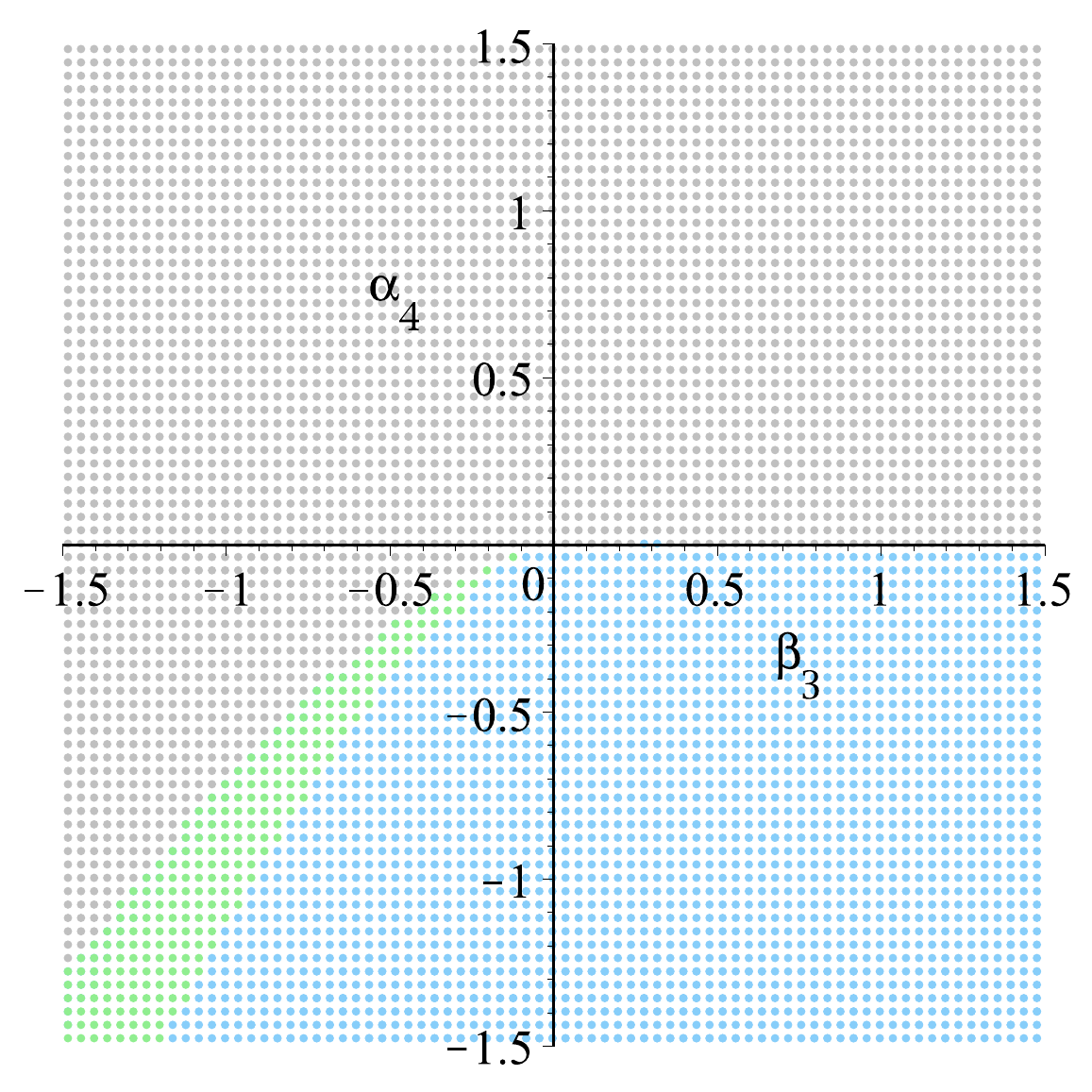}
\\
\end{tabular}
\caption{\textbf{Number of Critical Points as a function of the GQG couplings in $d=4$} (colour online). \textit{Top Left}: The number of critical points
for fixed electric charge ($e=1)$ as a function of $\alpha_4$ and $\beta_3$
for  $k=1$. The physically admissible region, shown in green colour, has only a single physical critical point for each value of the coupling.  For   $\gamma^2<0$ there are either one (blue), two (dark blue) or three
(purple) critical points for a given value of $\beta_3$ and $\alpha_4$. \textit{Top right}: The analogous plot for $d=4$ and $k=-1$ but with $e^2=0.1$. Grey regions have  no critical points. \textit{Bottom Left}: The plot for $d=4$ and $k=1$ but with zero charge. Red and brown regions exhibit two and single critical points with negative entropy. \textit{Bottom Right}: The critical regions for $d=4$ and $k=-1$ for chargeless case.
}
\label{domain}
\end{figure*}

\begin{figure*}[htp]
\centering
\begin{tabular}{cc}
\includegraphics[scale=.25]{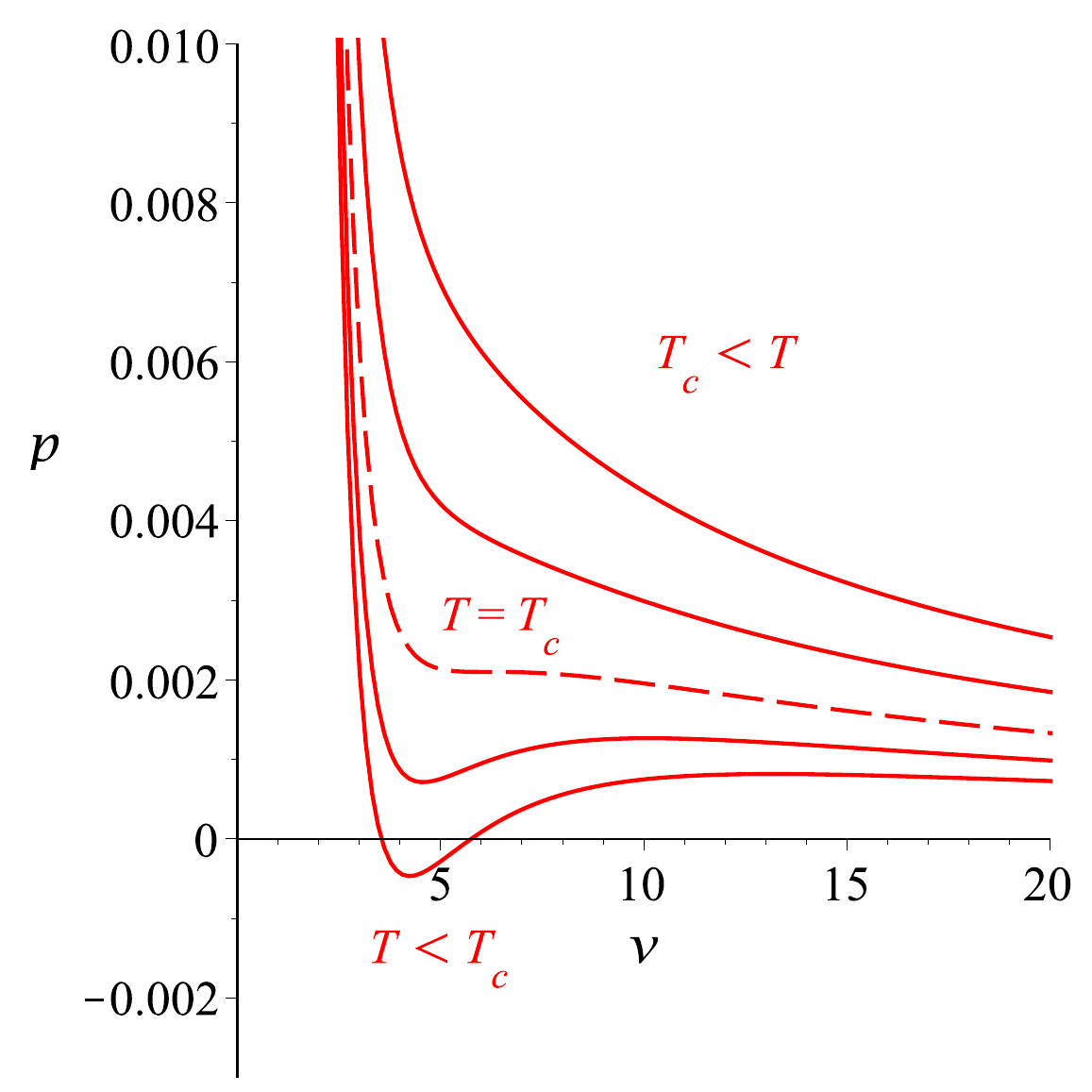}& \quad \quad\quad\quad
\includegraphics[scale=.25]{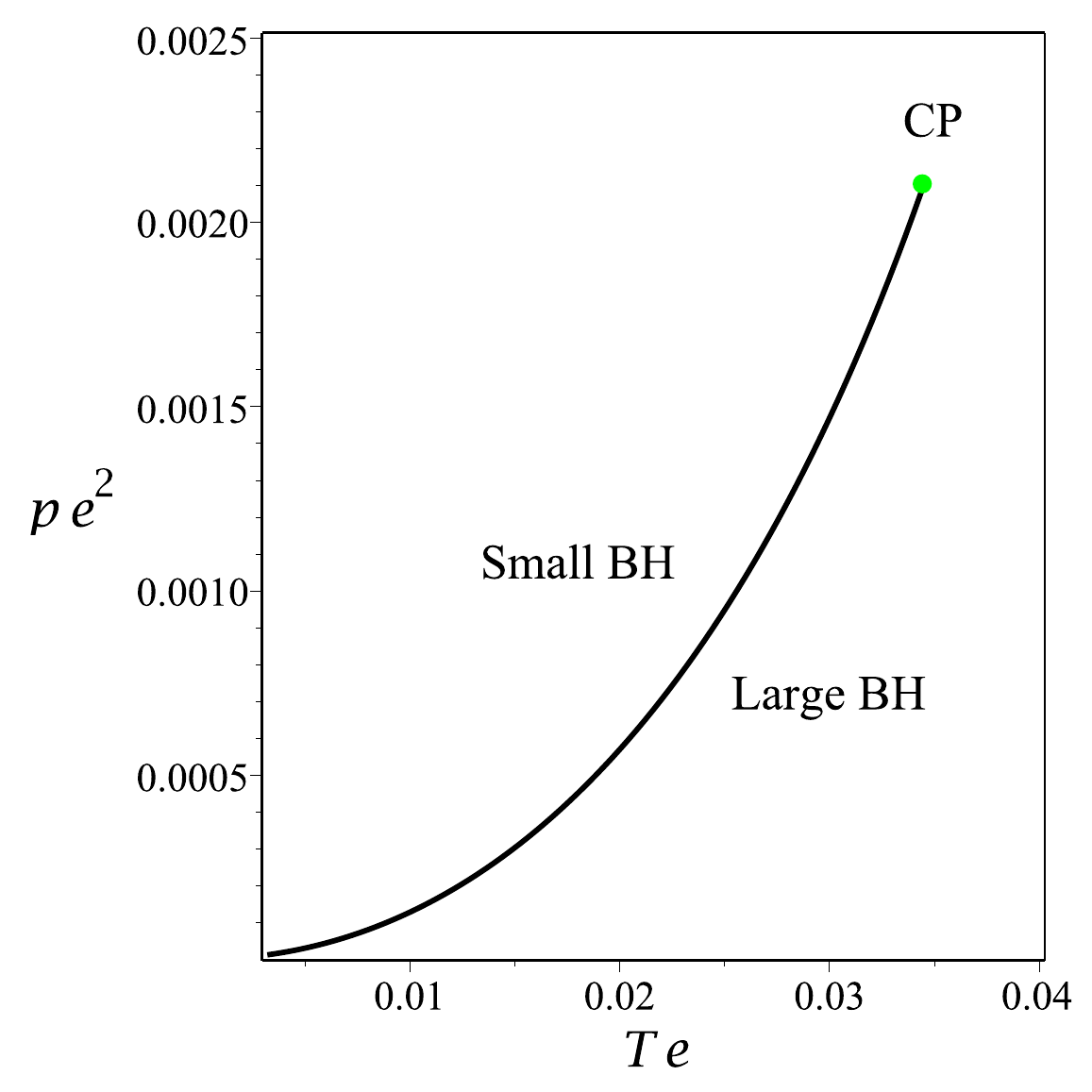}
\\
\end{tabular}
\caption{{\bf Van der Waals phase transition in four dimensions}(colour online): \textit{Left}:
The graph of pressure versus volume at various fixed temperatures for $d=4$ and $k=1$  shows the occurrence of  a first order phase transition with VdW behaviour.  The dashed line has  $T=T_c$, the solid red lines with $T<T_{c}$ are $T=0.8 T_c$ and  $0.65 T_c$, and the solid red lines with $T>T_{c}$ are $T=1.3 T_c, 1.7 T_c$.
\textit{Right}: The coexistence line in the pressure/temperature plane. In both graphs
 $e=1$, $\beta_3=-4 e^4$ and $\alpha_4=5 e^6$ with $T_c e\approx 0.03448$;  similar behaviour happens for any other values chosen from the physical domain given in figure~\ref{domain}. Appropriate factors of the electric charge parameter $e$ are employed to make the relevant quantities dimensionless.
}
\label{PT4d}
\end{figure*}

Due to the complexity of the equation of state, it is not possible to find an explicit bound on the couplings and the electric charge by applying the positivity constraints \reef{nghost} and \reef{gamma21}. However we can numerically  investigate whether these physical constraints are satisfied whilst varying the cubic and quartic couplings for a given fixed charge. The corresponding pattern is given in figure~\ref{domain}, where we also check for positivity of the entropy \reef{sratio}  as well.

The physical critical domain is the part of the parameter space for which the physical constraints discussed in section \ref{constraints} are satisfied, with the property that a phase transition occurs.  Looking at the left part of
figure~\ref{domain},
for $k=1$, this region has $\beta_3 < 0$ (or $\mu<0$) for all values of $\alpha_4$,  with the exception of the axis $\beta_3 = 0$ , where only for $\alpha_4\ge 0$ are the associated phase transitions physical (the positive axis is
green).  The point
 $\beta_3 = \alpha_4 = 0$ is also green, recovering the result that in  the limit of vanishing cubic and quartic couplings, a charged AdS black hole still has physical critical points \cite{Kubiznak:2012wp}. On the vertical axis $\alpha_4<0$, we have $\gamma^2<0$.  The entire physical region has only one critical point, whereas in the unphysical region it is possible to have either one, two, or three critical points depending on the given values of
 $(\beta_3,\alpha_4)$. As we make the fixed value of $e$ smaller (see for example the top right diagram in figure~\ref{domain}), we find that there are at most two possible critical points but only one of them is physical.
Summarizing, in the presence of charge, critical points exist for all $(\beta_3<0,\alpha_4)$ (see figure~\ref{domain}).
The center of the parameter space is the $k=1$ Reissner-Nordstrom-AdS solution for which all constraints are satisfied
in any dimension.

Even for $e=0$, there are regions in the $(\beta_3,\alpha_4)$ plane containing critical points
(bottom left in figure~\ref{domain})  quite unlike the situation in Einstein gravity, where there are no critical points for uncharged black holes. However there are also large regions of parameter space with no critical points.
The center of the parameter space is the $k=1$ Schwarzschild-AdS solution for which constraints are satisfied but which has no critical points.

For $k=-1$,  we see from the upper left diagram in figure~\ref{domain} that single physical critical points exist provided  both the $\beta_3$ cubic and $\alpha_4$ quartic couplings are nonzero and negative. To our knowledge, this phenomenon has not been previously observed
for   hyperbolic black holes in four dimensions.
No critical points exist if $\alpha_4>0$.  The point at the origin of the parameter space is also not a critical point, even in the presence of charge.

Whenever critical points exist we find that the  critical exponents\footnote{ The critical exponents quantify how physical quantities behaves in the vicinity of a critical point \cite{Kubiznak:2012wp}. For $t=T/T_c-1$, the exponent $\alpha$ characterizes the behaviour of the specific heat, while keeping volume constant
 \beqa
 C_V=T \frac{\partial S}{\partial T}\Big|_V\propto |t|^{-\alpha}. \nonumber
\eeqa
The exponent $\beta$ denotes a difference between the volume of a large black hole $V_l$ and the volume of a small black hole $V_s$ on the isotherm process
\beqa
V_l-V_s\propto |t|^{\beta}. \nonumber
\eeqa
The behaviour of the isothermal compressibility  $\kappa_T$ is given by exponent $\gamma$
\beqa
 \kappa_T=-\frac{1}{V} \frac{\partial V}{\partial P}\Big|_T\propto |t|^{-\gamma}.
 \nonumber
\eeqa
The exponent $\delta$ characterizes the following difference on the critical isotherm $T=T_c$
\beqa
|P-P_c|\propto |V-V_c|^{\delta}.  \nonumber
\eeqa
}
are
\beqa
\alpha=0, \quad \beta=\frac{1}{2}, \quad \gamma=1 ,\quad \delta=3,
\label{exponents}
\eeqa
which are the standard values from mean field theory, even when both numerator and denominator
vanish in \eqref{Tc4Q}. These are typically obtained by considering the equation of state near the critical point \cite{GunasekaranEtal:2012}, writing
\beqa
v = v_c(\phi + 1) \, , \quad T = T_c (\tau + 1),
\eeqa
and expanding in powers of $(\phi,\tau)$.
Since here we do not have a closed form for the critical quantities, we insert numerical values for parameters into equation of state to obtain critical values for $T_c$ and $v_c$ and we obtain
\beqa
\frac{P}{P_{c\ \pm}}&=&1+A \tau-B \tau \phi-C \phi^3+\cO(\tau \phi^2,\phi^4),
\label{expcoeff}
\eeqa
yielding \eqref{exponents}.

We explicitly illustrate the occurrence of the phase transition by drawing a $P-v$ graph in figure~\ref{PT4d} for parameters for which there is a physical critical point by setting
 $e=1$, $\beta_3=-4 e^4$ and $\alpha_4=5 e^6$.  We see clear Van der Waals behaviour, with two distinct phases for $T<T_c$ that   coalesce at $T=T_c$ and  become indistinguishable for $T>T_c$. For sufficiently low temperature, the curve tends to negative pressures;  however only positive values of the pressure  are physical.
The coexistence line is plotted in the right half of figure~\ref{PT4d}, illustrating the critical point at the end of a line of first-order phase transitions between large and small black holes.

\begin{figure*}[htp]
\centering
\begin{tabular}{cc}
\includegraphics[scale=.3]{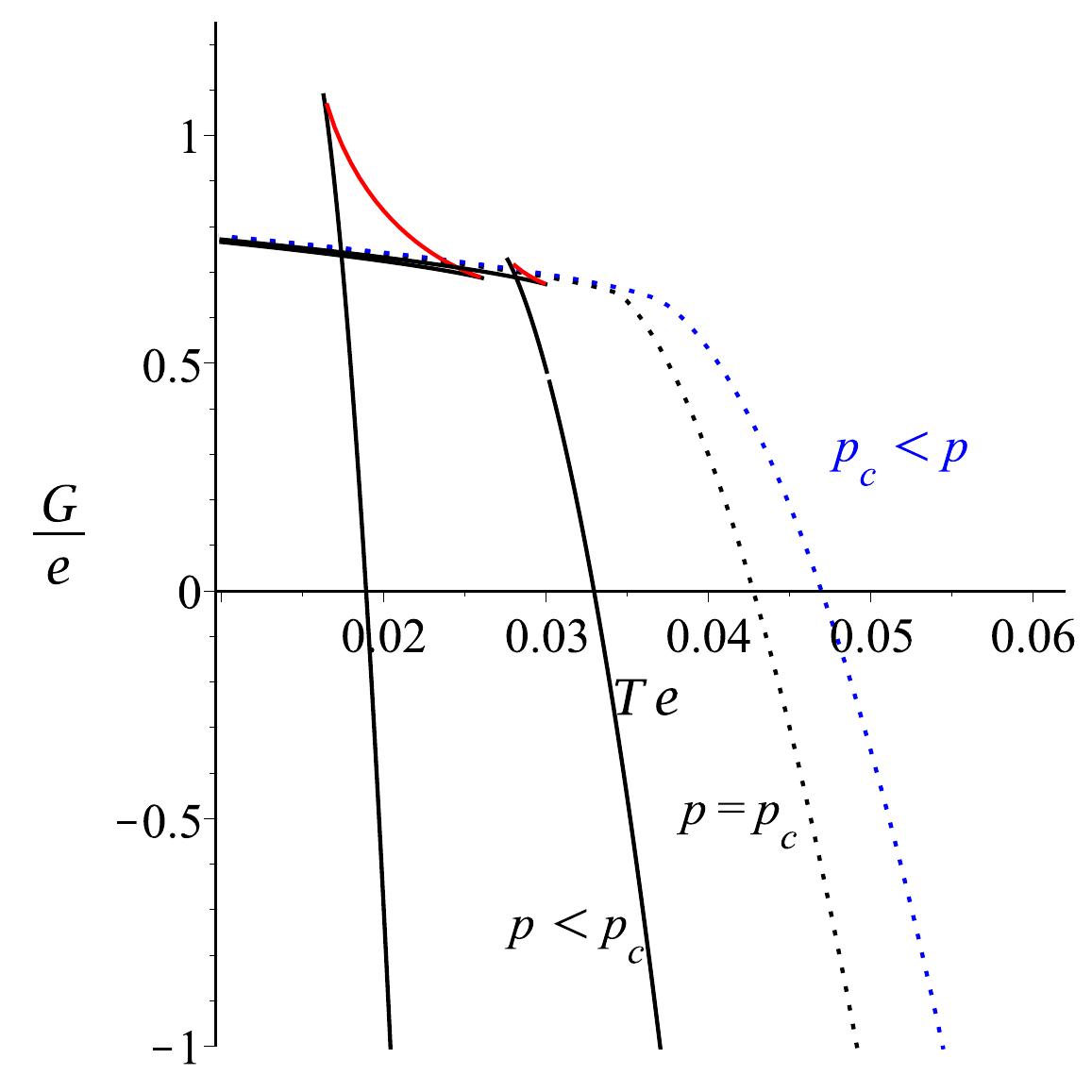}
&\quad\quad\quad\quad
\includegraphics[scale=.3]{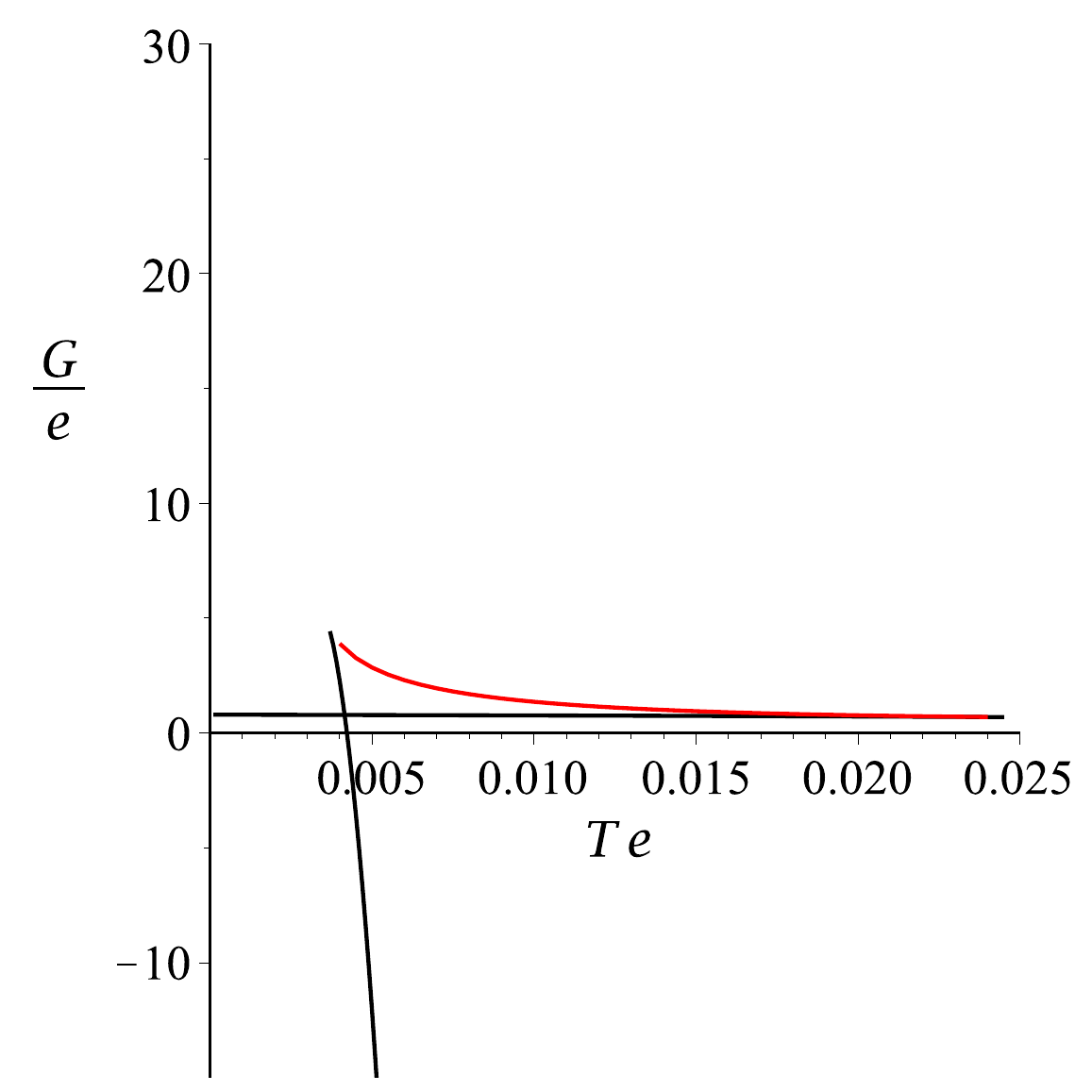}
\\
\end{tabular}
\caption{\textbf{Free energy} (color online). We select $e=1$, $\beta_3=-4 e^4$ and $\alpha_4=5 e^6$;  physical conditions are fulfilled with $P_c e^2\approx 0.00210$. {\it Left:} A plot of the Gibbs free energy versus temperature for $d=4$ and $k=1$, for various values of the pressure: $P =1.2 P_c$ (dotted, blue curve), for $P =P_c$ (dotted, black curve), for $P =0.6 P_c$ and $P =0.2 P_c$ (solid black and red lines).  {\it Right:}
Plot for $P =0.01 P_c$. In each plot, the red lines indicate the parts of the curves for which
the specific heat is negative; quantities are rescaled by appropriate values of  the electric charge $e$ to obtain dimensionless quantities.
}
\label{GTd4}
\end{figure*}

The Gibbs free energy as a function of temperature is shown in figure~\ref{GTd4}, exhibiting the typical swallowtail characteristic of Van der Waals behaviour.  It is notable that for quite small values of the pressure we still observe a swallowtail shape whose size grows rapidly. Computing the specific heat
\beqa
C_P = - T \frac{\partial ^2 G}{\partial T^2} \, ,
\eeqa
we find that two stable branches of black holes exist, with the physical one at the global minimum of $G$.

For the allowed regions of parameter space in figure~\ref{domain} for $k=-1$, the phase diagrams are qualitatively the same as in figures~\ref{PT4d} and~\ref{GTd4}.

\subsection{Critical behaviour in five dimensions}

In this section we consider five dimensional solutions. The equation of state becomes
\beqa
P =\frac{T}{v}-\frac{2 k}{3 \pi  v^2}+\frac{\beta_3 T^3}{v^3}+\frac{6 \beta_3 k^2 T}{35 \pi ^2 v^5}+\frac{e^2}{v^6}-\frac{244 \beta_3 k}{945 \pi ^3 v^6}-\frac{16 \alpha_4 k^2 T^2}{3  \pi ^2 v^6 },
\eeqa
and the critical temperature is
\beqa
T_c&=&\frac{1}{3}\Big(490 \pi ^5 \beta_3 k v_c^{11}+\pi   \left(6615 \pi^5 \beta_3 e^2-2128 \pi^2 \beta_3^2 k\right)v_c^7+\pi   \big(1003520 \alpha_4^2 k+1464 \beta_3^3 k\nonumber\\&&\left.-5670 \pi^3 \beta_3^2 e^2 k^2\big)v_c^3\Big)\Big/\Big(245 \pi^6 \beta_3 v_c^{12}-420 \pi^4 \beta_3^2 k^2 v_c^8+7840 \pi^3 \alpha_4 \beta_3 k v_c^6\right.\nonumber\\
&&\left.+  \left(313600 \pi^2 \alpha_4^2 k^2+180 \pi^2 \beta_3^3 k^2\right)v_c^4+ \left(105840 \pi ^4 \alpha_4 \beta_3 e^2 k^2-27328 \pi  \alpha_4 \beta_3^2 k\right)v_c^2\right. \nonumber\\
&&\left.+53760 \alpha_4^2 \beta_3 k^2\Big),\right.\label{5dTceq}
\eeqa
The critical volume satisfies the following equation
\beqa
-30240 \pi  \alpha_4 k^2 T_c^2+ \left(540 \pi  \beta_3 k^2 v_c-630 \pi^3 v_c^5\right)T_c-1464 \beta_3 k+5670 \pi^3 e^2+420 \pi^2 k v_c^4=0 \nonumber\\
\label{5dvceq}
\eeqa
and,  as before,  finding an explicit closed form for both the critical temperature and volume is not feasible.

However for vanishing cubic coupling, there is a considerable simplification; solving
\reef{dPd2p} with   $\beta_3=0$ for the corresponding critical temperature \reef{5dTceq} yields
we find
\beqa
t_c=\frac{16 k}{15 \pi v_c},
\label{tc0}
\eeqa
 where
$v_c$ satisfies
\beqa
-30k\pi^3v_c^6+675e^2\pi^4v_c^2-4096\alpha_4 =0,\label{nuc5d}
\eeqa
and we must have $k=1$ so that $t_c > 0$.
 This equation is a cubic polynomial in   $v_c^2$ and can be solved exactly. At most there are two real solutions for any given choice of parameters that are physically acceptable \footnote{ In general, we found that in the five dimensional quartic theory (with $\beta_3=0$), there are at most two physical critical points.}.

\begin{figure*}[htp]
\centering
\begin{tabular}{cc}
\includegraphics[scale=.3]{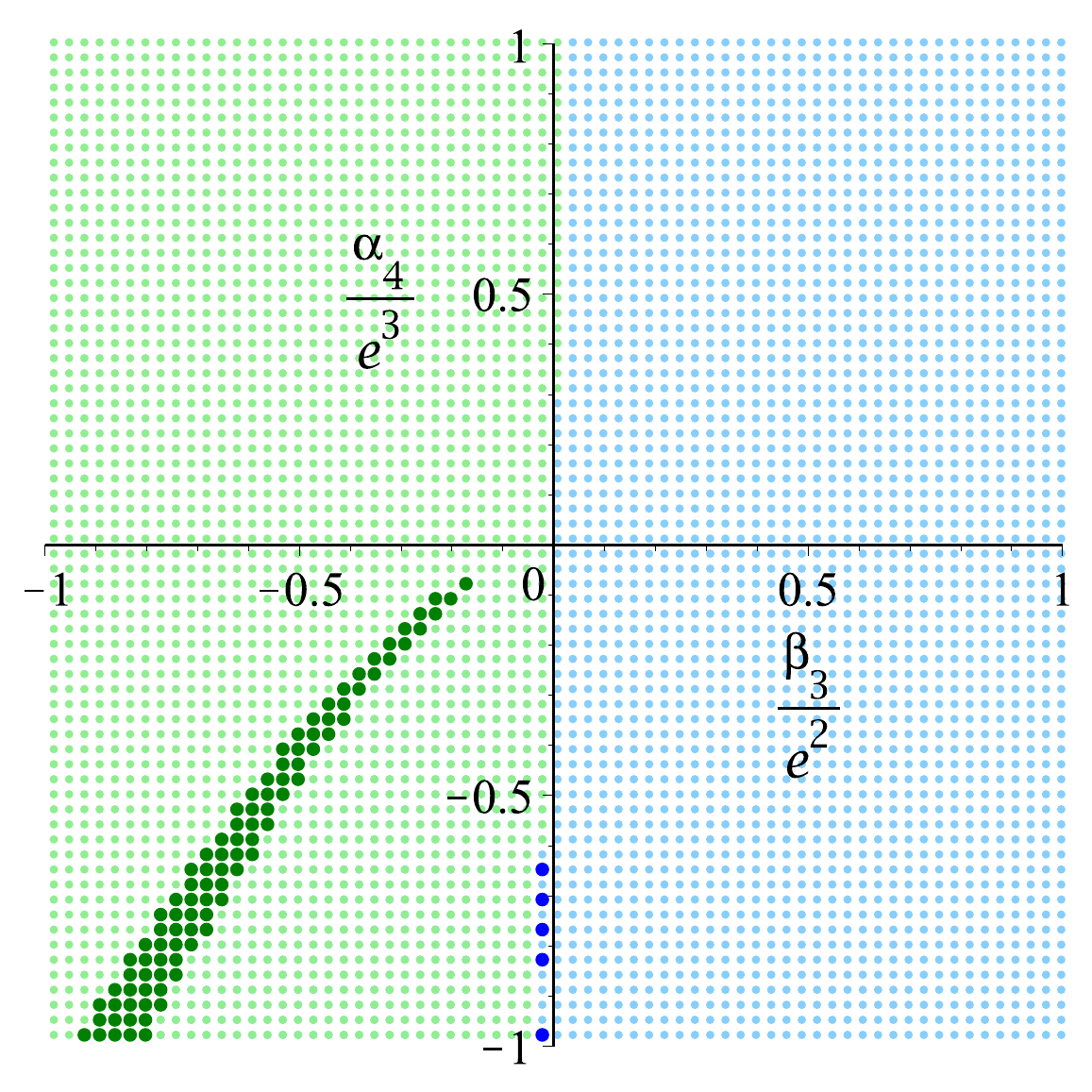}&\quad \quad\quad\quad
\includegraphics[scale=.3]{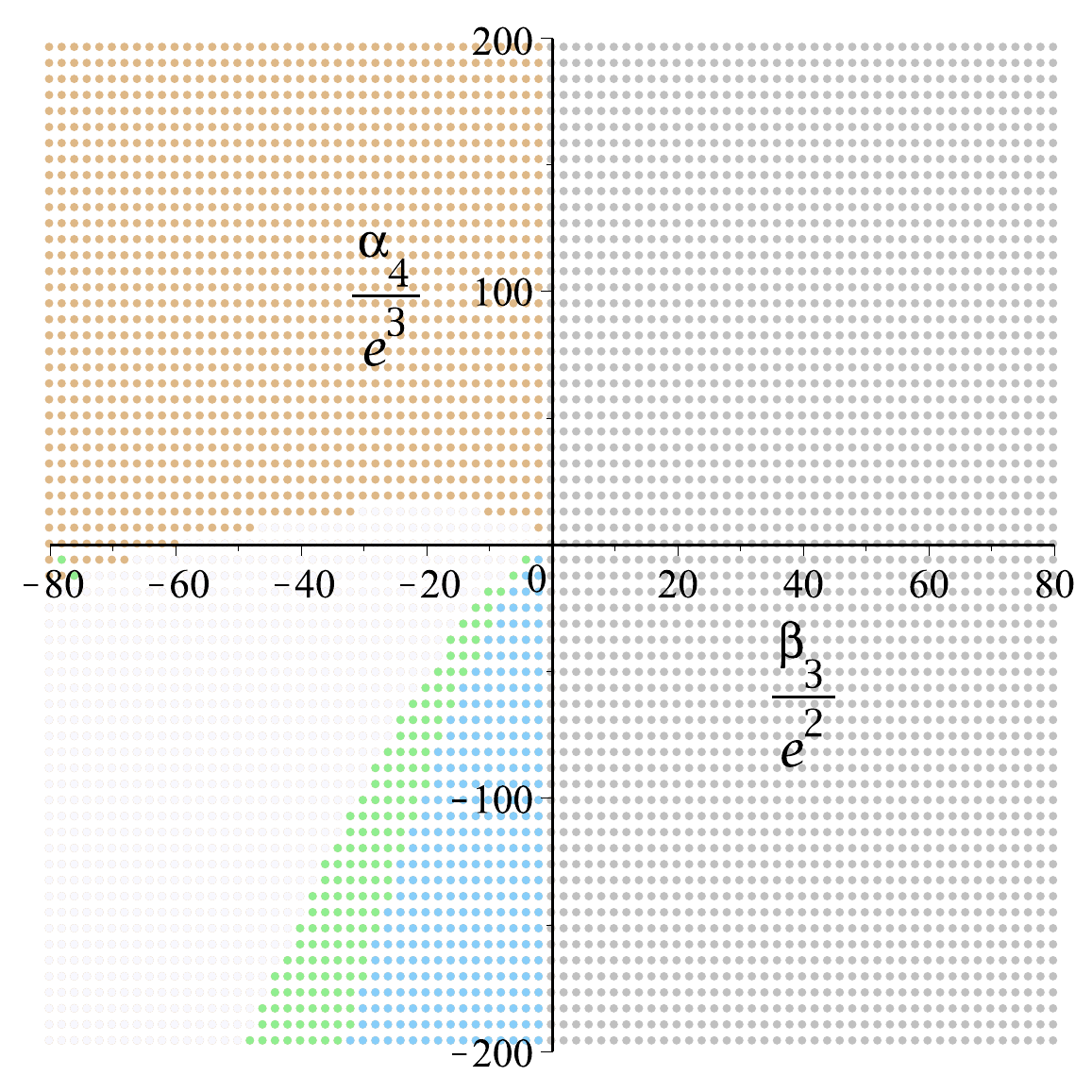}
\\
\includegraphics[scale=.3]{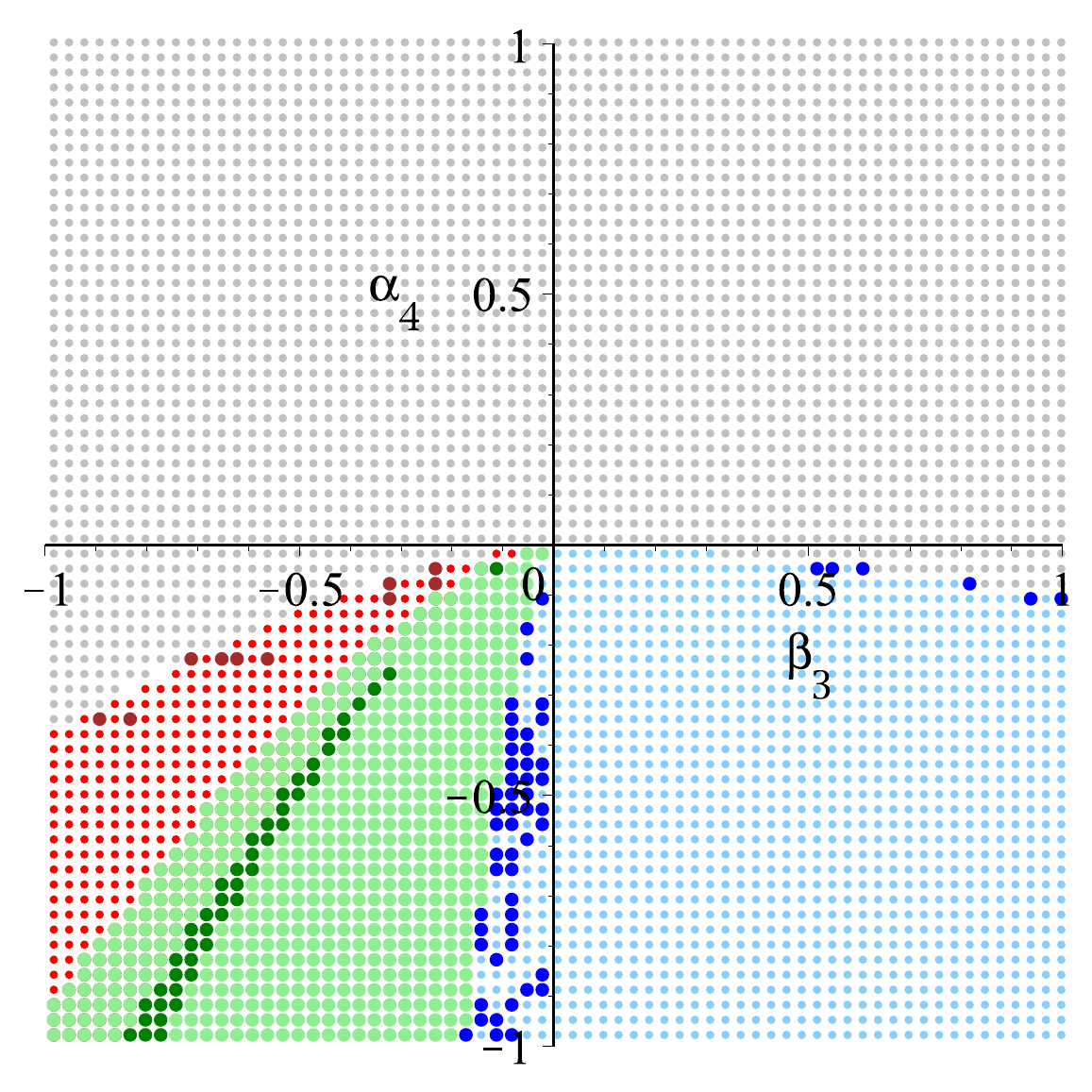}&\quad \quad\quad\quad
\includegraphics[scale=.3]{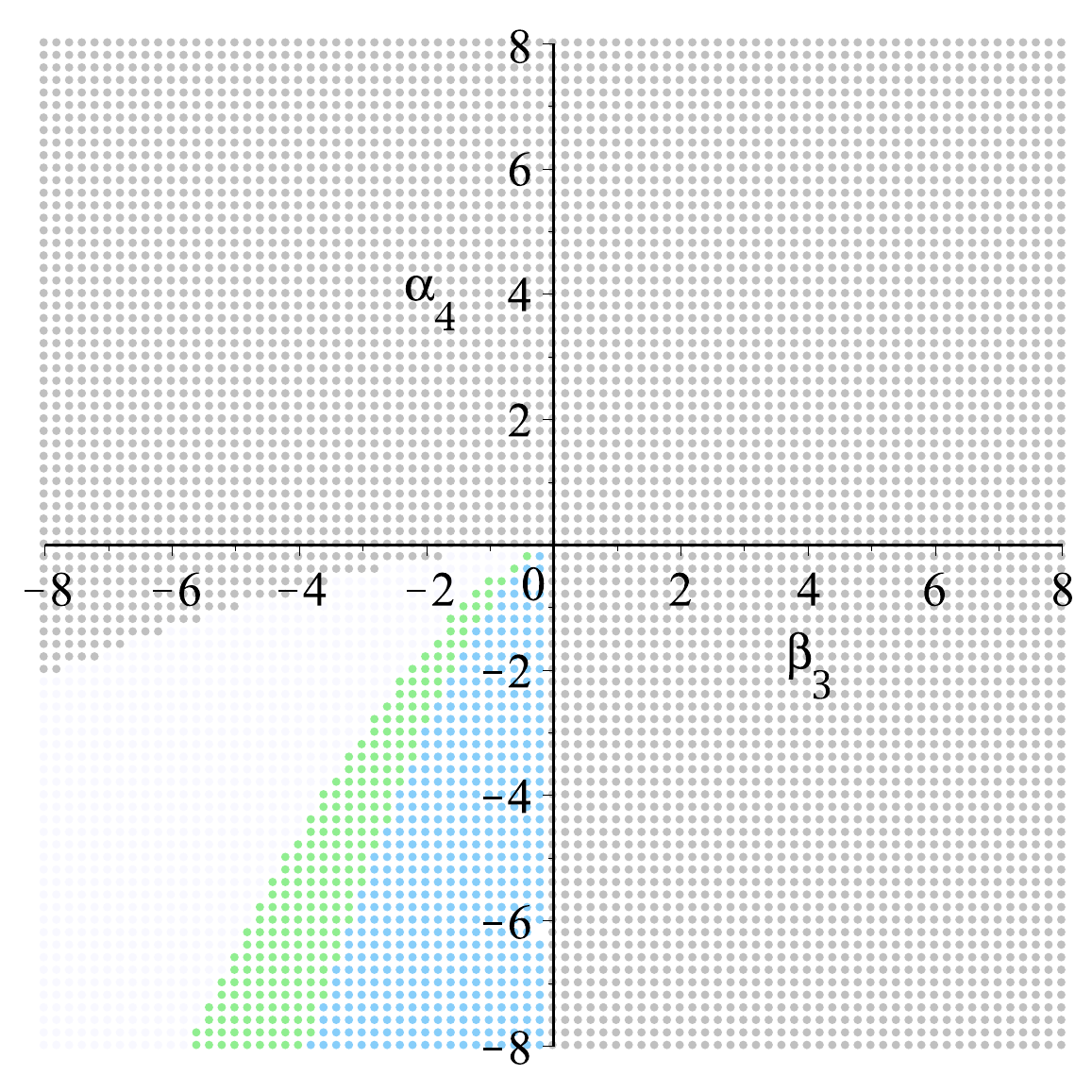}
\\
\end{tabular}
\caption{\textbf{Number of Critical Points as a function of couplings in $d=5$} (colour online). \textit{Top Left}: For $k=1$ and $e=1$ we see a broad (green) region in $(\beta_3,\alpha_4)$ parameter space having only one physical critical point with a band (dark green) where two physical critical points exist.  There are two and  single critical point with $\gamma^2<0$ in the dark blue and blue regions.   \textit{Top right}: For $k=-1$ and $e^2=0.1$  the only region having
(single) physical critical points is the green band in the lower-left quadrant; the light brown region has a single critical point with an unphysical imaginary asymptotic value for $f_{\infty}$; the  grey regions do not have critical points, and in the white region the mass is negative.
 \textit{Bottom left}: For $k=1,~e=0$, the  brown and red points respectively demonstrate  the existence of  one and two critical points with $S<0$; \textit{Bottom right}: For $k=-1,~e=0$, there is still an allowed region (green) of physical critical points; in the white region the mass is negative.
 }
\label{domain5d}
\end{figure*}

Figure \ref{domain5d} plots the number of critical points as a function of $(\beta_3,\alpha_4)$ with fixed charge.  For $k=1$, unlike in 4 dimensions, we see that only if both couplings are non-zero  we get two physical critical points
in a certain region of parameter space, shown in dark green.  The occurrence of two physical critical points for spherical black hole in five dimensions has to our knowledge not been seen previously.
On the axes $\beta_3<0,~\alpha_4=0$ and $\beta_3=0$  (\ie on the vertical axis) only for positive values of $\alpha_4$ (or $\lambda < 0$) greater than some specific lower bound for the quartic coupling  are the critical points physical,  whereas for  $\beta_3>0,~\alpha_4=0$ and $\beta_3=0,~\alpha_4<0$ they are unphysical, having $\gamma^2<0$.
Physical critical points exist for most of the region  $\beta_3<0$, except for small values of $|\beta_3|$ and large enough large values of $|\alpha_4|$ where $\gamma^2<0$.
The case
$\alpha_4 = 0$ is discussed in  \cite{mir:2018mmm}.

\begin{figure*}[htp]
\centering
\begin{tabular}{cc}
\includegraphics[scale=.3]{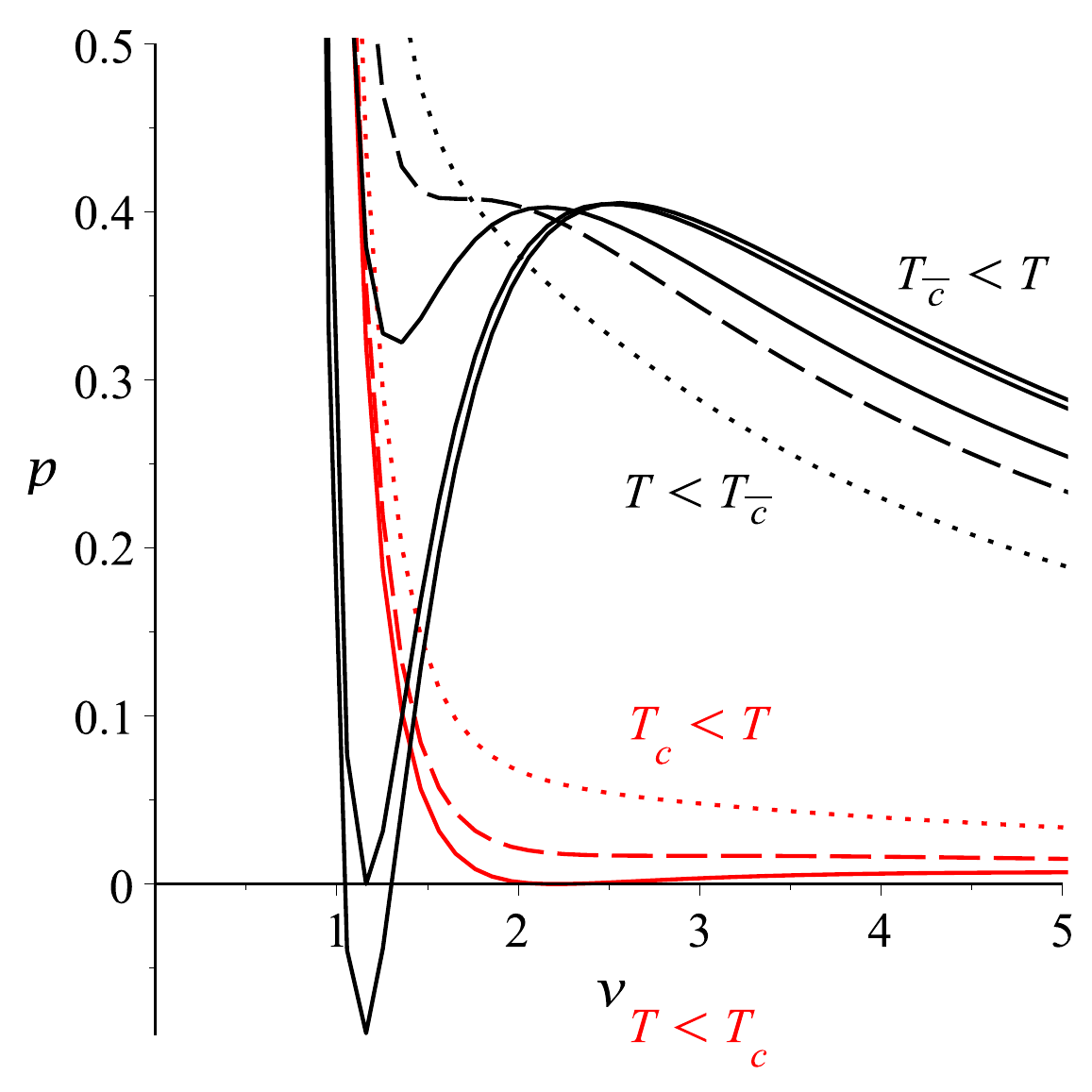}&\quad \quad\quad\quad \includegraphics[scale=.3]{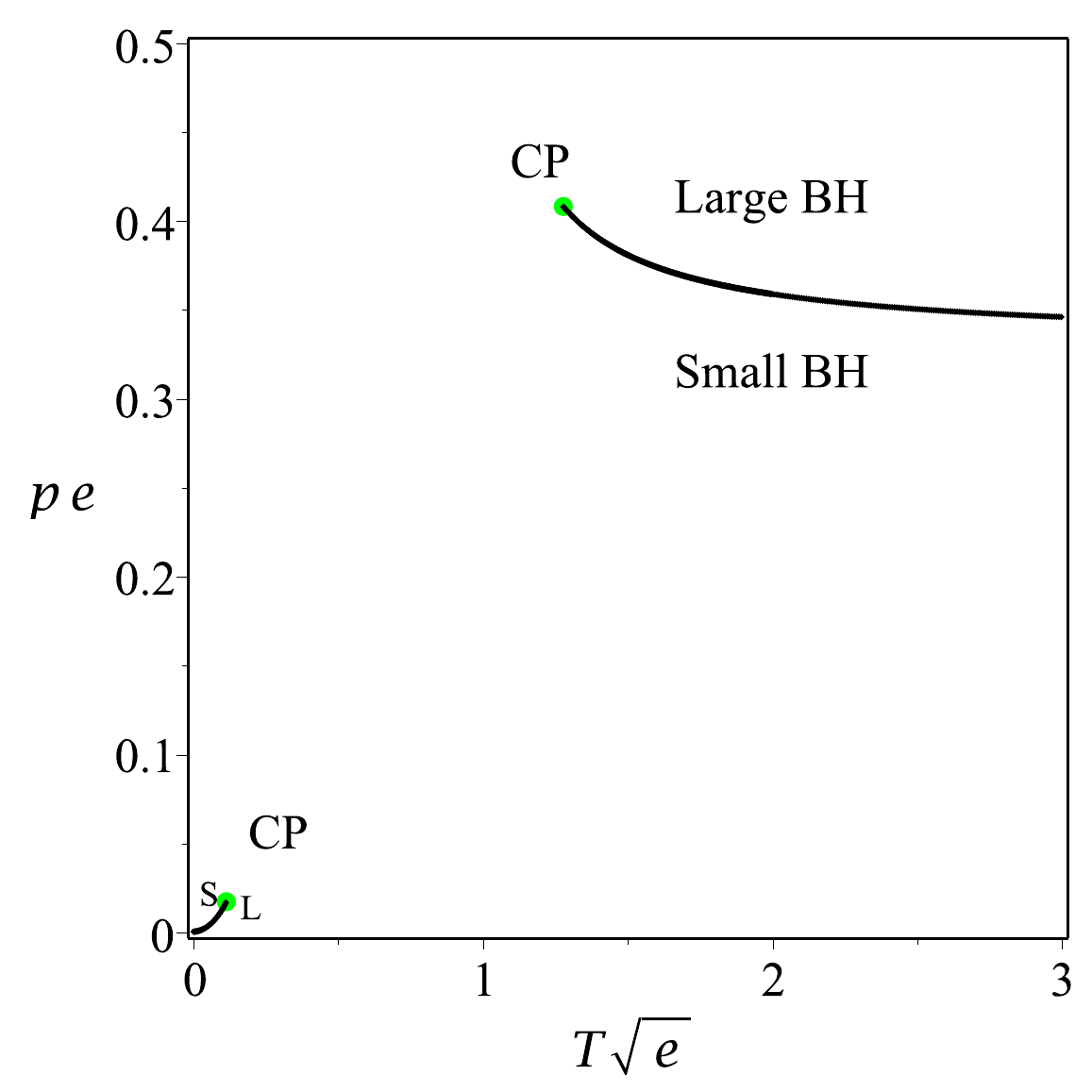}
\\
\end{tabular}
\caption{\textbf{Van der Waals and reverse Van der Waals phase transition in $d=5$ and $k=1$.} (color online).  We choose $e=1$, $\beta_3=-4/5 e^2$ and $\alpha_4=-4/5 e^3$.    Dimensionless critical quantities are $T_c \sqrt{e}\approx 0.11730$, $P_c e\approx 0.01686$, $T_{\overline{c}} \sqrt{e}\approx 1.28065$, and $P_{\overline{c}} e\approx 0.40771$. \textit{Left:} The behaviour of pressure versus volume for temperatures in the neighbourhood of the  critical point at smaller $T=T_c$.  We depict the critical curve  (dashed red line), $T\approx 0.65639 T_c$ (solid red line), and $T=1.8 T_c$ (dotted red line).  We also plot the behaviour in the neighbourhood of the second critical point  at larger $T=T_{\overline{c}}$.  We depict the critical curve  (dashed black line),  $T=0.8 T_{\overline{c}}$ (dotted black line), and $T=1.1 T_{\overline{c}}, 1.24282 T_{\overline{c}}, 1.27 T_{\overline{c}}$ (solid black lines).  For $T > 1.24282 T_{\overline{c}}$ the pressure becomes negative and the spacetime is
no longer asymptotically AdS.
\textit{Right:} Coexistence curves for five dimensional $k=1$ charged black holes, with standard
VdW behaviour at the lower left and reverse VdW behaviour in the upper right.   The critical points are the green points.
  }
\label{PT5dk1}
\end{figure*}
\begin{figure*}[h]
\centering
\begin{tabular}{cc}
\includegraphics[scale=.3]{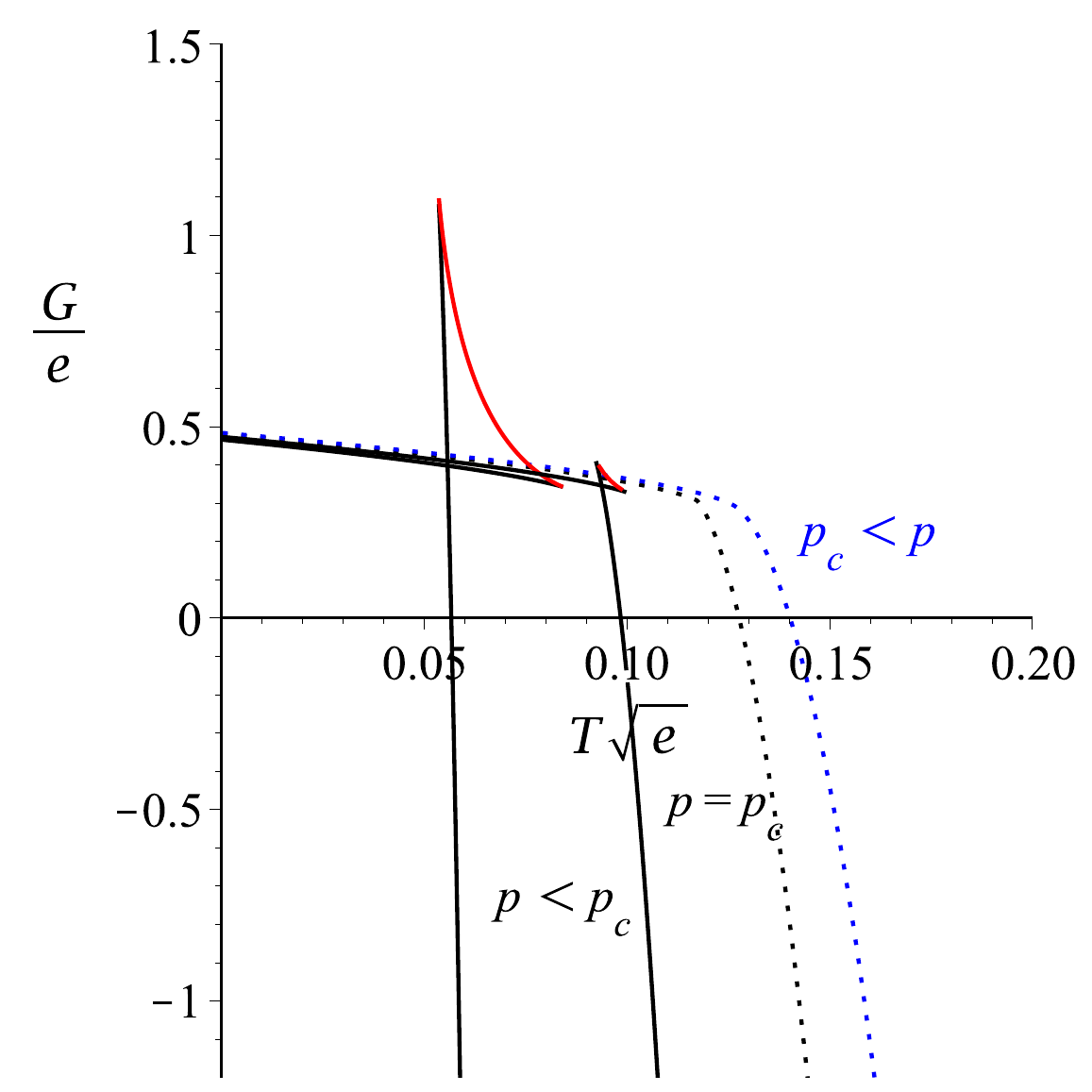}&\quad\quad\quad\quad
\includegraphics[scale=.3]{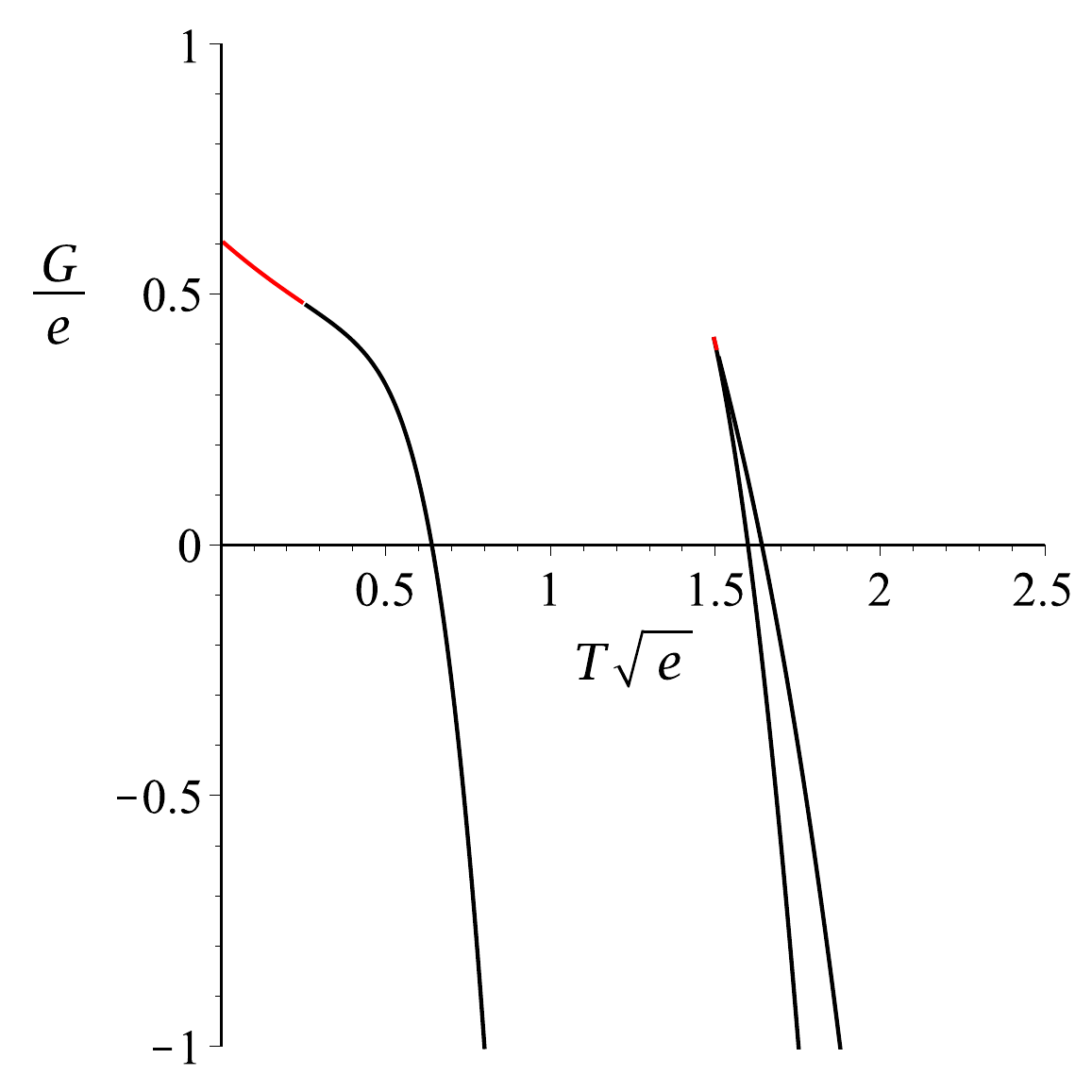}\\
\includegraphics[scale=.3]{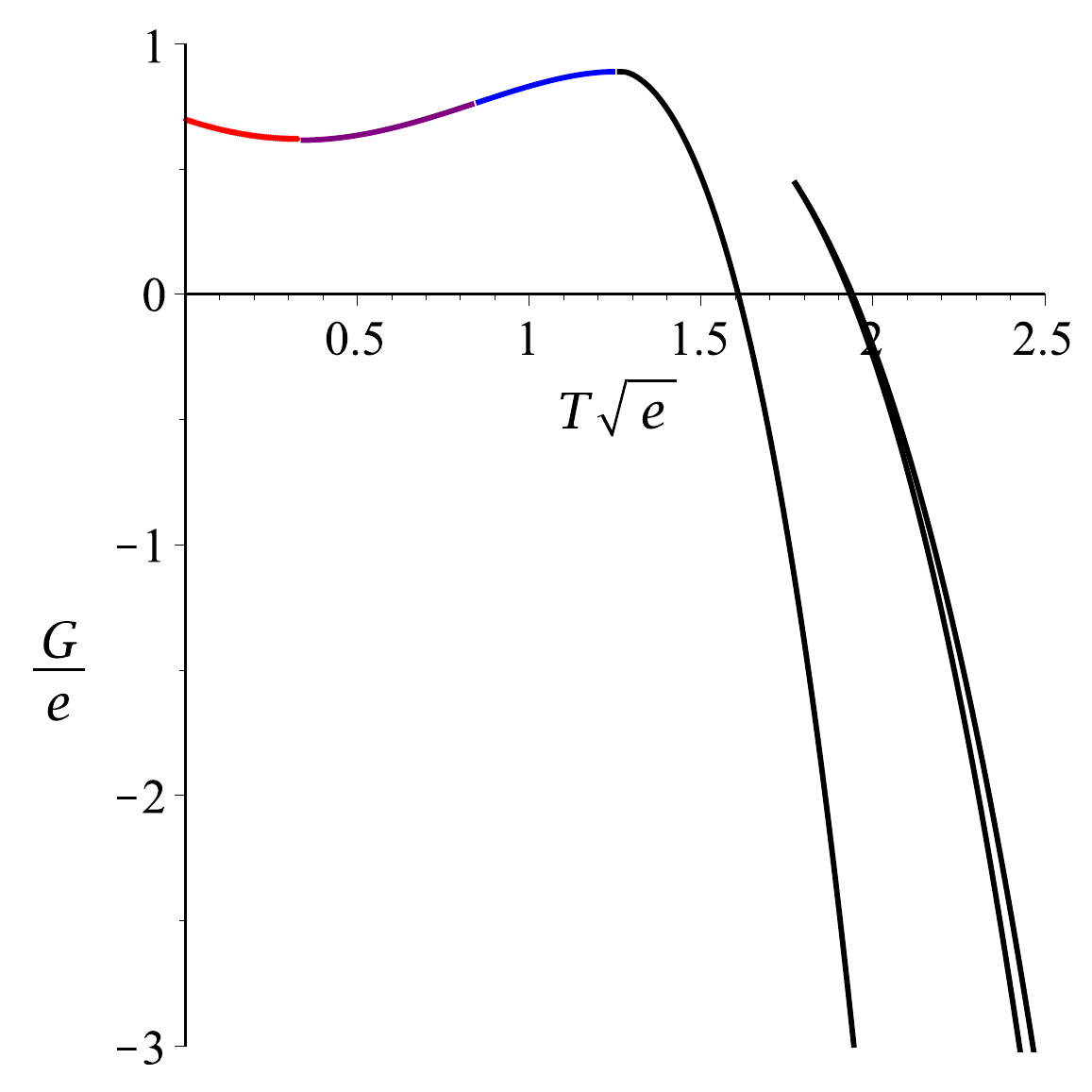}&\quad\quad\quad\quad
\includegraphics[scale=.3]{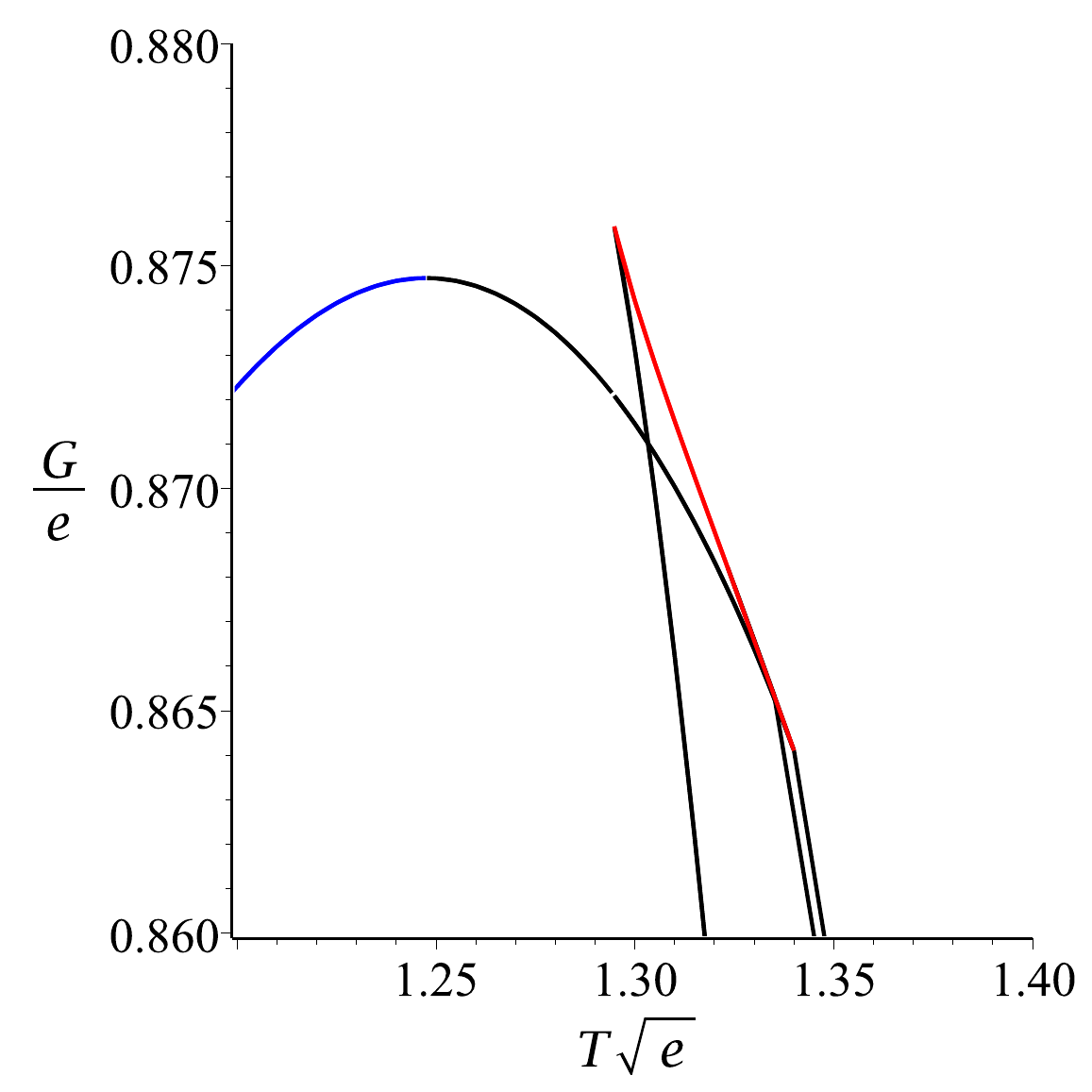}
\\
\end{tabular}
\caption{\textbf{Free energy as a function of temperature for $d=5,~k=1$} (color online).
We set $e=1$, $\beta_3=-4/5 e^2$ and $\alpha_4=-4/5 e^3$.   Dimensionless critical quantities are $T_c \sqrt{e}\approx 0.11730$ and $P_c e\approx 0.01686$, $T_{\overline{c}} \sqrt{e}\approx 1.28065$ and $P_{\overline{c}} e\approx 0.40771$, and appropriate powers of the electric charge parameter $e$ are used to render the relevant quantities dimensionless. In each plot,   red lines depict the parts of the curves for which the specific heat is negative, blue lines indicate negative entropy, and the purple line indicates that both specific heat and entropy are negative. \textit{Top left:} The low temperature region (for which there is a standard VdW phase transition), with  $P =1.2 P_c$ (dotted, blue curve),  $P =P_c$ (dotted, black curve) and $P =0.6 P_c, 0.2 P_c$ (solid, black and red curve) each plotted.
The remaining graphs pertain to the high temperature region (for which there is a reverse VdW phase transition).
\textit{Top right:} $P =0.5 P_{\overline{c}}$, \textit{Bottom left:}  $P = P_{\overline{c}}$  \textit{Bottom right:} $P = 0.99 P_{\overline{c}}$. }
\label{GTd5}
\end{figure*}
\begin{figure*}[htp]
\centering
\begin{tabular}{cc}
\includegraphics[scale=.3]{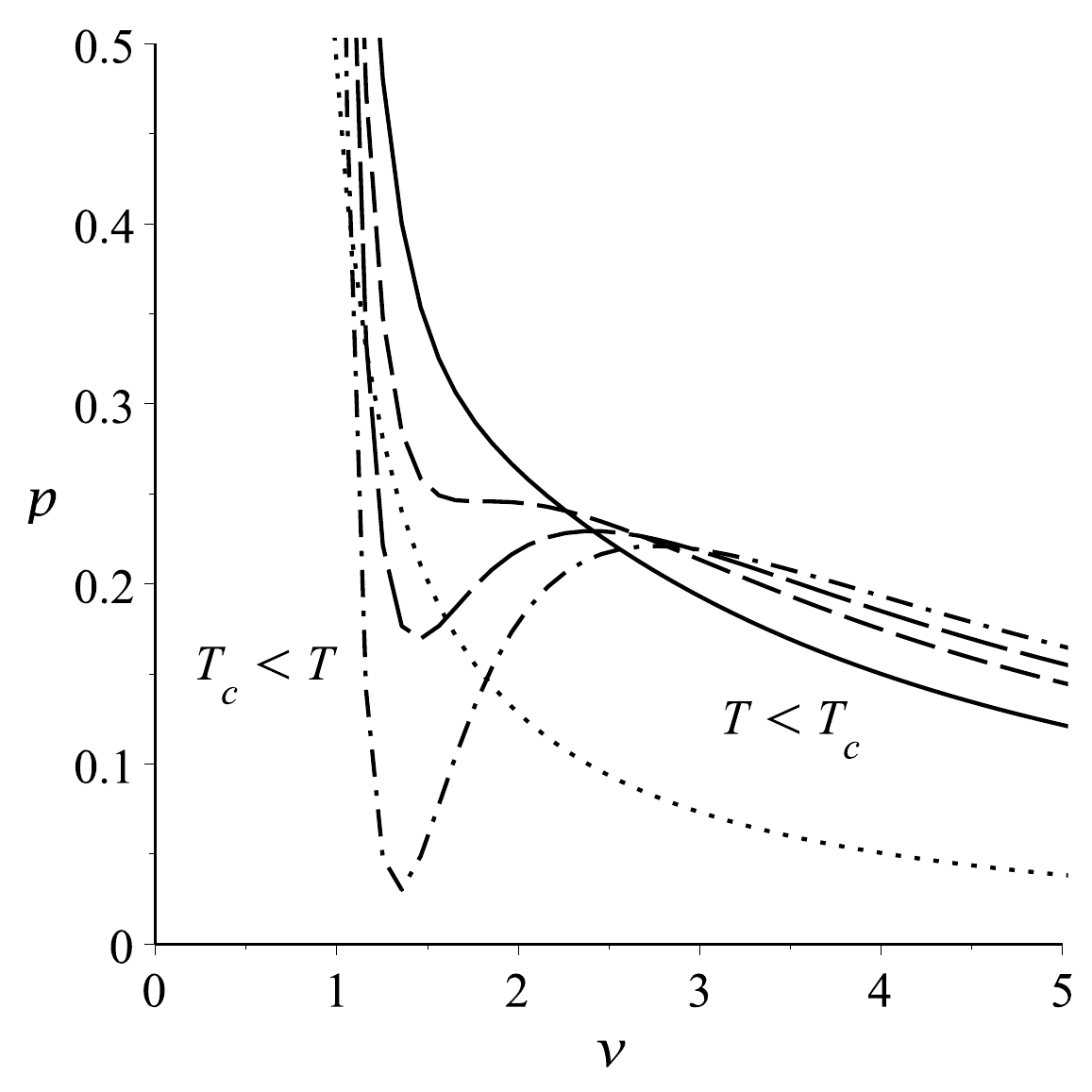}&\quad \quad\quad\quad \includegraphics[scale=.3]{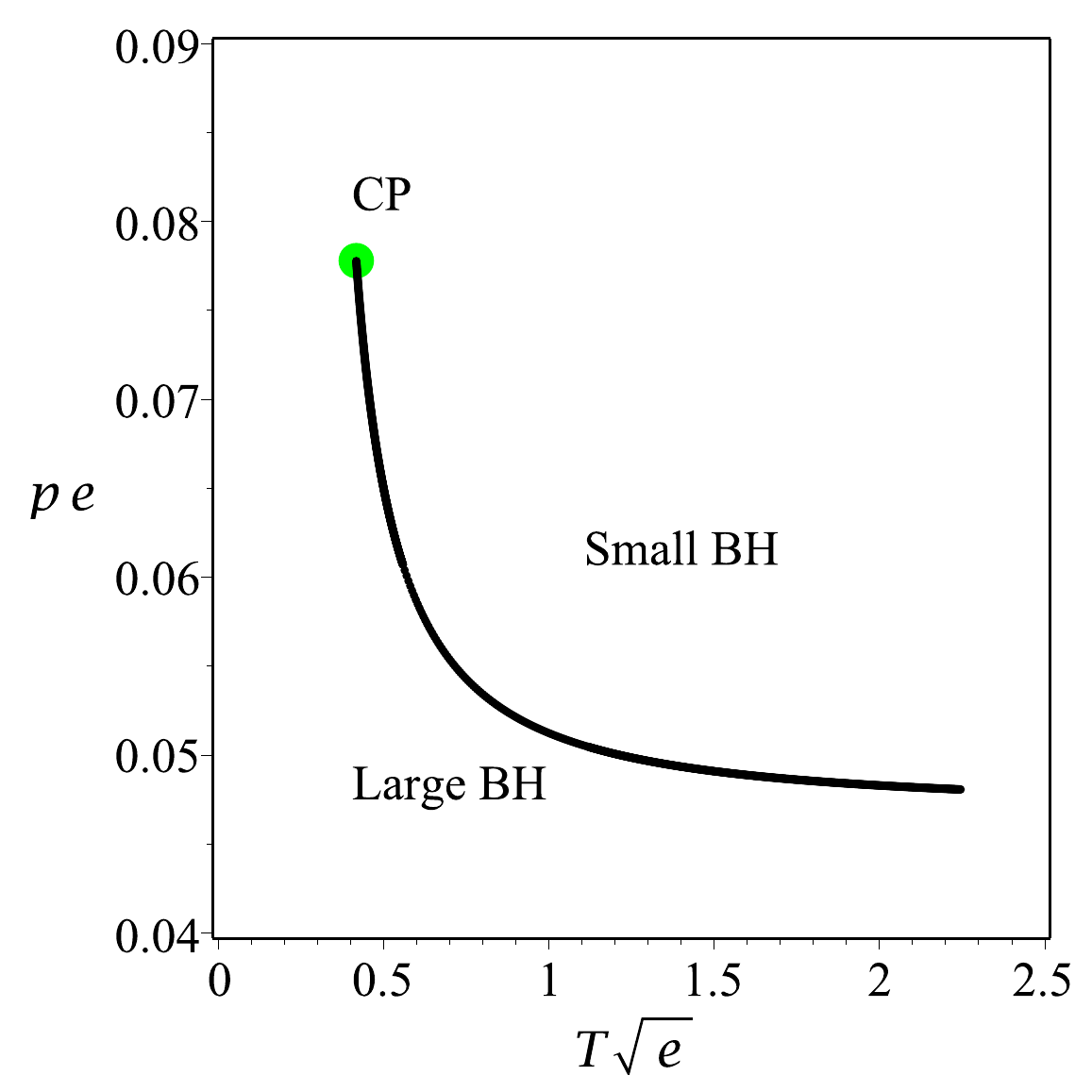}
\\
\end{tabular}
\caption{ \textbf{Reverse Van der Waals phase transition in $d=5,~k=-1$} (color online).  We choose $e^2=1/10$, $\beta_3=-4 e^2$ and $\alpha_4=-6 e^3$.    Dimensionless critical quantities are $T_c \sqrt{e}\approx 0.42157$ and $P_c e\approx 0.07772$. \textit{ Left:} The behaviour of pressure versus volume for $d=5$ and $k=-1$ for temperatures at critical point $T=T_c$ (dashed black line), $T\approx 0.8 T_c$ (solid black line), $T=0.2 T_c$ (dotted black line), $T=1.1 T_c$  (long dashed black line), $T=1.2 T_c$  (dash-dotted blue line).
\textit{Right:} Coexistence curves for five dimensional $k=-1$ charged black holes is depicted, with reverse
VdW behaviour.   The critical points are denoted by the green points.
}
\label{PT5dkm1}
\end{figure*}
\begin{figure*}[h]
\centering
\begin{tabular}{cc}
\includegraphics[scale=.25]{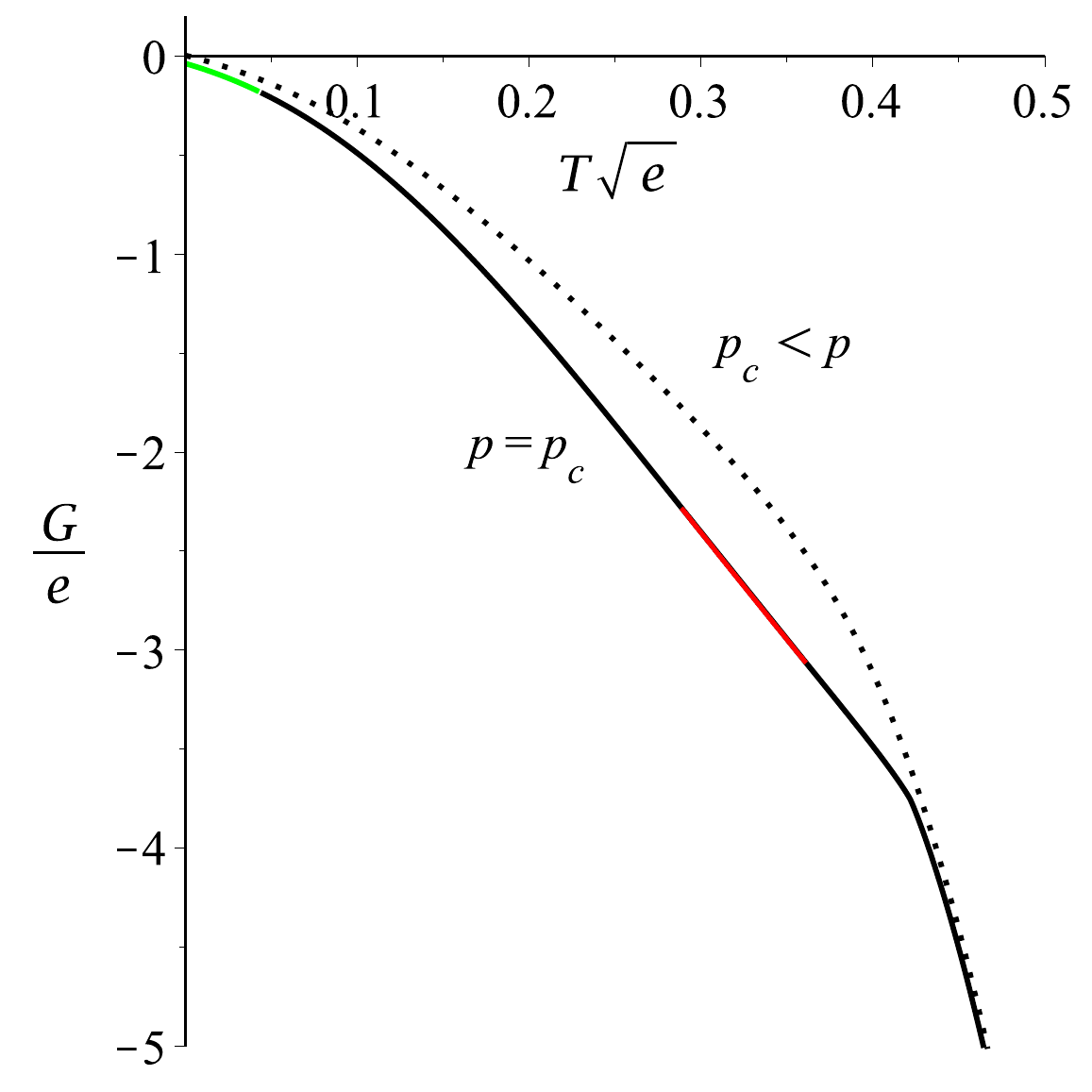}&
\includegraphics[scale=.25]{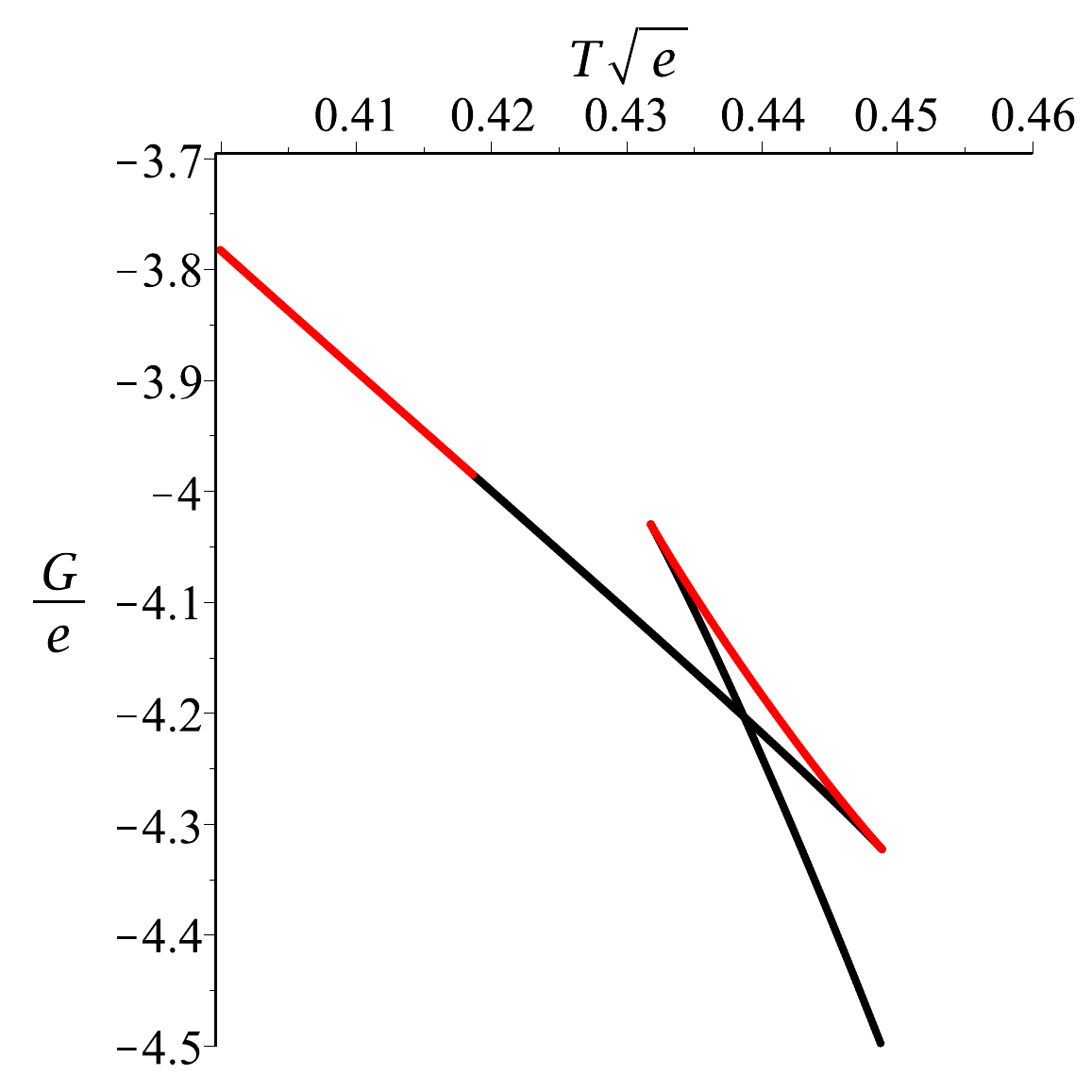}
\includegraphics[scale=.25]{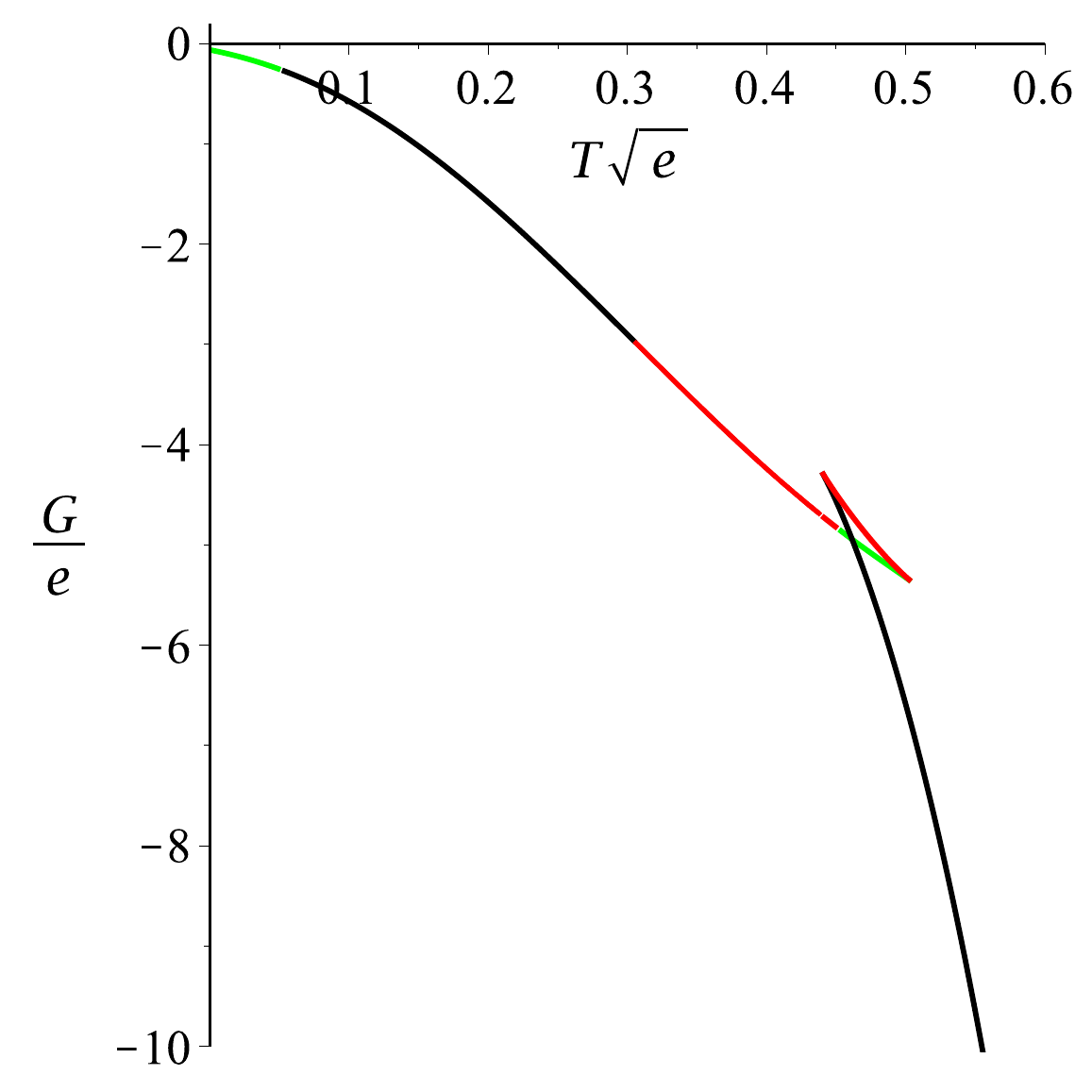}
\\
\includegraphics[scale=.25]{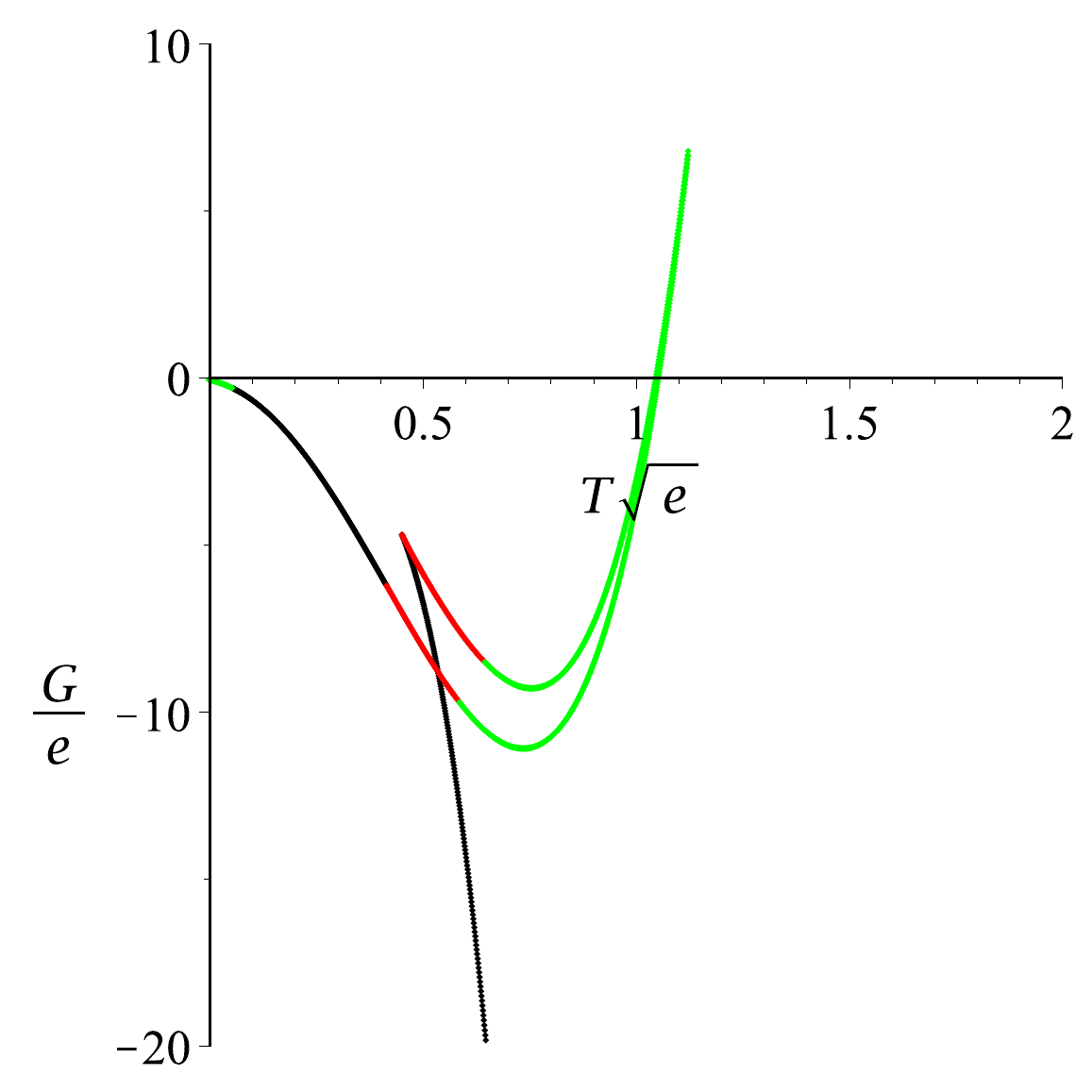}&
\includegraphics[scale=.25]{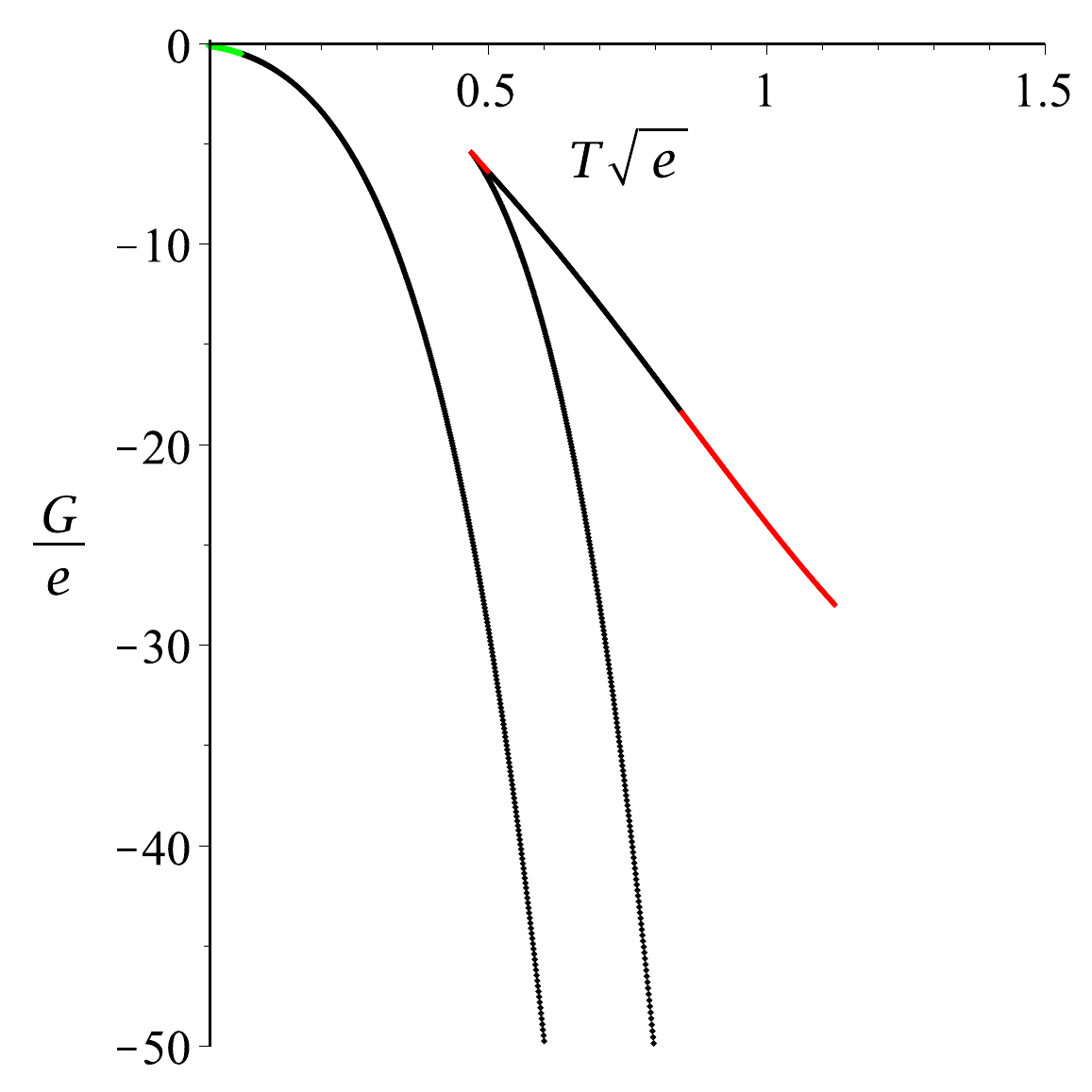}
\includegraphics[scale=.25]{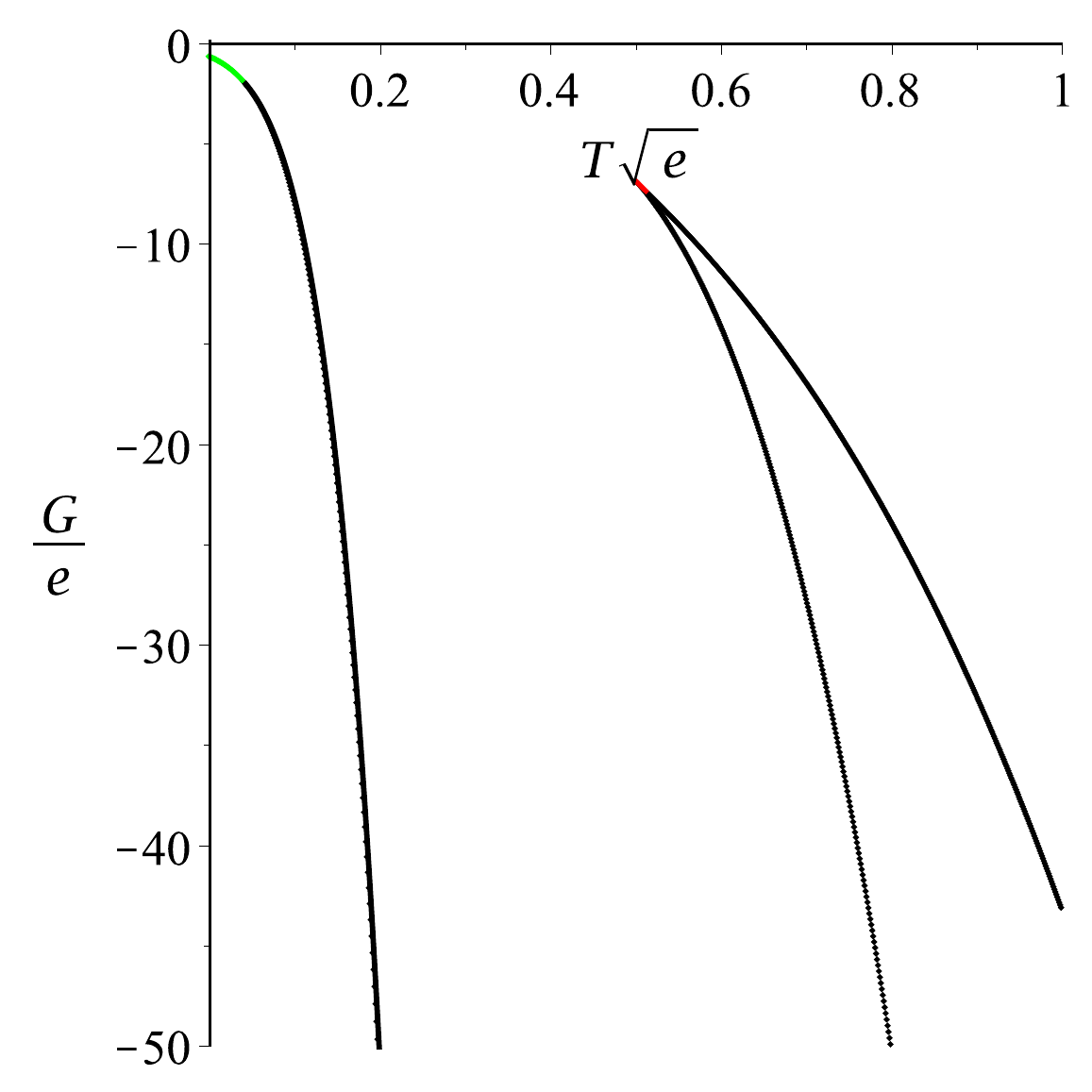}
\\
\end{tabular}
\caption{ \textbf{Free energy as a function of temperature for $d=5$ and $k=-1$} (color online).
We set $e^2=1/10$, $\beta_3=-4 e^2$ and $\alpha_4=-6 e^3$, with dimensionless critical quantities $T_c \sqrt{e}\approx 0.42157$ and $P_c e\approx 0.07772$.  Red lines correspond to negative specific heat and green lines denote negative mass.
\textit{Top left:} A plot of the Gibbs free energy for  $P =1.2 P_c$ (dotted, black curve) and $P =P_c$ (solid, black and red and green curve). \textit{Top center}:  For $P =0.95 P_c$ a reverse VdW phase transition occurs.
\textit{Top right:} $P =0.9 P_c$, \textit{Bottom left:}  $P =0.8 P_c$,
\textit{Bottom center:}  $P =0.6 P_c$,   \textit{Bottom right:} $P =0.2 P_c$.
The latter sequence shows that the swallowtail develops unphysical branches and then vanishes entirely.
 }
\label{GTd5km1}
\end{figure*}

Summarizing we have observed for the first time the occurrence of two  physical critical points for spherical ($k=1$)
black holes and one critical point for  hyperbolic ($k=-1$) black holes.  We must have  both couplings nonzero for this to take place.

 Again we see that even if $e=0$ there are physical critical points. For $k=1$ we get regions with either one or
two physical critical points,   and only one for  $k=-1$. In the latter case, there are some regions having negative mass (white region) and both couplings must be non-zero in order to get physical critical points.

For regions of parameter space having only a single critical point, five dimensional spherical black holes ($k=+1$), have a   first order VdW transition behaviour similar to that in the four dimensional case.  The critical exponents are the mean field theory values.

However in regions  of parameter space having two physical critical points, there is new behaviour.  We illustrate this in figure~\ref{PT5dk1},  which shows that there are two first order phase transitions.    The  transition at $T= T_c$ is standard VdW behaviour. But the second phase transition at $T = T_{\overline{c}}$ is that of
`reverse VdW' behaviour: it is a transition from one phase for $T < T_{\overline{c}}$ to two distinct small/large black hole phases for
$T > T_{\overline{c}}$.  Note that for a sufficiently large temperature the pressure becomes negative and the asymptotic structure of the spacetime is no longer AdS.   Consequently there is an upper bound on the temperature of AdS black holes. We illustrate this in figure~\ref{PT5dk1}.
 Note that curves having $P<0$ over a finite range of $T$
can be given physical meaning via the equal-area law \cite{Smailagic.2013}. Referring to figure~\ref{PT5dk1}, we see that although $P<0$ corresponds to a different asymptotic structure from that of AdS,  the equal area law implies that $P$ never actually attains these negative values, but rather remains constant and positive as the phase transition takes place.
The coexistence line of  the two distinct phase small/large  has a  critical point
at a minimal value of $T$ in contrast to that of a standard  VdW phase transition.
This phenomenon has been previously observed  for black branes~\cite{Hennigar:2017umz} and in cubic gravity \cite{mir:2018mmm}.

The existence of a standard VdW phase transition followed by a  reverse VdW transition at higher temperature is shown in figure~\ref{PT5dk1}. One might anticipate that with an appropriate choice of parameters that
these two critical points (illustrated in the top right diagram of figure~\ref{PT5dk1}) could merge, yielding
an isolated critical point  (see the discussion for the six dimensional case in the next section).  We find that this only occurs for either negative entropy and/or negative mass; these unphysical conditions persist in a neighbourhood of the merged critical point.

In figure~\ref{GTd5}, we illustrate the behaviour of the Gibbs free energy in both the low-temperature region containing
a standard VdW transition and in the high-temperature region containing a reverse VdW transition.  In the former case, we obtain the standard swallowtail behaviour for $P < P_c$.  In the latter case, for $P$ sufficiently smaller than  $P_{\overline{c}}$ there are no phase transitions, as the upper right part of figure~\ref{GTd5} indicates.  As
$P \to  P_{\overline{c}}$ we obtain swallowtail behaviour, shown in the lower right part of figure~\ref{GTd5}.

For $P =P_{\overline{c}}$, in addition to  negative specific heat we get regions with negative entropy (shown by the blue curve) and with both negative specific heat and negative entropy  (shown by the purple curve); these are all unstable. For larger temperature the black hole solutions become stable (black line).
 Figure~\ref{GTd5} shows that in addition to a reverse VdW transition,  at high temperatures there are three black hole solutions with positive specific heat, with one having a minimal Gibbs free energy.

  For the $k=-1$ hyperbolic black hole in  $d=5$ we get only a reverse VdW transition, in which higher temperatures
 have two distinct phases,  as depicted in figure \ref{PT5dkm1}.  The   Gibbs free energy in figure \ref{GTd5km1} exhibits swallowtail behaviour for pressures a bit less than the critical pressure.  However  for larger values of pressure one of the branches corresponds to solutions with negative mass (depicted by the green curve) and the phase transition is no longer physical. For even lower pressure (bottom left diagram in figure~\ref{GTd5km1}) we see that unstable black holes with negative specific heat have lower free energy than those with positive specific heat.  It is reasonable to expect that  there is now a zeroth order phase transition between the two black curves in this figure.
 For even smaller $P$, the unphysical parts of the  branches shrink and (apart from a small red region)
become stable. As in the $k=1$ case, there are again three black hole solutions with positive specific heat, with one having a minimal Gibbs free energy.

\subsection{Critical behaviour in six dimensions}

Turning now to six dimensions, we note from  \reef{asympf} that for the linearized field equations  the contribution of the cubic term  drops out\footnote{In  eight dimensions the quartic term drops out.}~\cite{Bueno:2016xff}.
This yields some simplification, but the analysis is still somewhat complicated.
The equation of state takes the following form
\beqa
 P=\frac{T}{v}-\frac{3 k}{4 \pi  v^2}+\frac{2 \beta_3 T^3}{v^3}+ \left(\frac{3 \beta_3 k}{2 \pi  v^4}+\frac{3 \alpha_4 k^2}{2 \pi ^2 v^6}\right)T^2-\frac{\alpha_4 T^4}{v^4}-\frac{\beta_3 k}{8 \pi ^3 v^6}+\frac{e^2}{v^8},
\eeqa
with the explicit  expression
\beqa
T_c&=&\Big(3\big((36\pi^7\beta_3^3k+24\alpha_4^2\pi^7k){v_c}^{18}
+180\alpha_4\pi^6{v_c}^{16}\beta_3^2k^2+(162\pi^5{\beta_3}^4k
+276\alpha_4^2\pi^5{\beta_3}k){v_c}^{14}\nonumber\\
&&\left.+(256\alpha_4^2\pi^8e^2
+1476{\alpha_4}\pi^4{\beta_3}^3k^2+960\pi^8{\beta_3}^3e^2-144
{\alpha_4}^3k^2\pi^4){v_c}^{12}+(2880{\alpha_4}\pi^7{\beta_3}^2k e^2\right.\nonumber\\
&&\left.
+2988\pi^3{\beta_3}^2{\alpha_4}^2k-108\pi^3{\beta_3}^5k){v_c}^{10}
+(-1152{\alpha_4}^3k^2\pi^2{\beta_3}-459{\alpha_4}k^2\pi^2{\beta_3}^4
\right.\nonumber\\
&&\left.+2304\pi^6{\beta_3}^4k^2e^2-384\pi^6{\beta_3}{\alpha_4}^2e^2k^2){v_c}^8
+(-3888{\alpha_4}^4k\pi+11040{\alpha_4}k\pi^5{\beta_3}^3e^2\right.\nonumber\\
&&\left.-3072
{\alpha_4}^3k\pi^5e^2){v_c}^6+(612{\alpha_4}^3k^2{\beta_3}^2+7776\pi^4
{\beta_3}^2{\alpha_4}^2k^2e^2+10240{\alpha_4}\pi^8{\beta_3}^2e^4){v_c}^4
\right.\nonumber\\
&&\left.-12288{\alpha_4}^3k\pi^3{v_c}^2{\beta_3}e^2-10368{\alpha_4}^4k^2\pi^2e^2
-16384{\alpha_4}^3k^2\pi^6e^4\big)\Big)\right.\nonumber\\
&&\left.
\times 1\Big/\Big(\pi {v_c}^3\big((72{\alpha_4}^2\pi^7+144
\pi^7{\beta_3}^3){v_c}^{16}+612{v_c}^{14}\pi^6{\beta_3}^2k{\alpha_4}\right.\nonumber\\
&&\left.+(324\pi^5
{\beta_3}^4k^2+792{\alpha_4}^2\pi^5{\beta_3}k^2){v_c}^{12}+(3096\pi^4{\beta_3}^3
k{\alpha_4}-864\pi^4{\alpha_4}^3k){v_c}^{10}\right.\nonumber
\eeqa
\beqa
&&\left.
+(7146\pi^3{\alpha_4}^2k^2
{\beta_3}^2+192\pi^7{\beta_3}^2e^2{\alpha_4}+486\pi^3{\beta_3}^5k^2){v_c}^8
+(3072\pi^6{\alpha_4}^2k{\beta_3}e^2\right.\nonumber\\
&&\left.-1512\pi^2{\alpha_4}^3k{\beta_3}
-8640\pi^6{\beta_3}^4e^2k+2187\pi^2{\beta_3}^4k{\alpha_4}){v_c}^6+
(-9720\pi{\alpha_4}^4k^2\right.\nonumber\\
&&\left.-38880\pi^5{\beta_3}^3e^2{\alpha_4}k^2+36
\pi{\alpha_4}^2k^2{\beta_3}^3-9216\pi^5{\alpha_4}^3k^2e^2){v_c}^4\right. \\
&&\left.+
(-1536\pi^4{\alpha_4}^2k{\beta_3}^2e^2-2916{\beta_3}^2k{\alpha_4}^3)
{v_c}^2+51840\pi^3{\beta_3}e^2{\alpha_4}^3k^2+16384{\alpha_4}^2\pi^7
{\beta_3}e^4\big)\Big) \right.\nonumber
\eeqa
for the  critical temperature and
\beqa
 &&(-24{v_c}^{10}\pi^5{\beta_3}{\alpha_4}-72{v_c}^8\pi^4{\beta_3}^3k-324{v_c}^6\pi^3
{\beta_3}^2{\alpha_4}k^2+432{\alpha_4}^3k^4\pi {v_c}^2){T_c}^2\nonumber\\
&&\left.
+(-12{v_c}^5\pi^2{\beta_3}^2{\alpha_4}k+24{v_c}^9\pi^4{\beta_3}{\alpha_4}k
+256{v_c}^3\pi^5{\beta_3}{\alpha_4}e^2+24{v_c}^{11}\pi^5{\beta_3}^2
\right.\nonumber\\
&&\left.-72{\alpha_4}^2k^2\pi^3{v_c}^7)T_c+768{\alpha_4}^2k^2e^2\pi^3
-36{\alpha_4}^2k^3{\beta_3}{v_c}^2-18{v_c}^10\pi^4{\beta_3}^2k
\right.\nonumber\\
&&\left.+27{v_c}^6\pi^2{\beta_3}^3k-480{v_c}^4\pi^5{\beta_3}^2e^2
+72{\alpha_4}^2k^3\pi^2{v_c}^6 =0 \right.
\eeqa
being the equation yielding the critical volume.

Setting the cubic coupling to zero, the associated critical temperature is
\beqa
t_c&=&\frac{1}{3}\Big(-1296 \pi  \alpha_4^2 e^2 k^2-2048 \pi^5 \alpha_4 e^4 k^2+ \left(-486 \alpha_4^2 k-384 \pi^4 \alpha_4 e^2 k\right)v_c^6\nonumber\\
&&\left.+ \big(32 \pi^7 e^2
-18\pi^3 \alpha_4 k^2\big)v_c^{12}+3 \pi^6 k v_c^{18}\Big)\right.\nonumber\\
&&\left.\times 1\Big/ \Big(\pi v_c^7 \big(-135 \alpha_4^2 k^2-128 \pi^4 \alpha_4 e^2 k^2-12 \pi^3 \alpha_4 k v_c^6+\pi^6 v_c^{12}\big)\Big), \right.\label{tcd6}
\eeqa
and the critical volume obeys the relation
\beqa
36 \alpha_4 k^2 t_c^2  v_c^2+64 \pi ^2 e^2+6 \pi  k  v_c^6-6 \pi ^2 t_c  v_c^7  = 0 \; .
\eeqa
There is a singularity in the critical temperature at
\beqa
 v_c^6=\frac{6 \alpha_4 k}{\pi^3}+\frac{k \sqrt{  171 \alpha_4^2+128 \pi^4 \alpha_4 e^2 }}{\pi^3},
\eeqa
 however it is possible to remove this singularity via a
suitable choice of $\alpha_4$ in \reef{tcd6}.  For $k=\pm1$ the solutions to these equations indicate that for $\alpha_4=-130.39277\ e^2,\ -225.86056\ e^2$ this takes place for any values of electric charge. However, none of these values yield physical critical points;  if $\alpha_4<0$ we get $\gamma^2<0$  in the first case, and the critical temperature or critical volume becomes negative in the second case.

\begin{figure*}[htp]
\centering
\begin{tabular}{cc}
\includegraphics[scale=.3]{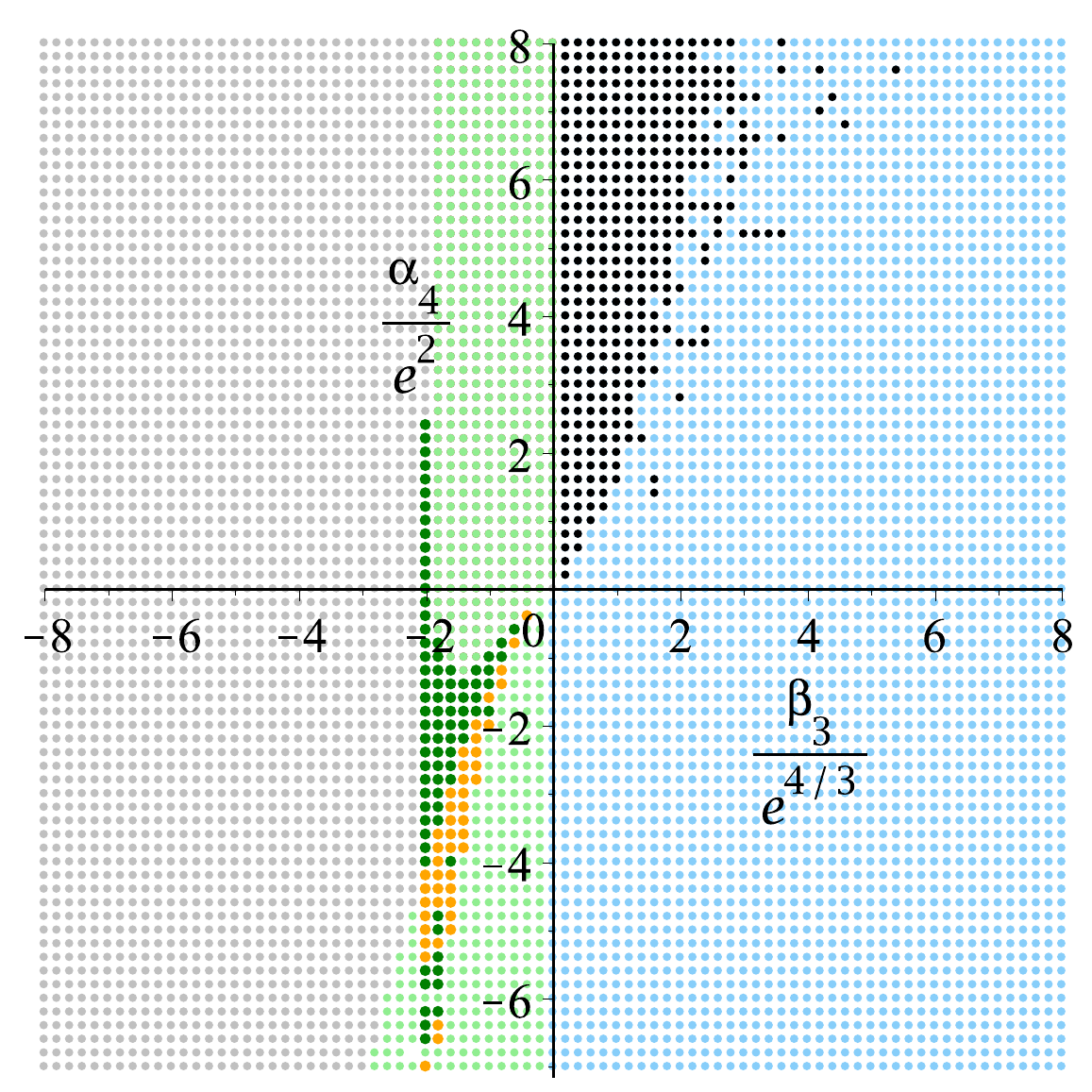}&\quad\quad\quad\quad
\includegraphics[scale=.3]{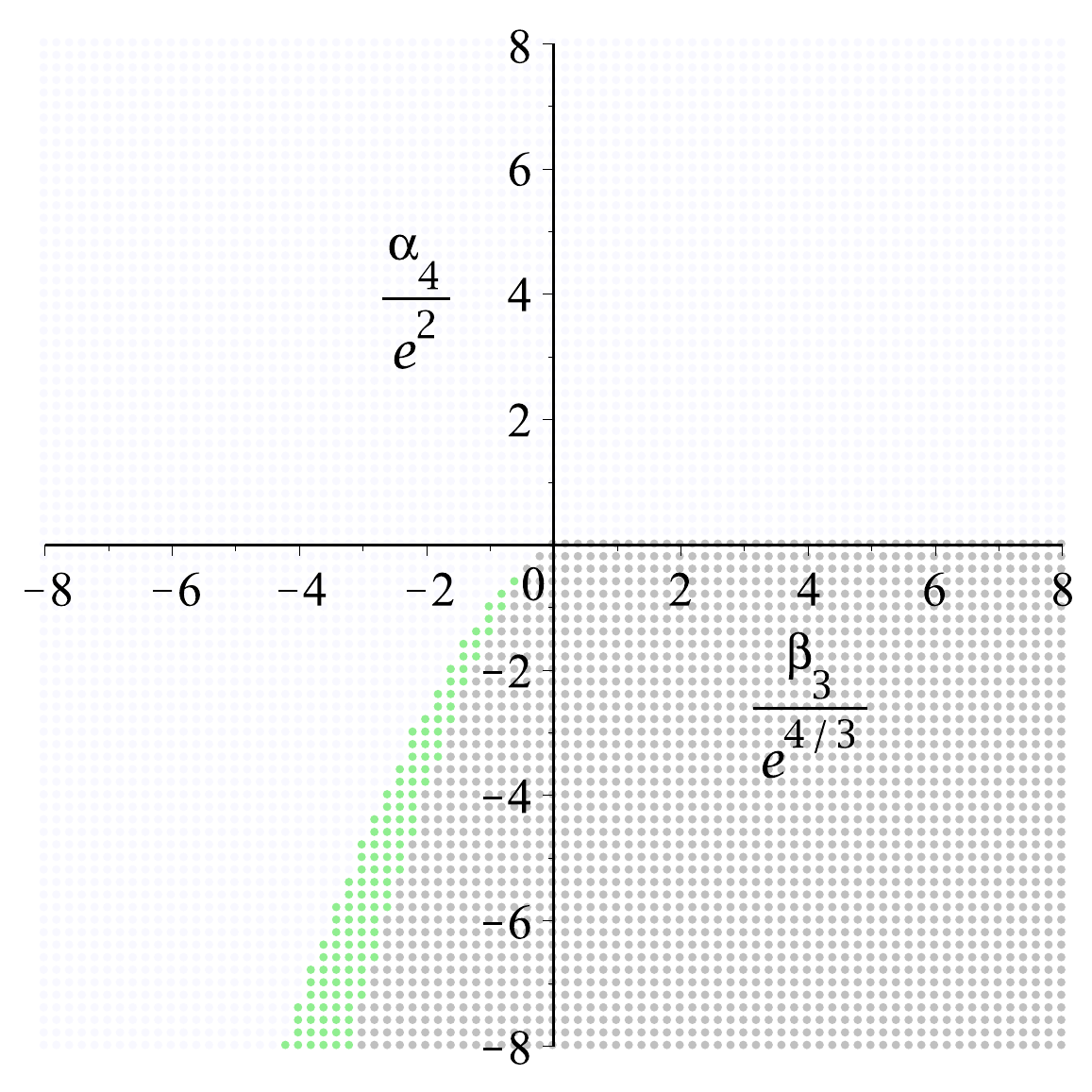}
\\
\includegraphics[scale=.3]{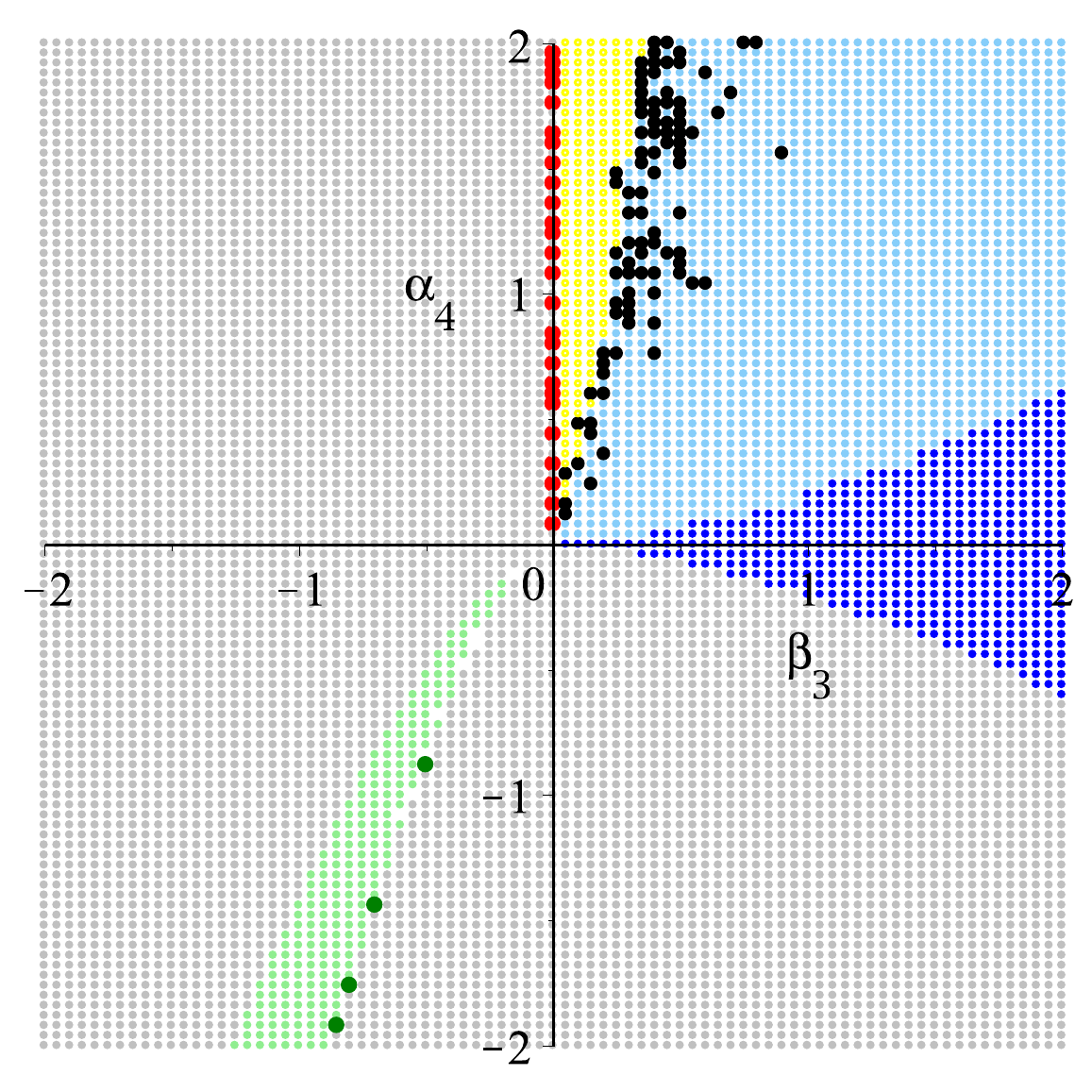}&\quad\quad\quad\quad
\includegraphics[scale=.3]{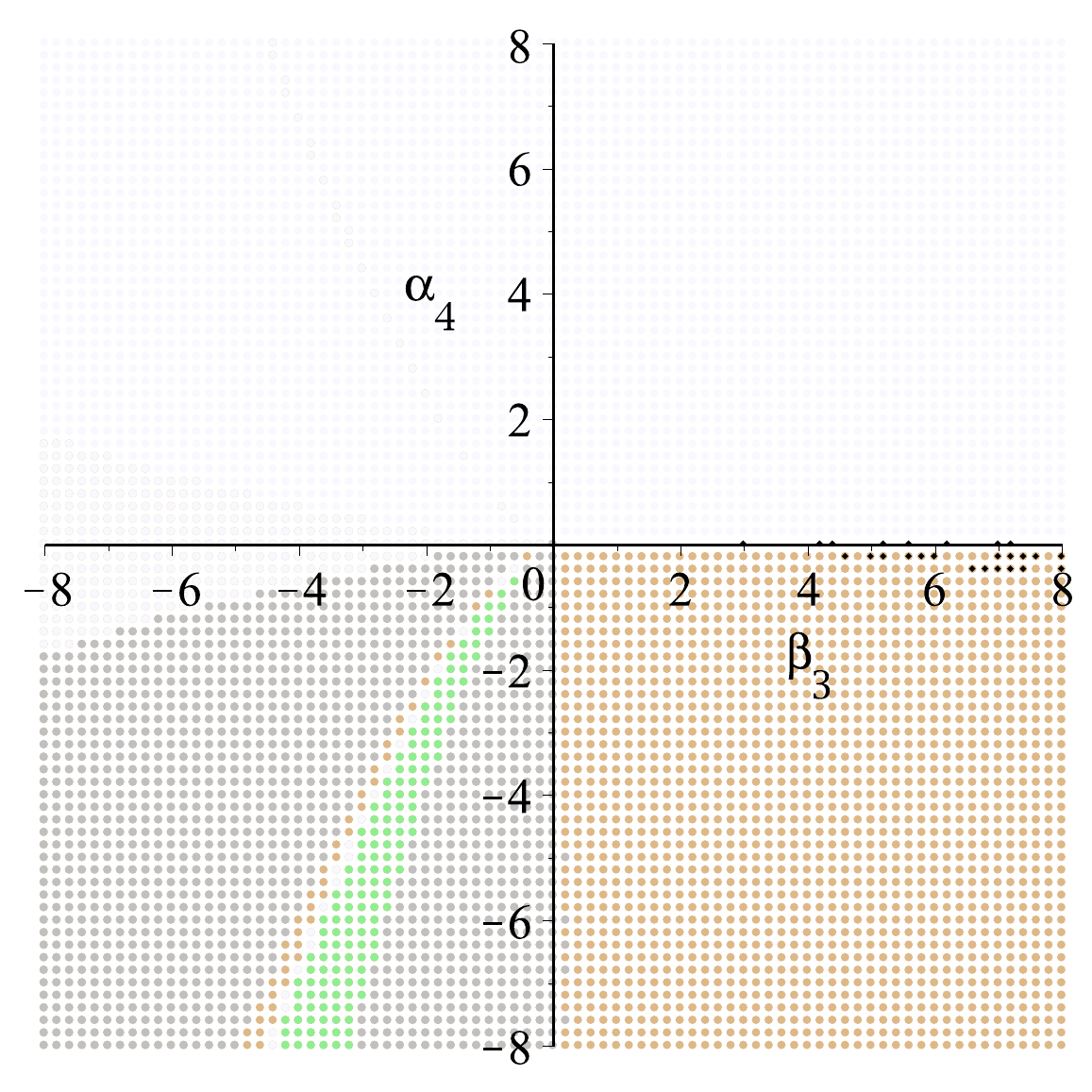}
\\
\end{tabular}
\caption{{\bf Number of Critical Points as a function of couplings in $d=6$} (color online). At top left is the spherical $k=1$ case, at top right the hyperbolic $k=-1$ case; at bottom left is the spherical $k=1$, $e=0$ case,  at bottom right the hyperbolic $k=-1$, $e=0$ case,
 with colour coding  given in table \ref{table:nonlin}.
Green, dark green, and orange regions respectively indicate one, two, and three physical critical points.
Blue  regions indicate single critical points with  $\gamma^2<0$, and  black indicate single critical points  both  $S<0$ and $\gamma^2<0$. Dark blue,
red, and yellow regions respectively indicate $\gamma^2<0$,  $S<0$, and  both $S<0$ and $\gamma^2<0$, but with
two critical points. Light brown regions have solutions with no asymptotic positive real value for $f_{\infty}$.
Grey regions have no critical points, and white regions indicate negative mass.}
\label{domain6d}
\end{figure*}

We plot in  figure \ref{domain6d} the   possible critical points in the $(\beta_3,\alpha_4)$ plane. Physical critical points appear only for $\beta_3 < 0$, and there can be as many as three for certain ranges of
$(\beta_3,\alpha_4)$ if $k=1$  but only one for the hyperbolic ($k=-1$)  case.
Other possible critical points have one or both of $\gamma^2<0$ and $S<0$.  For vanishing charge we also have a region of at most two critical points for both $k=1$ and just one critical point for $k=-1$;  for vanishing $\alpha_4$ there are still two physical critical points if $k=+1$,  studied in detail in \cite{mir:2018mmm}.

 Clearly the maximal number of critical points for a given value of
$(\beta_3,\alpha_4)$  depends on both  horizon geometry and  dimension.   We do not need to have both couplings non-zero to obtain physical critical points. However in the absence of charge or when $k=-1$ (both with or without charge) both couplings must be nonzero.


\begin{figure*}[h]
\centering
\begin{tabular}{cc}
\includegraphics[scale=.25]{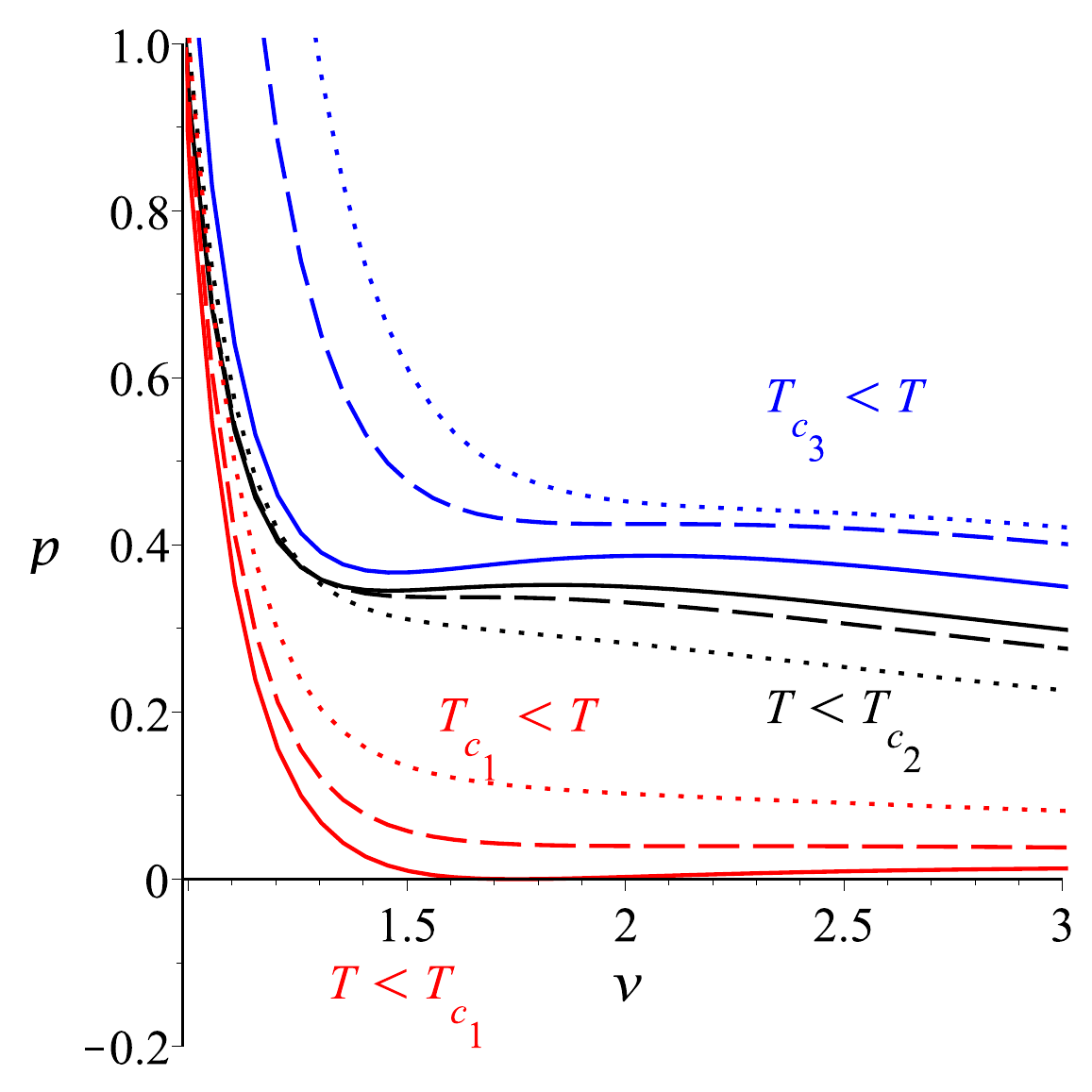}&
\includegraphics[scale=.25]{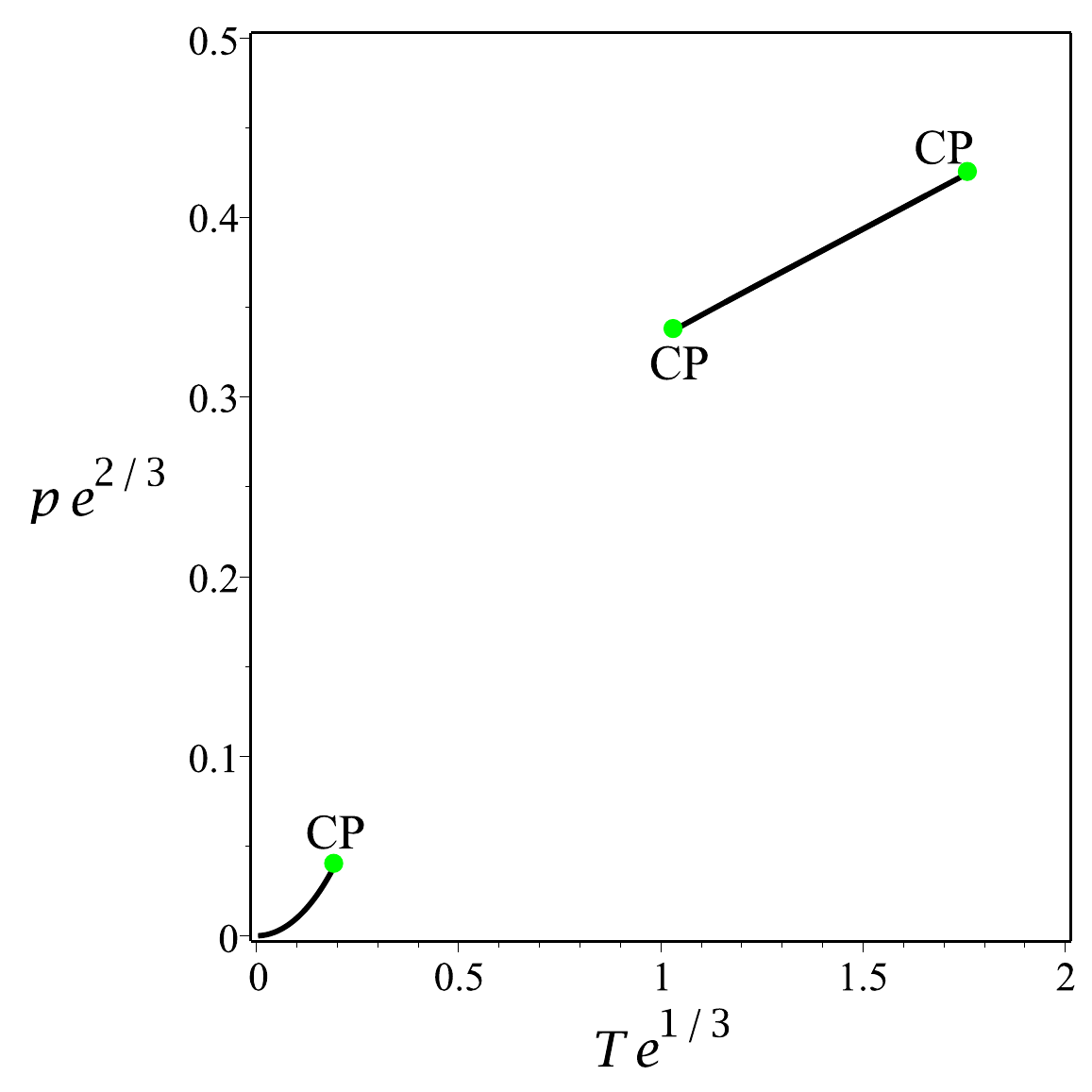}
\includegraphics[scale=.25]{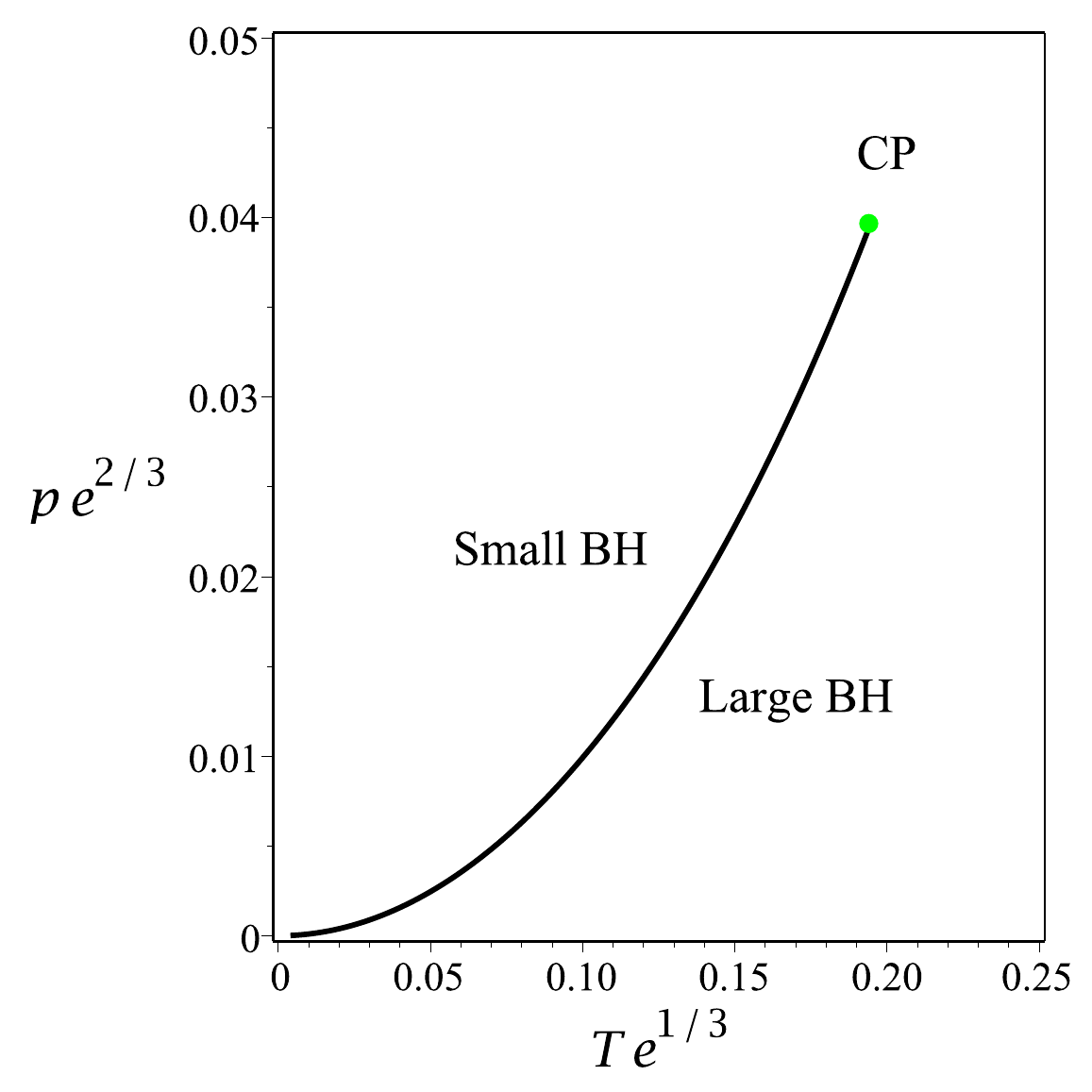}
\\
\includegraphics[scale=.25]{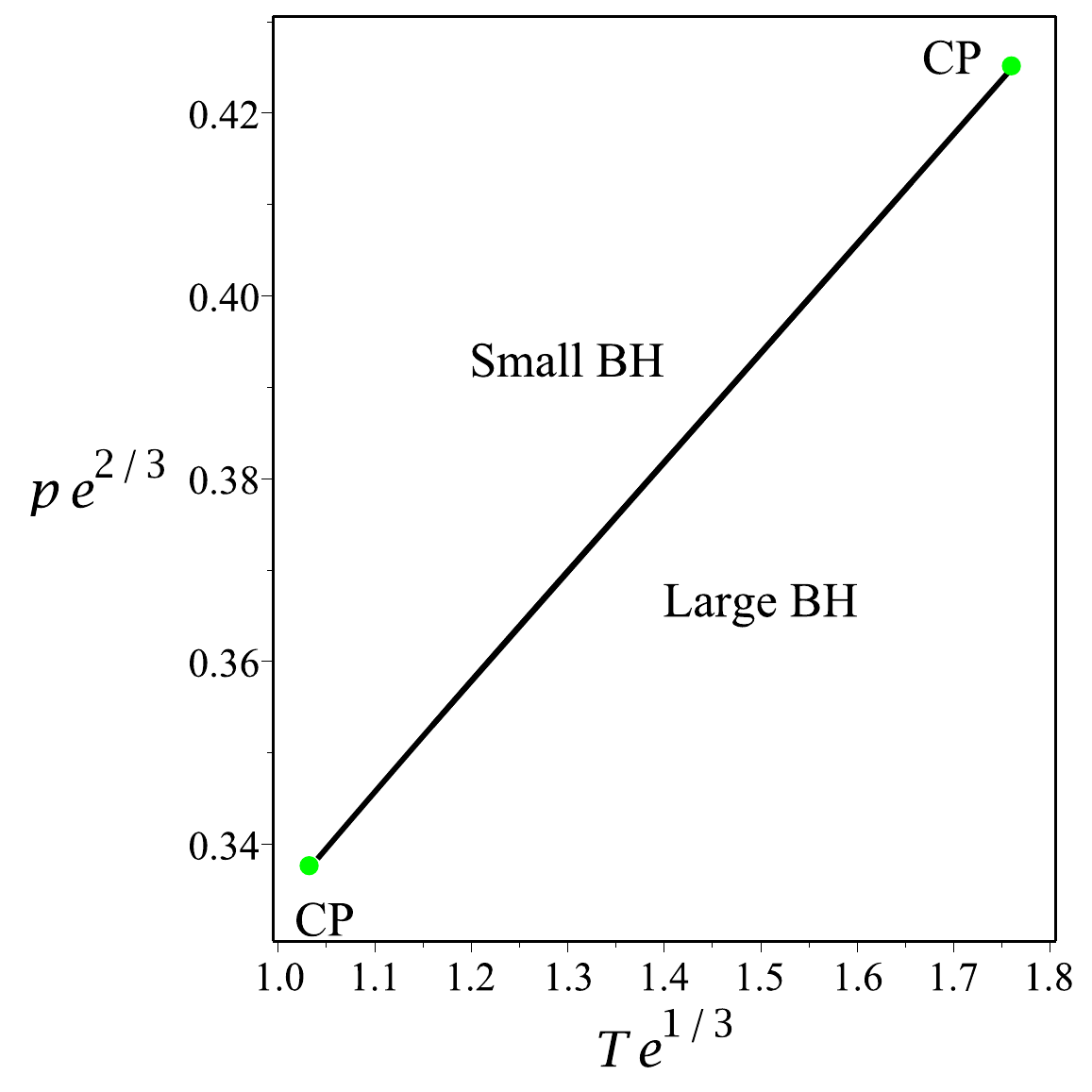}&
\includegraphics[scale=.25]{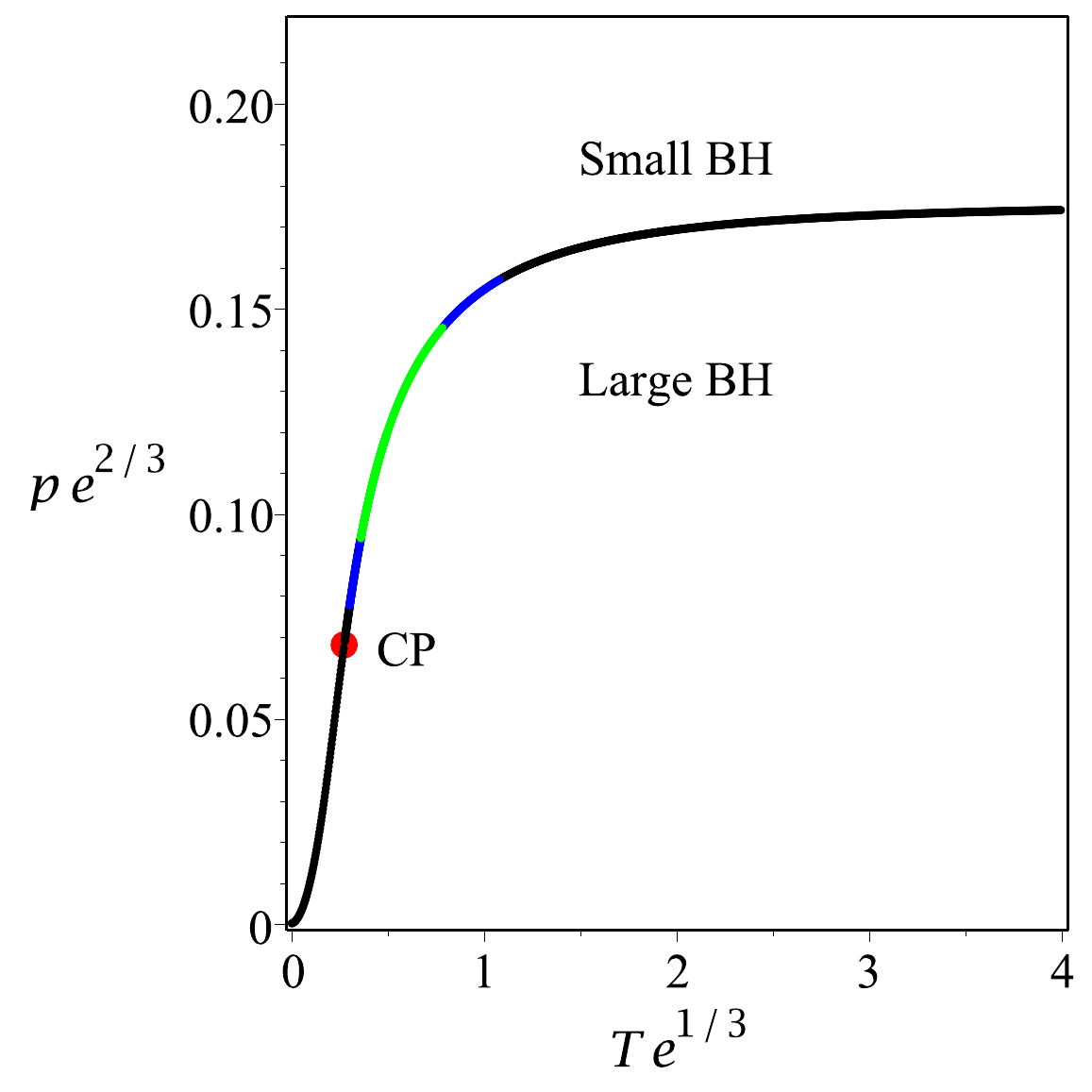}
\includegraphics[scale=.25]{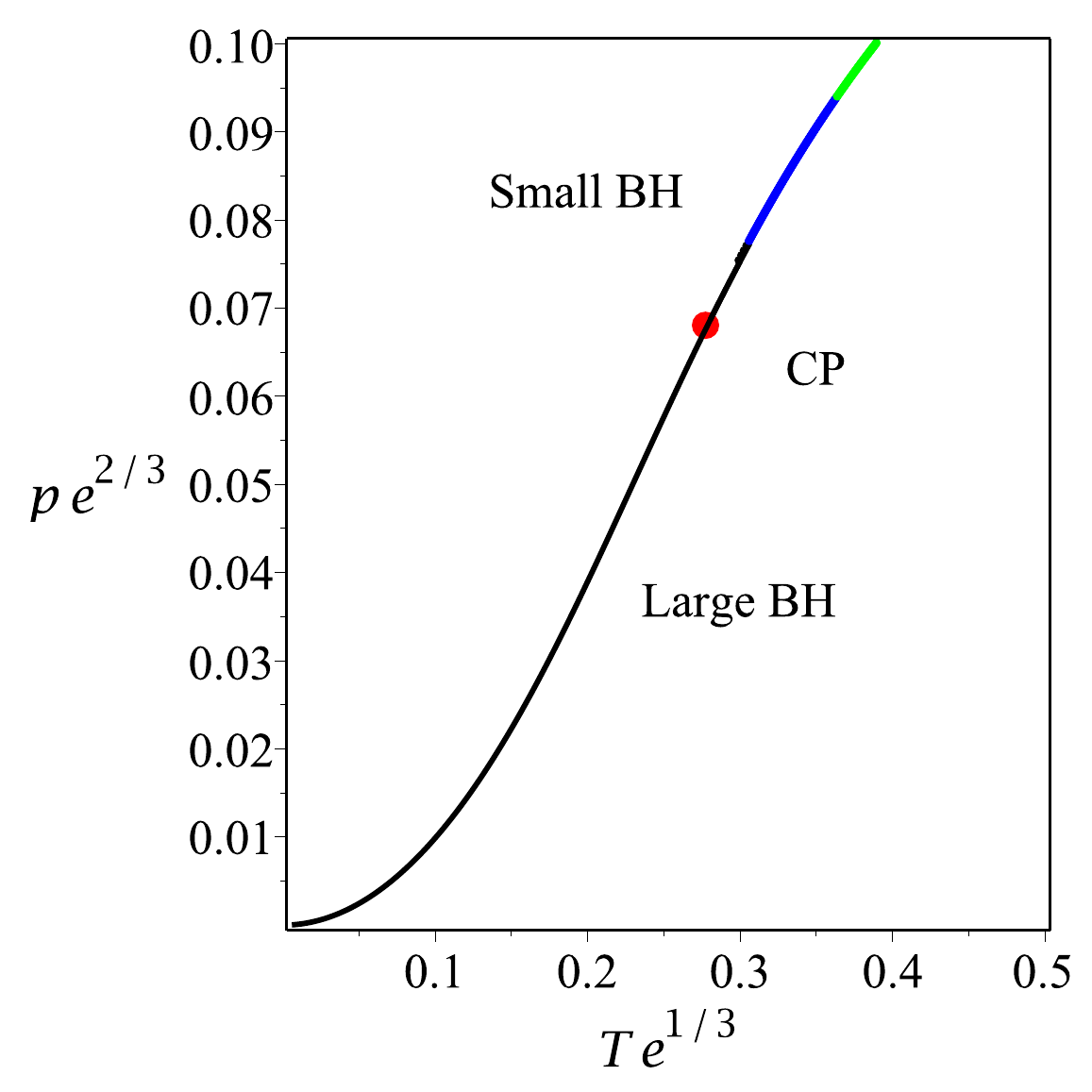}
\\
\end{tabular}
\caption{\textbf{Three first order phase transition in six dimensions for $d=6$ and $k=1$}(color online).
 \textit{Top left }:  The reverse VdW transition occurs at intermediate temperatures between two standard
VdW transitions, with curves below (solid), at (dashed), and above (dotted) critical temperature displayed for each.
 For the cold VdW transition, $T=T_{c_1}$ (dashed red line), $T=0.60613 T_{c_1}$ (solid red line) $T=1.7 T_{c_1}$ (dotted red line); for the reverse VdW one  $T=0.8 T_{c_2}$ (dotted black line), $T=T_{c_2}$ (dashed black line), $T=1.1 T_{c_2}$ (solid black line); for the hot VdW transition $T=0.8 T_{c_3}$ (solid blue line), $T=T_{c_3}$ (dashed blue line) and $T=1.1 T_{c_3}$ (dotted blue line). \textit{Top center}:  The phase diagram, with green dots denoting the critical points and black lines indicating  three first order phase transitions for temperatures smaller and larger than the three illustrated critical points; we have chosen $e=1$, $\beta_3=-3/5 e^{4/3}$ and $\alpha_4=-4/5 e^2$, with $T_{c_1}e^{1/3}\approx0.19440$, $P_{c_1}e^{2/3}\approx0.03960$ and $T_{c_2}e^{1/3}\approx1.03337$, $P_{c_2}e^{2/3}\approx0.33752$ and $T_{c_3}e^{1/3}\approx1.76108$, $P_{c_3}e^{2/3}\approx0.42503$.   \textit{Top right and bottom left}: Close ups of the curves of the top center plot. \textit{Bottom center}: For  $e=1$, $\beta_3\approx -2.04428 e^{4/3}$ and $\alpha_4=-4/5 e^{2}$ we obtain an isolated critical point (in red); numerically we find that the critical temperature and pressure are $T_c e^{1/3} \approx 0.27826$ and $P_c e^{2/3} \approx 0.06790$. Blue and green lines show negative entropy and negative mass respectively. \textit{Bottom right}: A magnification of the bottom center plot close to the isolated critical point.
}
\label{PT6d}
\end{figure*}

The appearance of three physical critical points in the  $d=6,~k=+1$ case stands in contrast to previous studies. This remarkable feature   has not been observed before, and its  occurrence   is related to the
conjunction of  electric charge,  cubic,  and quartic couplings all being nonzero. Figure~\ref{PT6d} shows that
there is a reverse VdW transition in between two standard ones, one at cold temperatures $T<T_{c_1}$ and the other at high temperatures $T >T_{c_3}$, with the critical temperature  $T=T_{c_2}$ of the reverse transition  $T_{c_1}< T_{c_2} < T_{c_3}$.

The  curves in figure~\ref{PT6d} correspond to phase transitions that obey  Maxwell's equal area law \cite{Smailagic.2013}
with the actual pressure remaining positive during the phase transition (despite the curve indicating $P$ becomes negative over a finite range of $T$) as noted earlier.
At sufficiently low temperatures the $P-v$ curves cross the horizontal axis, yielding unphysical behaviour since  the asymptotic structure is no longer AdS.

Critical points and coexistence lines are also displayed  in figure~\ref{PT6d}.
Numerical analysis confirms that for typical values of the parameters, each of the three critical points are characterized by mean field theory critical exponents, a hallmark of the end point of a first order phase transition.
Note that the two critical points at high temperature are joined by a coexistence line.

For certain choices of the couplings and charge these two disjoint lines merge into each other, and
an  \textit{isolated critical point} appears at the merge point.
This new critical point is characterized by critical exponents that differ from the mean field theory ones. This phenomena was first observed in Lovelock and quasi-topological gravity \cite{Frassino:2014pha, Dolan:2014vba,Hennigar:2015esa},  where isolated critical points were found to occur for hyperbolic horizons and massless AdS black holes.  A thermodynamic singularity, at which
pressure remains constant for any temperature and isotherms cross and reverse,
 was also found to occur at this point. However for Lovelock and quasi-topological black holes accompanied with conformal scalar hair, isolated critical points have been observed in five and higher dimensions  for massive black holes without coinciding with a thermodynamic singularity \cite{Hennigar:2016ekz,Dykaar:2017mba}.
 More recently, in GQG cubic gravity only,
isolated critical points have been found  in six dimensions with no scalar hair for spherical black holes \cite{mir:2018mmm}.  In all of these cases there are initially only two critical points that   converge to one isolated critical point.

Here we see for the first time three critical points, two of which merge to form an isolated critical point. Furthermore these isolated critical points do not correspond to any thermodynamic singularity, $\ie$ the $P-T$ curve does not have a zero slope. From the relation \reef{expcoeff} we find that the coefficient $B$ vanishes and the associated critical exponents  are
\beq
\tilde{\alpha} = 0 \, , \quad \tilde{\beta} = 1 \, , \quad  \tilde{\gamma} = 2 \, , \quad \tilde{\delta} = 3 \, .\label{nonstanexp}
\eeq
which differ  from the standard critical exponents appearing in
\reef{exponents} but  are in accord with previous studies \cite{Frassino:2014pha, Dolan:2014vba,Hennigar:2015esa}. On either side of the isolated critical point the black holes satisfy all physical constraints, as the bottom right diagram in  figure~\ref{PT6d} indicates.

\begin{figure*}[htp]
\centering
\begin{tabular}{cc}
\includegraphics[scale=.3]{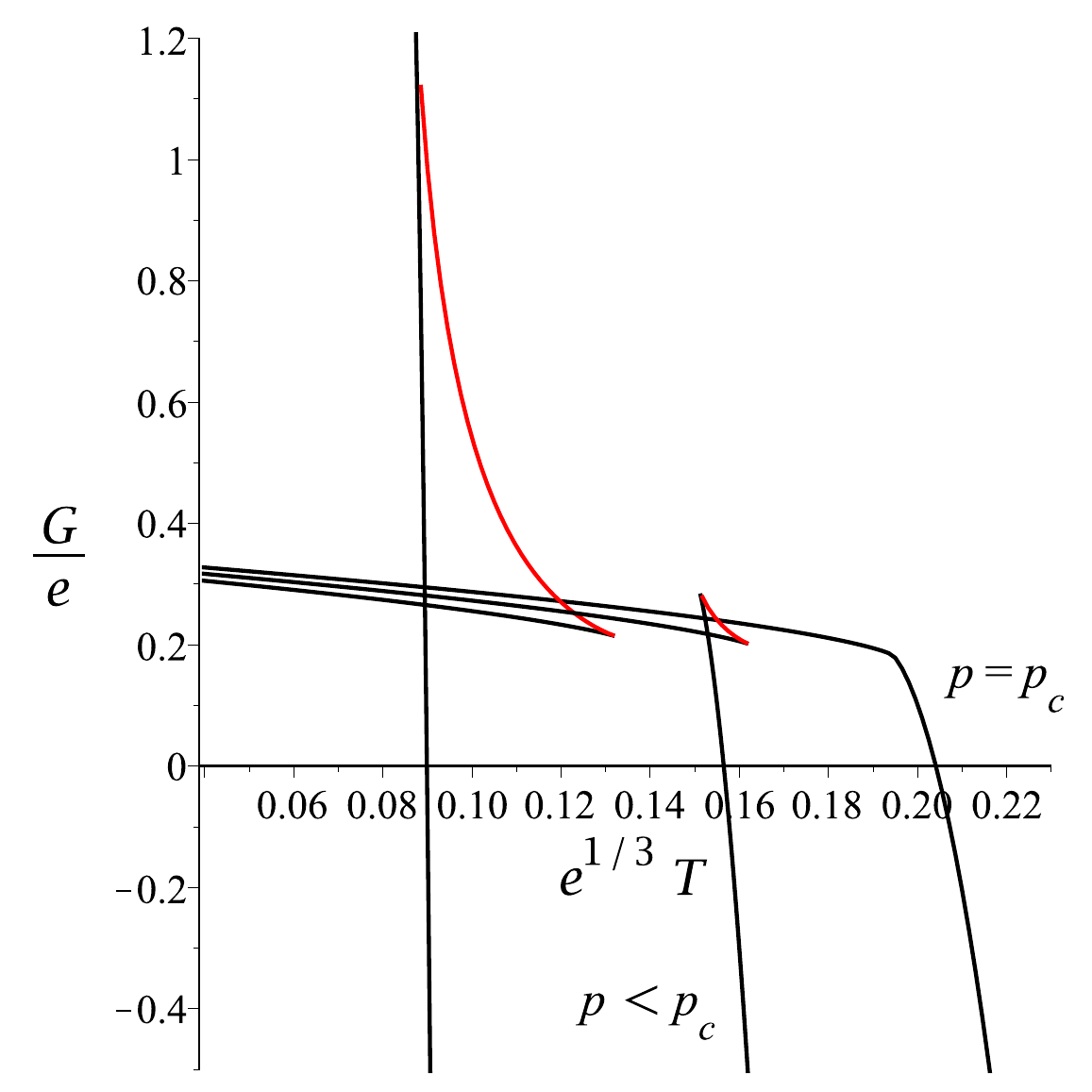}&\quad\quad\quad\quad
\includegraphics[scale=.3]{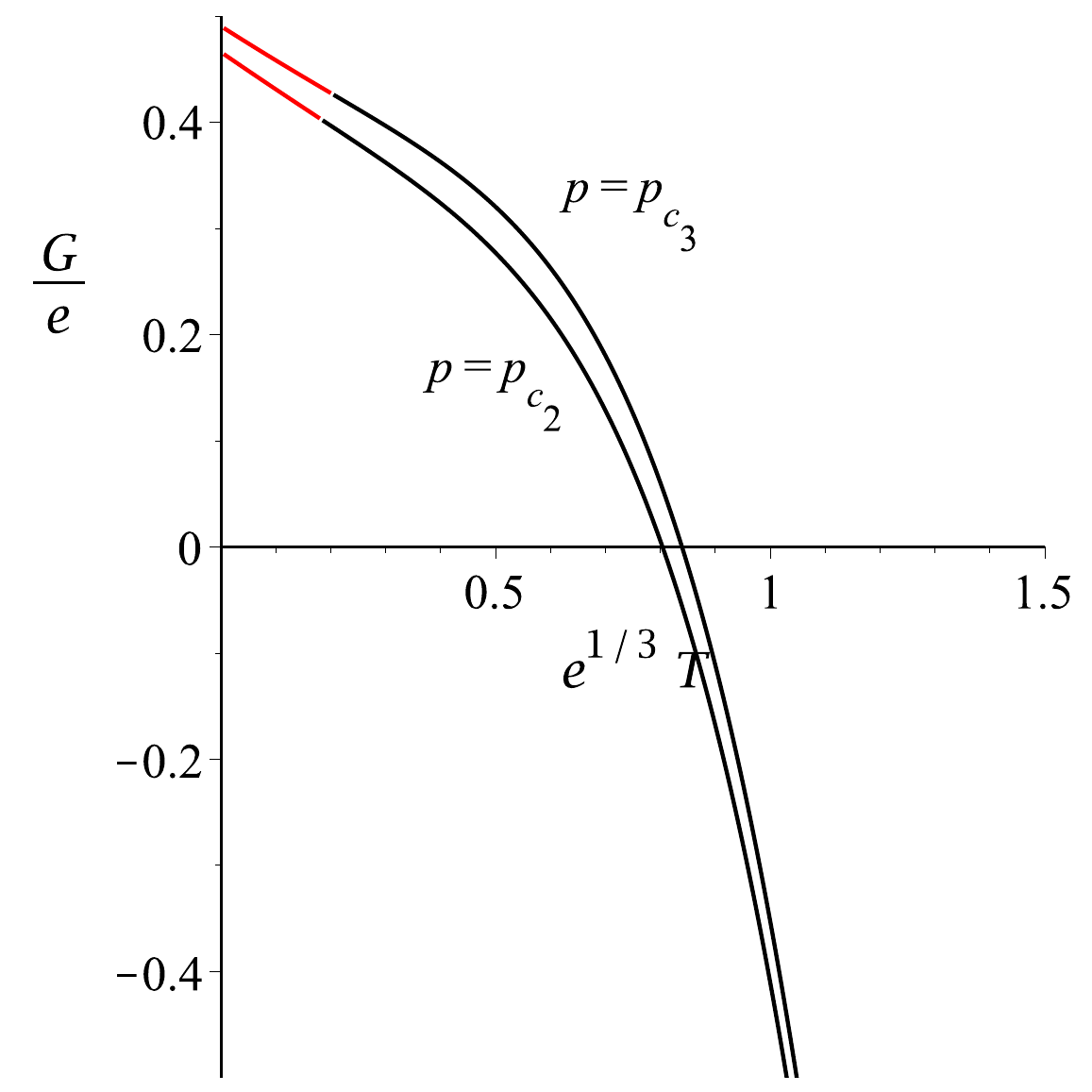}\\
\includegraphics[scale=.3]{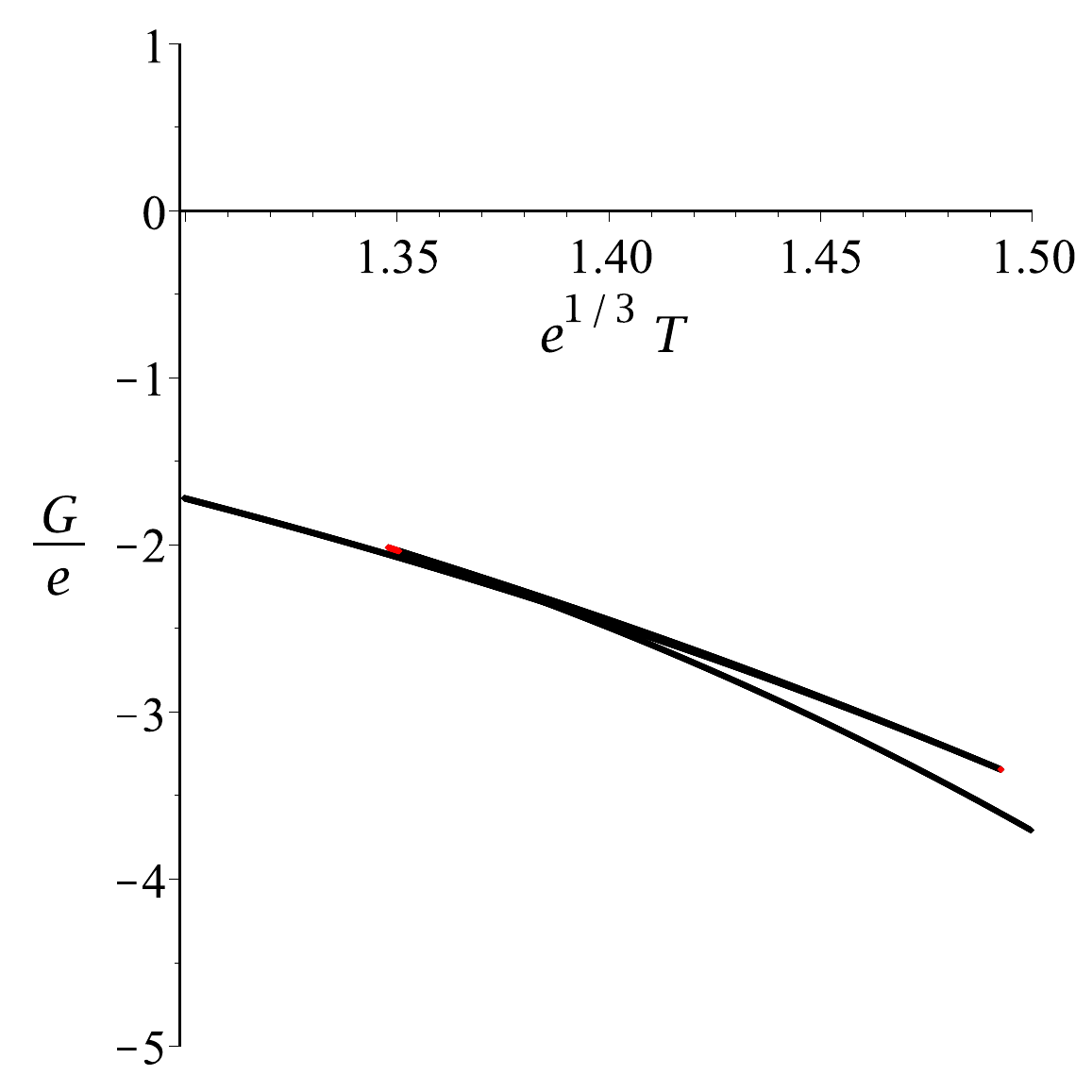}&\quad\quad\quad\quad
\includegraphics[scale=.3]{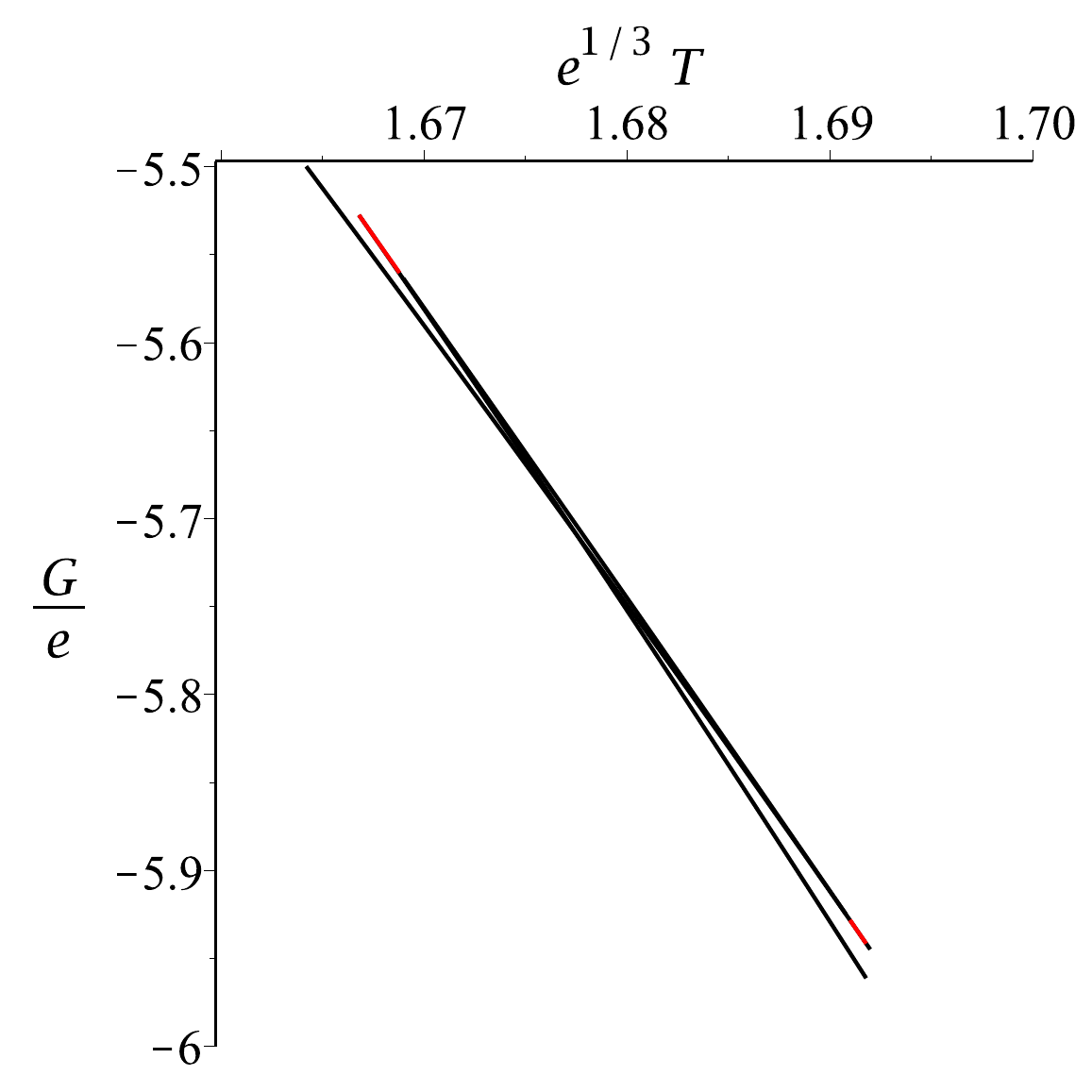}
\\
\end{tabular}
\caption{\textbf{Free energy of six dimensional spherical black holes} (colour online).  Plots of the Gibbs free energy  for $P =P_{c}$, $0.6 P_{c}$ and $0.2 P_{c}$ (top left), for  $P = P_{c_2},\ P_{c_3}$ (top right),  for $P = 1.12585 P_{c_2}$ (bottom left) and for $P =0.97639  P_{c_3}$ (bottom right). The latter two cases exhibit very sharp swallowtails.
In all cases red lines indicate negative specific heat.
For the choice of $e=1$, $\beta_3=-3/5 e^{4/3}$ and $\alpha_4=-4/5 e^{2}$, we have $P_{c_1} e^{2/3} \approx 0.03960$, $P_{c_2}e^{2/3}\approx 0.33752$ and $P_{c_3}e^{2/3}\approx 0.42503$.
}
\label{GT6d}
\end{figure*}

The behaviour of free energy with respect to temperature for spherical black holes is illustrated in figure~\ref{GT6d}.  At
low temperatures, for  $P =P_c$ the solution is stable and for $P < P_c$ a standard VdW transition occurs. For the two other critical points $P_{c_2},~P_{c_3}$ there is a region of the curve that has negative specific heat at low temperatures (red lines). For pressures  $P_{c_2} < P < P_{c_3}$ there are   reverse and standard transitions that are shown by the rather sharp swallowtails depicted  in the bottom graphs of figure~\ref{GT6d} from left to right respectively. In contrast to the lower temperature  ($P<P_c$)
swallowtails  depicted in the upper diagrams in  figure~\ref{GT6d}, these swallowtails
have positive specific heat everywhere apart from a few short segments.

\begin{figure*}[h]
\centering
\begin{tabular}{cc}
\includegraphics[scale=.25]{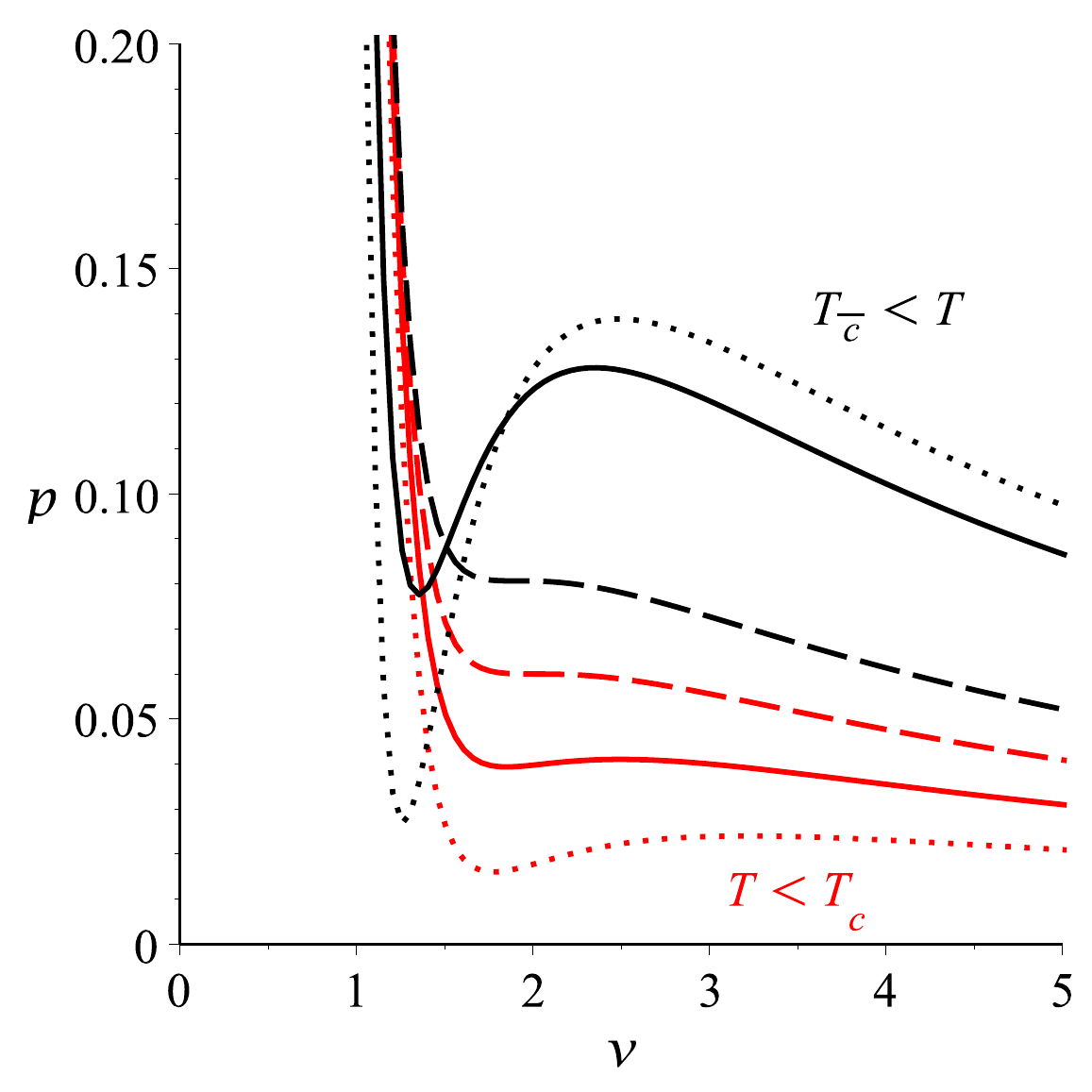}&
\includegraphics[scale=.25]{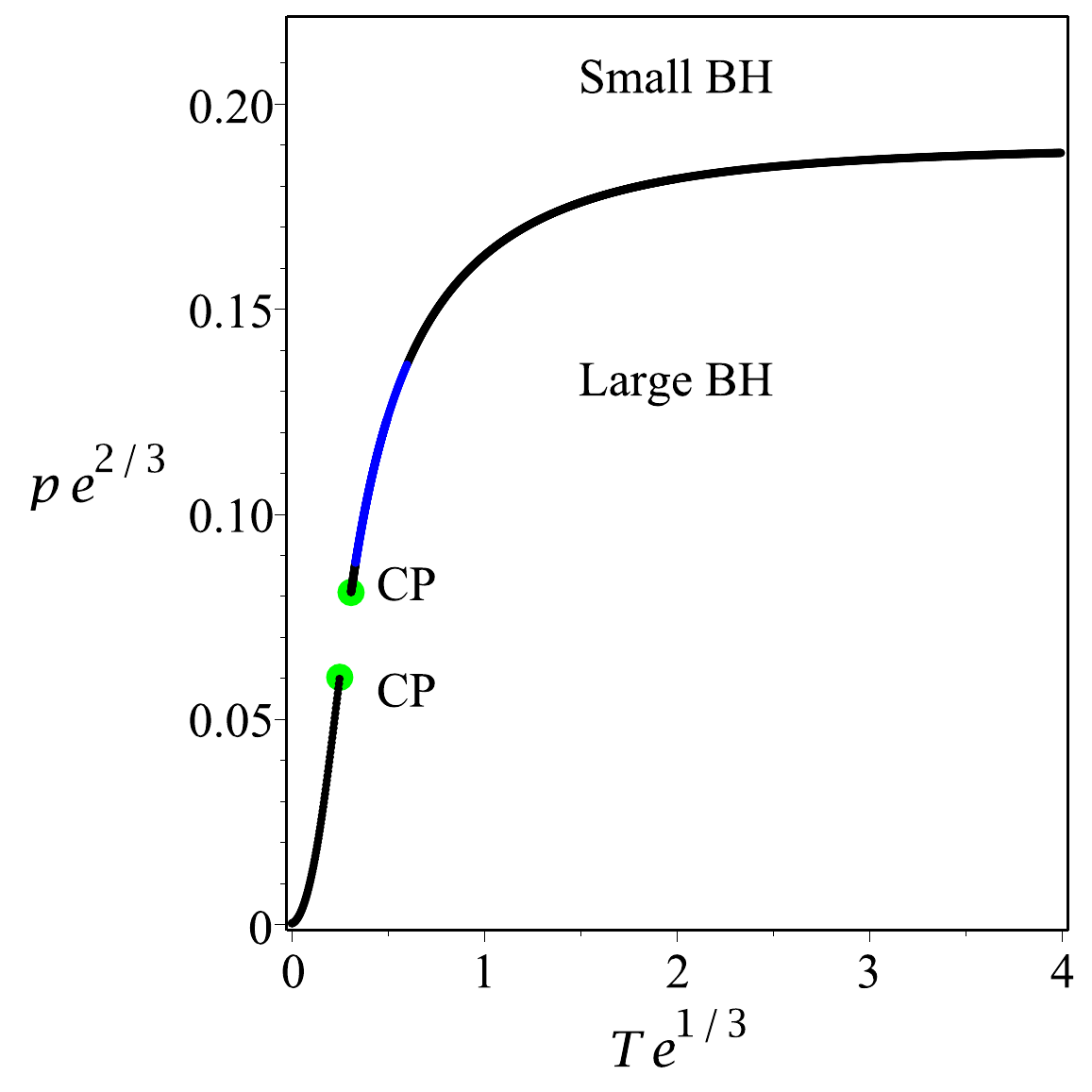}
\includegraphics[scale=.25]{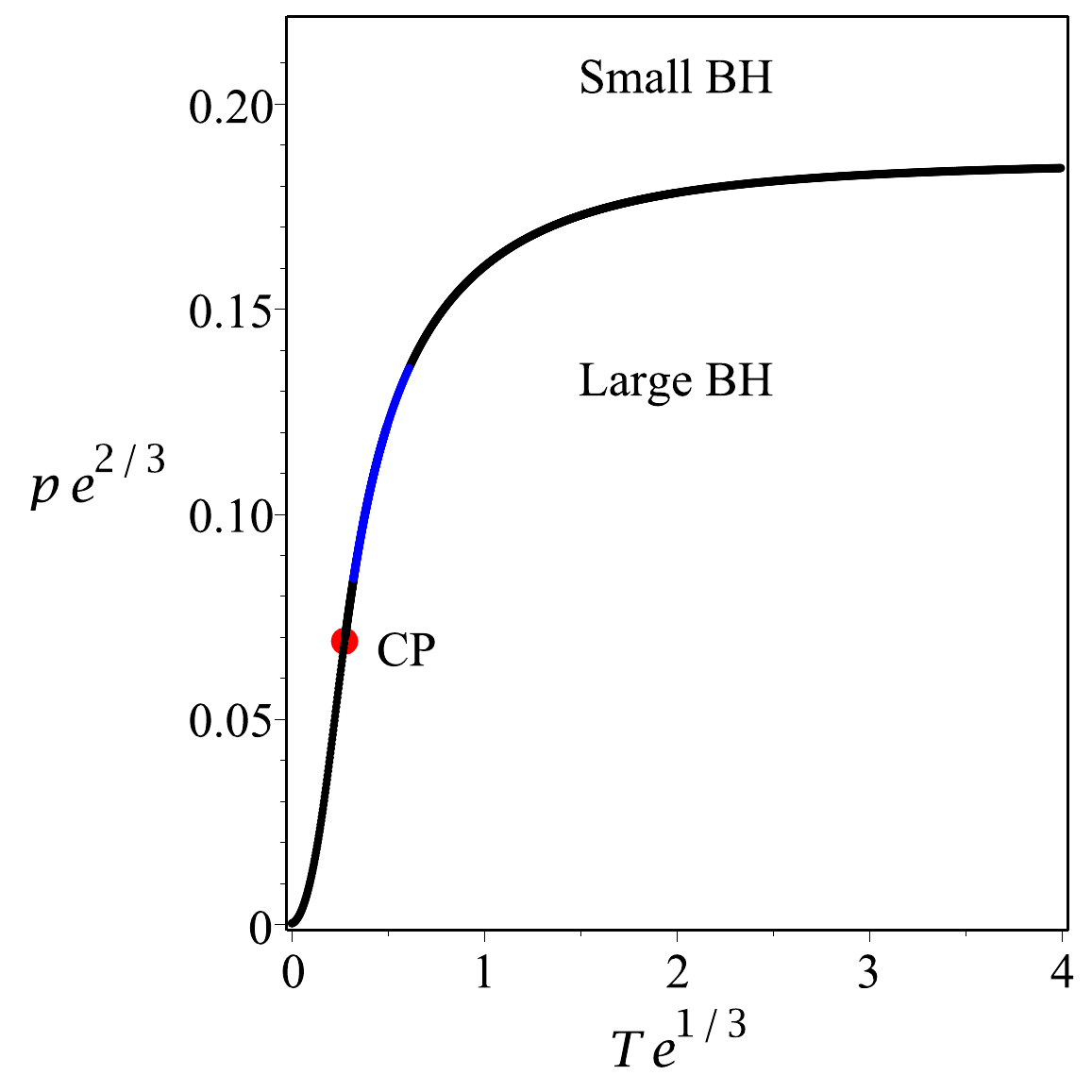}
\\
\end{tabular}
\caption{\textbf{Two first order phase transition in six dimensions for  $d=6$ and $k=1$} (color online). \textit{Left }: A low temperature VdW transition with critical temperature $T=T_{c}$ (dashed red line), $T=0.8 T_{c},~0.6 T_{c}$ (solid, dotted red lines), and a higher temperature reverse VdW transition with $T=T_{\bar{c}}$ (dashed black line), $T=1.6 T_{\bar{c}},~1.8 T_{\bar{c}}$ (solid and dotted black lines).  \textit{Center}:  The phase diagram for six dimensional spacetimes. The green dots denote the critical points and the black lines show that there are two first order phase transition for temperatures less and larger than the two critical points.
 Blue lines correspond to negative entropy.
We have chosen $e=1$, $\beta_3=-2 e^{4/3}$ and $\alpha_4=-9/5 e^{2}$ with $T_{c}e^{1/3}\approx0.31392$, $P_{c}e^{2/3}\approx0.08065$ and $T_{\bar{c}}e^{1/3}\approx0.25527$, $P_{\bar{c}}e^{2/3}\approx0.05998$.
\textit{Right}:  For  $e=1$, $\beta_3\approx -2.05912 e^{4/3}$ and $\alpha_4= -9/5 e^{2}$ 
we obtain an isolated critical point (red point); the approximate values at the conjoined critical temperature and pressure are $T_c e^{1/3} \approx 0.28047$ and $P_c e^{2/3} \approx 0.06876$.
}
\label{PT6d2pt}
\end{figure*}

 For regions of parameter space having two physical critical points, as depicted in figure \ref{PT6d2pt} there
is a  first order standard VdW phase transition between small and large black holes at low temperatures and then
a reverse VdW transition at higher temperatures.
 The right graph shows that for fixed charge and $\alpha_4$, and varying $\beta_3$ we obtain the appropriate value for the cubic coupling $\beta_3$ such that two coexistence lines meet, yielding an isolated critical point, with non-standard critical exponents  given in \eqref{nonstanexp}.
 Again we see that there exist a range of temperatures on either side of the isolated critical point for which the black holes satisfy all physical requirements.

Finally, in regions of parameter space with one physical critical point we get a standard VdW phase transition for $k=1$.
However if $k=-1$ then we find an reverse VdW transition that is similar what we described for the $d=5$  hyperbolic black hole.

\subsection{Critical behaviour in more than six dimensions}

Increasing the value of $d$ further, we find for seven dimensions that the qualitative features remain similar to $d=6$ spacetime. Quantitatively, however,  the  parameter regions with only a single critical point get larger, whereas regions with two or three physical critical points get smaller.
 No further features emerge and so we shall not consider this case further.

However for $d=8$ we find that we get up to three critical points. The structure of the associated phase transitions is similar to that we presented for $d=6$ in the previous subsection, so again we shall not consider this case further.

Finally, we compute
the ratio of critical quantities for any value of $d$. In general this must be done  numerically, but we can obtain an analytic expression for  small values of the  couplings  for which the physical constraints hold.
Here we shall set the cubic coupling to zero\footnote{See the corresponding results for the cubic case in \cite{mir:2018mmm}.}
and compare results with the critical behaviour in Einstein gravity for spherical black holes.
To leading order the critical quantities read
\beqa
T_c&=&\frac{4 (d-3)^2 }{\pi  (d-2) (2 d-5) v^{(0)}_c}
+\Big(128 \pi  (d-3)^4 (d-2)^2 (2 d-5) (2 d-3) \big(24 d^7-540 d^6\nonumber\\
&&\left.+4980 d^5-24292 d^4+67202 d^3-103983 d^2+80703 d-22140\big) e^2 {v^{(0)}_c}^6\right.\nonumber\\
&&\left.-128 (d-3)^5 \big(24 d^7-644 d^6+7012 d^5-40044 d^4+129678 d^3-238653 d^2\right.\nonumber\\
&&\left.+231021 d-90180\big) {v^{(0)}_c}^{2 d}\Big)
\Big/\Big(\pi ^5 (d-2)^7 (2 d-5)^5 (2 d-3) (3 d-16) e^2 {v^{(0)}_c}^{13}\right.\nonumber\\
&&\left.-\pi ^4 (d-3) (d-2)^4 (2 d-5)^4 (3 d-16) (4 d-9) {v^{(0)}_c}^{2 d+7}\Big)\alpha_4+\cO(\alpha_4^2),\right.\nonumber
\eeqa
\beqa
v_c&=&v^{(0)}_c-64 (d-3)^4 \Big(36 d^6-644 d^5+4428 d^4-14776 d^3+24357 d^2-16587 d+1260\Big)\nonumber\\
&&\left.\times 1\Big/\bigg(\pi ^3 (d-2)^3 (2 d-5)^3 (3 d-16) {v^{(0)}_c}^5 \Big(\pi  (d-2)^3 (2 d-5) (2 d-3) e^2 {v^{(0)}_c}^{6-2 d}\right.\nonumber\\
&&\left.+(d-3) (9-4 d)\Big)\bigg)\alpha_4+\cO(\alpha_4^2),\right. \nonumber
\eeqa
\beqa
P_c&=&\frac{(d-3)^2 }{\pi  (d-2)^2 {v^{(0)}_c}^2}+
\Big(-128 (d-3)^6 (4 d-9) \big(82 d^6-1594 d^5+12272 d^4-48254 d^3\nonumber\\
&&\left.+102927 d^2-113451 d+50580\big) {v^{(0)}_c}^{4 d}+128 \pi ^2 (d-3)^4 (d-2)^6 (2 d-5)^2 (2 d-3)\right.\nonumber\\
&&\left. \left(36 d^7-802 d^6+7294 d^5-34880 d^4+93686 d^3-138231 d^2+98067 d-21060\right)\right.\nonumber\\
&&\left.\times {e^4} {v^{(0)}_c}^{12}-768 \pi  (d-3)^5 (d-2)^3 (2 d-5) \big(24 d^8-616 d^7+6638 d^6-39082 d^5\right.\nonumber\\
&&\left.+136998 d^4-291119 d^3+362005 d^2-234726 d+56880\big) {e^2} {v^{(0)}_c}^{2 d+6}\Big)\right.\nonumber\\
&&\left.\times 1\Big/\bigg({v^{(0)}_c}^8 \Big(\pi ^6 (d-2)^{10} (2 d-5)^6 (2 d-3)^2 (3 d-16) {e^4} {v^{(0)}_c}^{12}-2 \pi ^5 (d-3) (d-2)^7 \right.\nonumber
\eeqa
\beqa
&&\left.\times(2 d-5)^5 (2 d-3) (3 d-16) (4 d-9) {e^2} {v^{(0)}_c}^{2 d+6}+\pi ^4 (d-3)^2 (d-2)^4 (2 d-5)^4 \right.\nonumber\\
&&\left.\times(3 d-16) (4 d-9)^2 {v^{(0)}_c}^{4 d}\Big)\bigg) \alpha_4+\cO(\alpha_4^2),\right. \nonumber
 \eeqa
where
 \beqa
 v^{(0)}_c= \left(\frac{(d-2)^2 (2 d-5) \pi e^2}{(d-3) }\right)^{\frac{1}{2 (d-3)}},
\eeqa
 is the critical volume in Einstein gravity.

We see a dimension-dependent deviation from the critical values in Einstein gravity.
The critical temperature and pressure  increase and the critical volume decreases relative
to Einstein gravity, except in four and six dimensions, where  critical pressure and temperature are  smaller and the critical volume is  larger.   In four dimensions these corrections are negligible.

 We finally obtain
 \beqa
 \frac{P_c  v_c }{T_c }&=&\frac{2 d-5}{4 (d-2)}
 +\Big(-8 \pi  (d-3)^7 (d-2)^2 (2 d-5) (4 d-9)^2 \big(96 d^8-2288 d^7+22748 d^6\nonumber\\
&&\left.-122532 d^5+386956 d^4-717526 d^3+722577 d^2-299793 d-8460\big) {v^{(0)}_c}^{4 d+1}\right.\nonumber\\
&&\left.+8 \pi ^4 (d-3)^4 (d-2)^{10} (2 d-5)^4 (2 d-3)^2 \big(48 d^9-1240 d^8+13664 d^7-84108 d^6\right.\nonumber\\
&&\left.+317348 d^5-754436 d^4+1108370 d^3-925455 d^2+341247 d-4860\big) {e^6} {v^{(0)}_c}^{19-2 d}\right.\nonumber\\
&&\left.-8 \pi ^3 (d-3)^5 (d-2)^7 (2 d-5)^3 (2 d-3) \big(576 d^{10}-16032 d^9+193720 d^8\right.\nonumber\\
&&\left.-1336844 d^7+5820852 d^6-16626688 d^5+31236380 d^4-37317081 d^3\right.\nonumber\\
&&\left.+25805259 d^2-7920864 d+36720\big) {e^4} {v^{(0)}_c}^{13}+8 \pi ^2 (d-3)^6 (d-2)^4 (2 d-5)^2 \right.\nonumber\\
&&\left.
\times(4 d-9) \big(576 d^{10}-15888 d^9+189992 d^8-1295464 d^7+5562612 d^6\right.\nonumber\\
&&\left.-15631748 d^5+28802548 d^4-33602712 d^3+22528269 d^2-6569739 d-57780\big)\right.\nonumber\\
&&\left.\times {e^2} {v^{(0)}_c}^{2 d+7}\Big)\Big/\bigg(
\pi ^4 (d-3)^2 (d-2)^6 (2 d-5)^4 (3 d-16) {v^{(0)}_c}^7 \Big(-4 d^2+\pi  \big(4 d^5\right.\nonumber\\
&&\left.-40 d^4+159 d^3-314 d^2+308 d-120\big) {e^2} {v^{(0)}_c}^{6-2 d}+21 d-27\Big) \Big(\pi  (d-2)^3 \right.\nonumber\\
&&\left.\times(2 d-5) (2 d-3) {e^2} {v^{(0)}_c}^6-(d-3) (4 d-9) {v^{(0)}_c}^{2 d}\Big)^2
\bigg)\alpha_4+\cO(\alpha_4^2),\right.
\label{pvtq}
\eeqa
and we see that in the limit of vanishing quartic coupling, these results reduce those of charged $k=1$ black holes  in Einstein gravity \cite{GunasekaranEtal:2012} .

The effect of the quartic curvature term on the Van der Waals ratio is to increase it
in any dimension above the Einsteinian value.
In four and five dimensions these corrections are negligible for small values of coupling and large enough charge. However in higher dimensions we see considerable deviation and the van der Waals ratio no longer is a ``universal" quantity as it is in Einstein gravity:  it depends on parameters of the theory as well as the dimension under considerations.

\section{Thermodynamics in grand canonical ensemble} \label{sec: thermofpe}

We now consider the grand canonical ensemble, in which we have  fixed potential instead of fixed charge.
From the viewpoint of AdS/CFT holography,  fixed potential on the gravity side is related to   fixed chemical potential on the CFT side. We first consider $d=4$ and then discuss  properties for generic dimensions,  employing the approach in ref.~\cite{ChamblinEtal:1999a}. We shall consider only the quartic term in the action;  the cubic case was studied in \cite{mir:2018mmm}.  We expect that when both couplings are nonzero a pattern  similar to that of the fixed
charge case
for the number of critical points
will emerge,  though we shall not perform that analysis here.

\subsection{Four dimensions}

For fixed potential  one needs to find   expressions for the mass and the temperature solving again the equations of motion for the metric function near the horizon, since this choice of ensemble alters how the equations depend on the horizon radius.  The first two leading order terms in the  expansion  result in the following formulas, parameterized by the quartic coupling and $r_+$:
\beqa
8\pi M &=&k r_+ +\frac{\Phi^2r_+}{4}+\frac{8\pi P r_+^3}{3}+ \frac{64\pi^4\lambda  T^4}{r_+} +\frac{128\pi^3k\lambda T^3 }{3 r_+^2},\nonumber\\
0&=&k-\frac{\Phi^2}{4}+8\pi P r_+^2-4\pi T r_+ +\frac{64\pi^4\lambda T^4}{3r_+^2}+\frac{128\pi^3k \lambda T^3}{3 r_+^3},\label{PTphi}
\eeqa
with the second equation yielding the equation of state
\beq
\label{eqn:gce_eos_4d}
P =\frac{T}{v}-\frac{k}{2\pi v^2}+\frac{\Phi^2}{8 \pi v^2}-\frac{128\pi^3\lambda T^4}{3 v^4}-\frac{512\pi^2k\lambda T^3}{3 v^5},
\eeq
where we defined $v = 2 r_+$.

The explicit form of the critical quantities from the equation of state~\eqref{eqn:gce_eos_4d} are
\beqa
T_c &=& -\frac{6^{\frac{2}{3}}}{24\pi\lambda}\Big(-\left(588384k^3\lambda -19008k^2\lambda \Phi^2-18k\lambda\Phi^4-\lambda\Phi^6+Z\right)\lambda^4\Big)^{\frac{1}{6}},\nonumber\\
v_c &=& \frac{20\times 6^{\frac{1}{3}}k\lambda}{3}\left(2358720k^3\lambda -78480k^2\lambda \Phi^2+180k\lambda\Phi^4+5\lambda\Phi^6+4 Z\right)\nonumber\\
&&\left.\times1/\Big(\left(-(588384k^3\lambda -19008k^2\lambda \Phi^2-18k\lambda\Phi^4-\lambda\Phi^6+Z)\lambda^4\right)^{\frac{1}{6}}\right.\nonumber\\
&&\left.\times(576576k^3\lambda -16272 k^2\lambda \Phi^2+28k\lambda\Phi^4+\lambda\Phi^6+Z)\Big),\right.\label{vcphi}
\eeqa
where
\beqa
Z=\sqrt{\lambda^2(\Phi^2-24k)(\Phi^2+156k)^2(\Phi^2-84k)^3} \label{Zeq}
\eeqa
and the formula for the corresponding critical pressure is given inserting the above relations for $T_c$ and $v_c$ into equation \reef{eqn:gce_eos_4d}. Note that we have the  constraint
\beqa\label{Phiconstraint}
(\Phi^2-24k)(\Phi^2-84k)^3>0,
\eeqa
so that critical quantities remain real.

\begin{figure*}[h]
\centering
\begin{tabular}{cc}
\includegraphics[scale=.25]{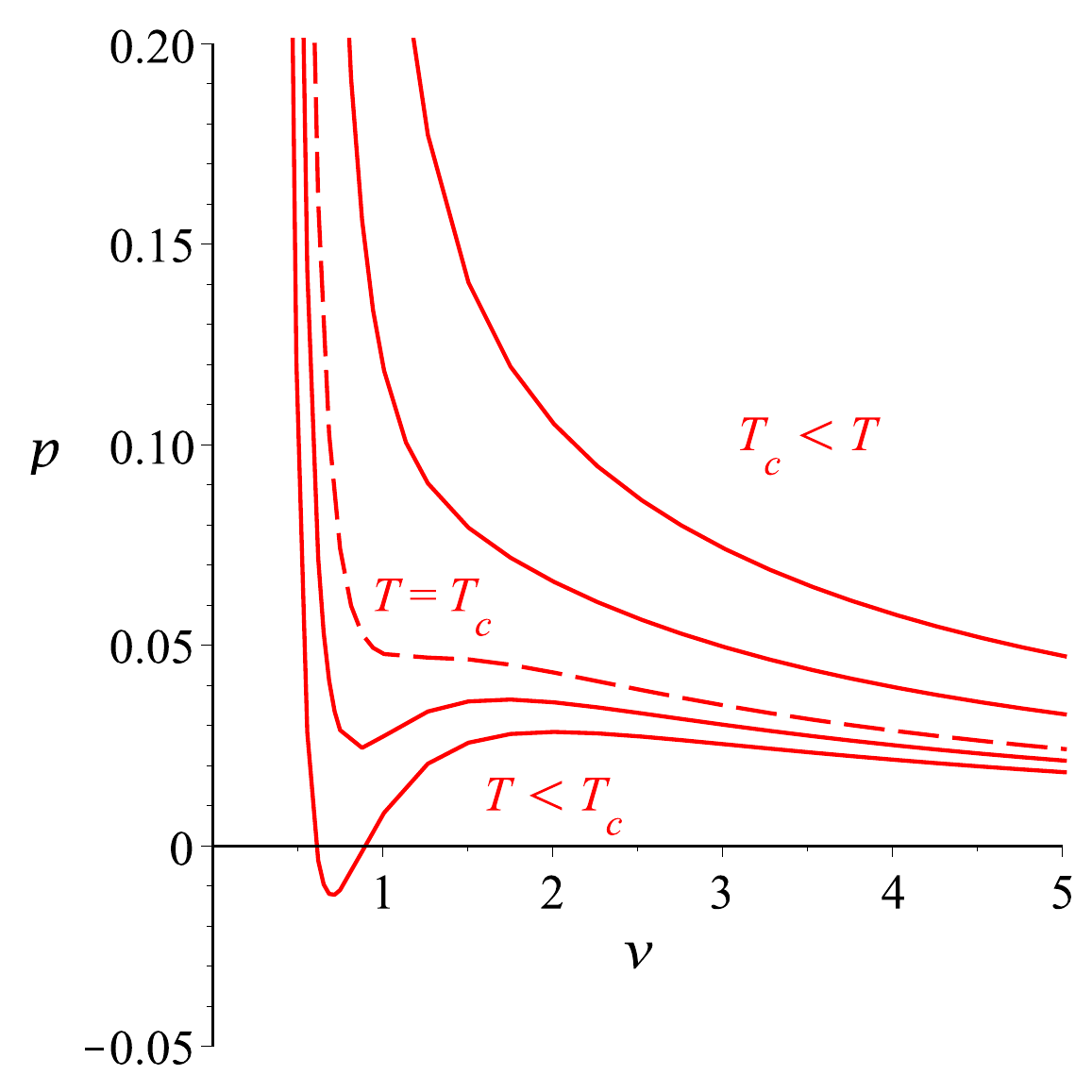}&\quad\quad\quad\quad\quad\quad
\includegraphics[scale=.25]{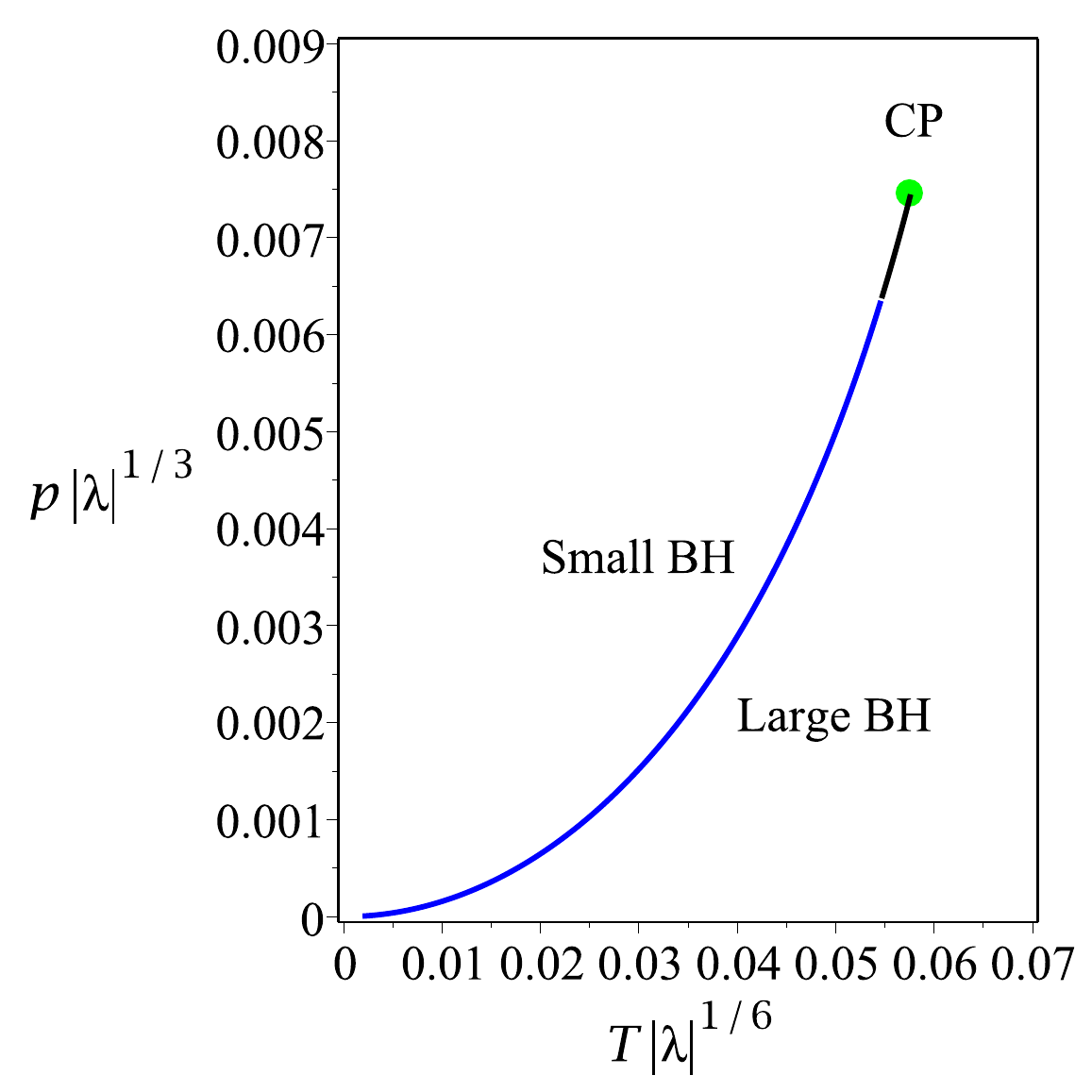}\\
\includegraphics[scale=.25]{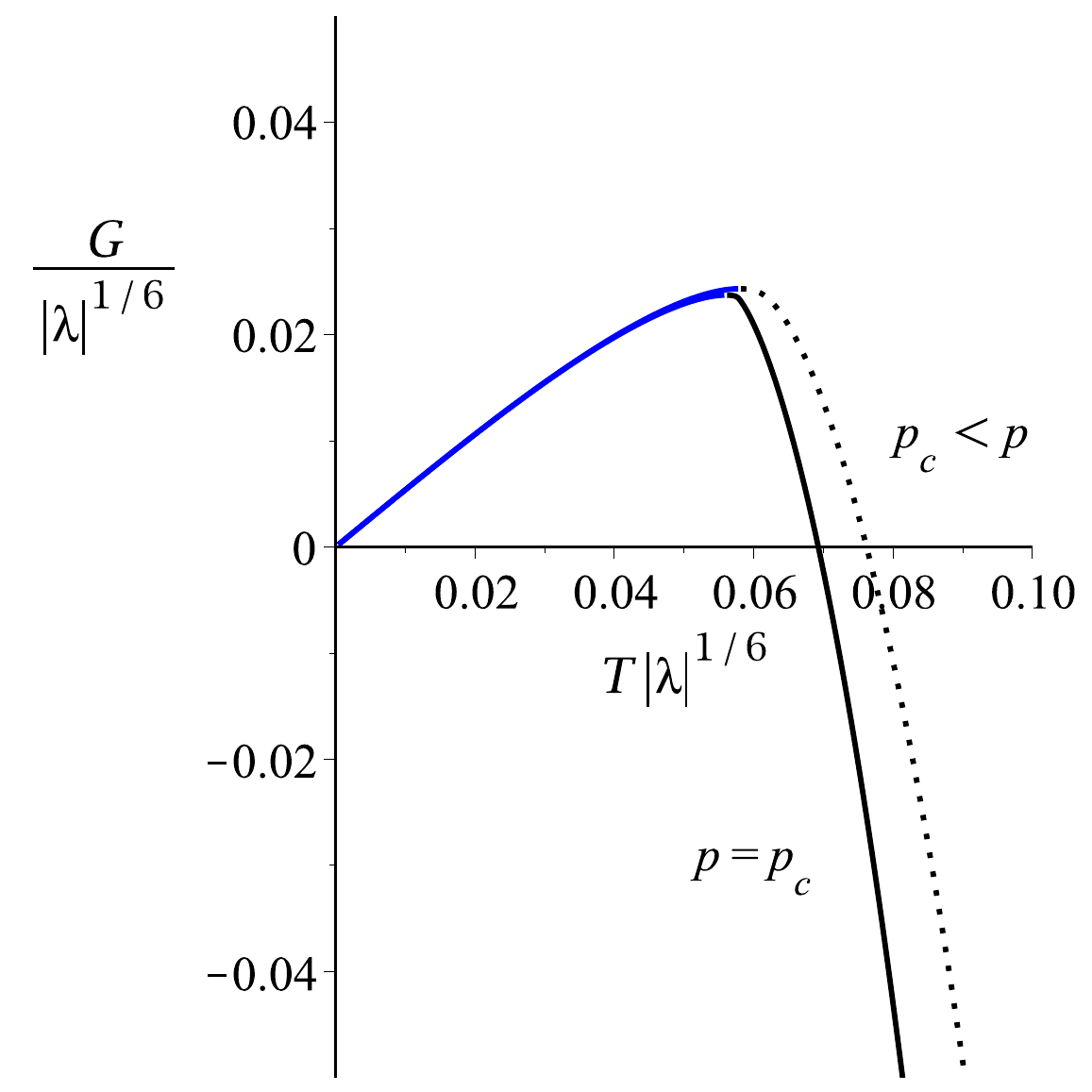}&\quad\quad\quad\quad\quad\quad
\includegraphics[scale=.25]{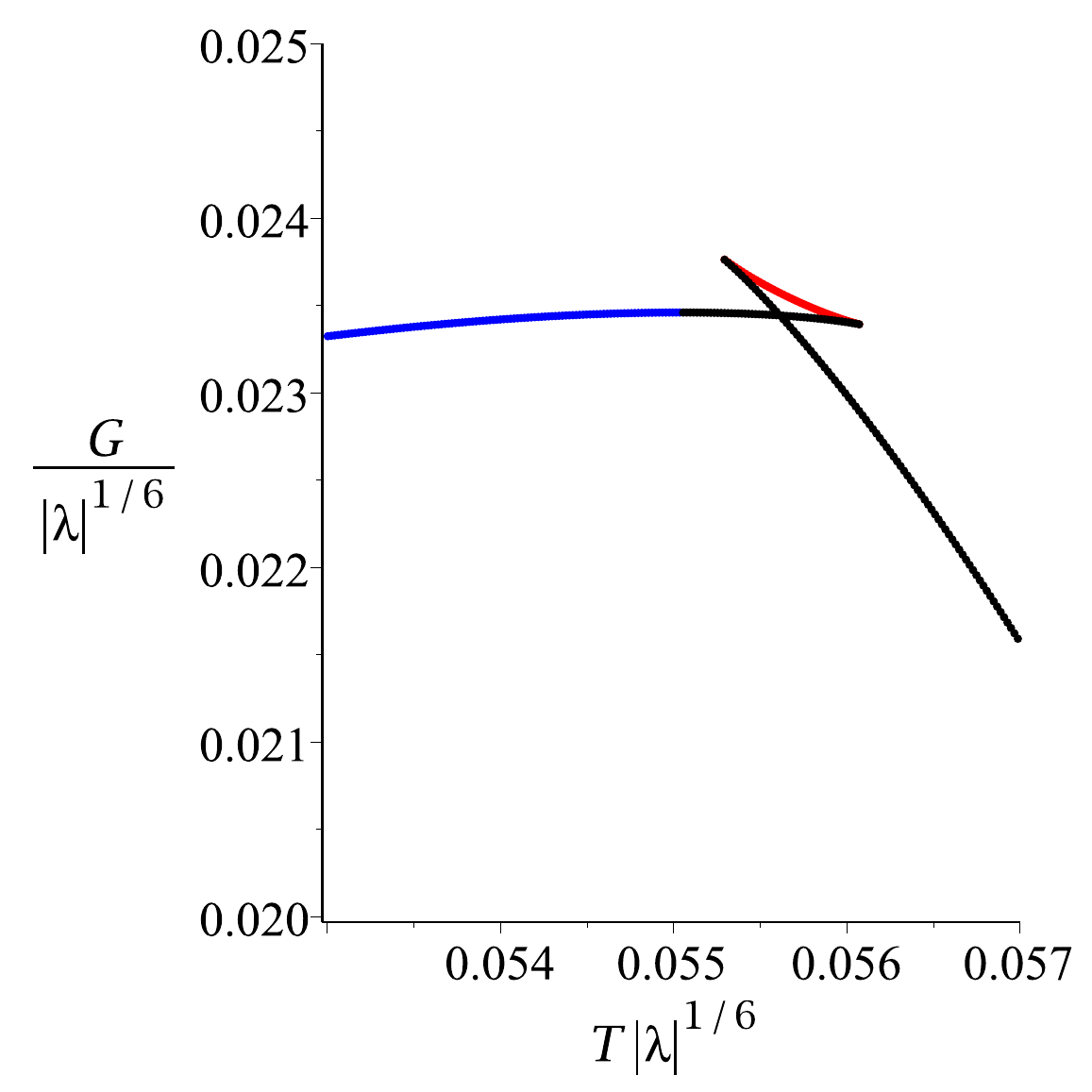}
\\
\end{tabular}
\caption{{\bf Van der Waals behaviour in grand canonical ensemble for $d=4$ and $k=1$}.  {\it Top left}:  Isotherms in the $P-v$ plane, showing Van der Waals oscillations.  The dashed red line corresponds to $T = T_c$; the solid red curves correspond to $T \neq T_c$.
{\it Top right}: Coexistence line in the $P-T$ plane for $\Phi = 1$. The green dot exhibits the critical point, and the black line is the line of coexisting phases for the first order phase transition. The solid line has  positive entropy, whereas the blue line indicates the occurrence of negative entropy. {\it Bottom left}: free energy vs temperature for various pressures. Solid black curves correspond to $P = P_c$, and the dotted black curve marks pressures $P > P_c$. The solid blue line has positive specific heat but negative entropy. {\it Bottom right}:  For $P =0.9 P_c$  and $\Phi = 1$,  the phase transition illustrated by swallowtail behaviour that grows by decreasing pressure. The red curves represent negative specific heat with positive entropy; the blue curves represent positive specific heat but negative entropy.
}
\label{fig:gce_4d}
\end{figure*}

Solving equations \reef{PTphi} to determine the explicit form for $M$ and $T$, and choose solutions that in the limit $\lambda \to 0$ approach the Einstein branch, we find that only for $k=+1$ (spherical) black holes  are there  physical critical points.

 Admitting a positive mass, while
imposing the condition \reef{asympf} with zero cubic coupling, we use the formula \reef{PMAX} to obtain an upper bound
\beq
0 < P \le\frac{9 \sqrt[3]{\frac{3}{2}} }{64 \pi |\lambda|^\frac{1}{3}},
\eeq
for the pressure, where the absolute value comes about because   $\lambda<0$  in four dimensions.
Expanding the equation of state about the critical point, we again obtain standard critical exponents \reef{exponents} from mean field theory.

The black hole entropy at the critical point becomes
\beq
S=\frac{{{r_+^2}_c}}{4}+\frac{6 ^{\frac{1}{3}}k X^{\frac{1}{3}}}{12\lambda {{r_+^2}_c}}-\frac{\sqrt X}{36\lambda^2 {r_+}_c},
\eeq
where
\beqa
X=\left(-588384 k^3\lambda+19008k^2\lambda \Phi^2 +18k\lambda\Phi^4+\lambda\Phi^6-Z\right)\lambda^4,
\eeqa
and ${r_+}_c=\frac{v_c}{2}$ with $v_c$ introduced in   \reef{vcphi} and  $Z$ given in \reef{Zeq}. Explicit numerical
computation for $k=+1$ and $\lambda < 0$ indicates that  the critical entropy is always positive; this will not hold for other values of these parameters.

In order to search for the phase structure of the black hole solutions, we obtain the formula for the free energy in the grand canonical ensemble
\begin{align}
G &= M - TS - \Phi Q \, ,
  \nn\\
  &= \frac{P v^3}{24}-\frac{ T v^2}{16}-\frac{(-4k+\Phi^2)v}{64\pi }-\frac{16\pi^3\lambda T^4}{3 v}-\frac{32\pi^2\lambda k T^3}{3 v^2},
\end{align}
which we plot in figure~\ref{fig:gce_4d}.
 We again observe standard Van der Waals behaviour.
 However we also see that the free energy is a
 decreasing function of the temperature for
 small $T$, vanishing at $T\to 0$, in contrast to
 the fixed charge case; however a large branch of the curve has negative entropy.

The $P-v$ diagram in figure~\ref{fig:gce_4d} illustrates that a phase transition happens for temperatures a bit less than the critical temperature. However for enough low temperatures, the pressure becomes negative, and we do not consider this unphysical case.   Another way to observe the phase transition is
via the coexistence line in the $P-T$ plane, shown in the top right of   figure~\ref{fig:gce_4d}.
For low enough temperatures the  entropy becomes negative,  denoted by the  blue solid line.

\begin{figure*}[htp]
\centering
\begin{tabular}{cc}
\includegraphics[scale=.3]{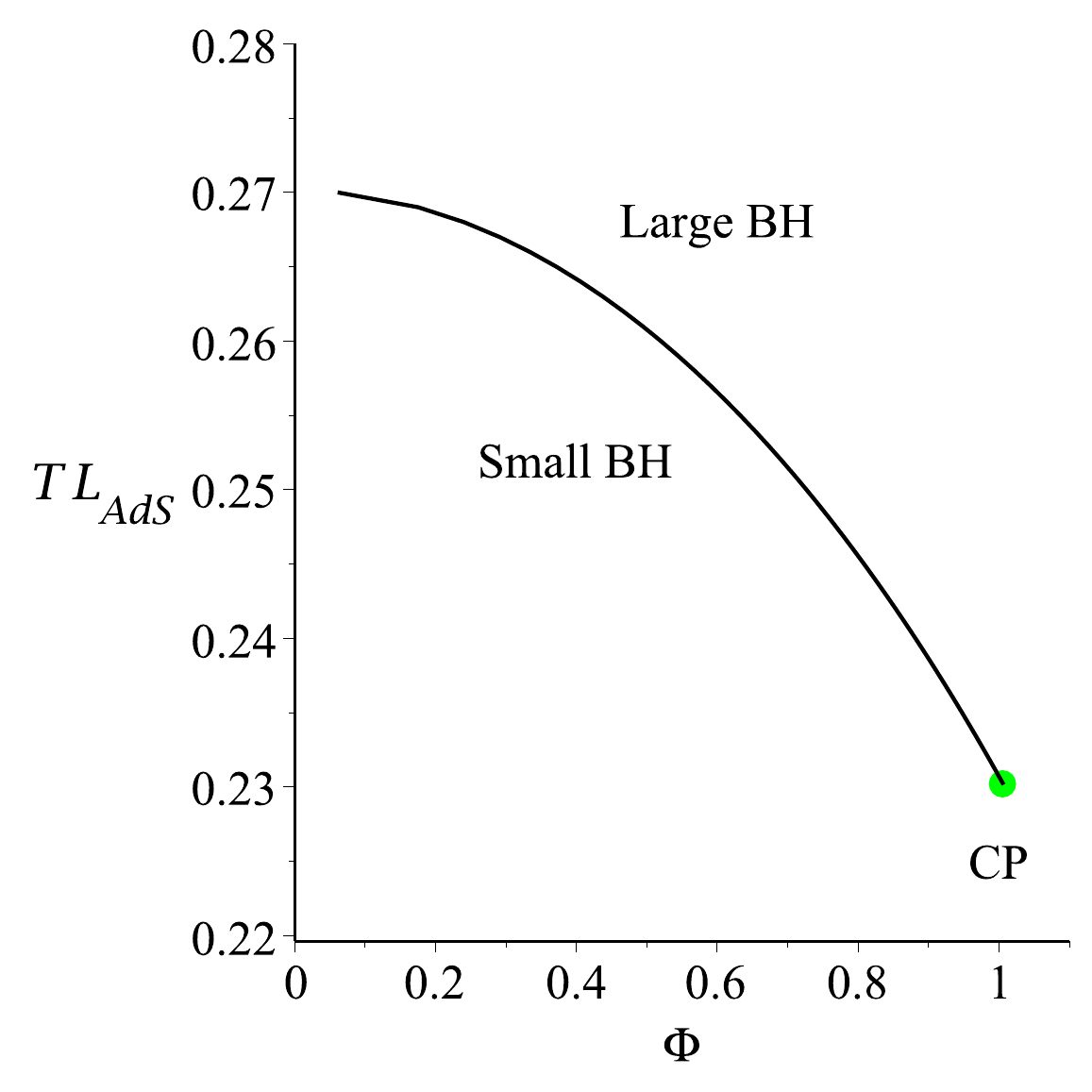}
\\
\end{tabular}
\caption{{\bf Phase diagram in $T$--$\Phi$ plane for $d=4$ spherical black holes}. We illustrate the phase diagram  in the $ T - \Phi$ plane for fixed pressure $P \ell^2 = 3/(8\pi)$.  The green dot shows the critical point, while the black line marks the coexistence curve for the first order phase transition. In this figure, the coupling was set to $\lambda/\ell^6 = 0.32$.
}
\label{fig:phiT_phase_4d}
\end{figure*}

Instead of finding the equation of state for fixed potential, we perform the analysis with fixed pressure,  commonly used in holography. The phase graph of temperature versus potential in figure~\ref{fig:phiT_phase_4d} shows it  again describes a first order phase transition between small and large black holes, and the coexistence line terminates at the critical point.

\subsection{Higher dimensions}

Here, we look into the thermodynamic properties of black hole solutions in the fixed ensemble in generic dimension. The equation of state in $d$ dimensions is
\beqa
P&=&\frac{(d-2)T}{4 r_+}+\frac{(d-3)[-2(d-2)k+(d-3)\Phi^2]}{32 \pi r_+^2}
-\frac{\pi^3 (d-2)(d-5)(3d-16)\lambda T^4}{3 r_+^4}\nonumber\\
&&\left.-\frac{4\pi^2 (d-5)(d-6)(d-2)\lambda
k T^3}{3 r_+^5}-\frac{\pi (d-2)(d-4)(d-7)\lambda k^2 T^2}{2 r_+^6},\right.
\eeqa

To compare results between the two  ensembles with the cubic coupling set to zero, we note that
in the fixed potential ensemble for spherical  black holes in quartic GQG we get physical critical points for $\alpha_4>0$
in four and five dimensions. However in six dimensions no physical critical points exist. This is in contrast  with the fixed charge ensemble, where ($\alpha_4>0$) we get single  physical critical points in $d=4,6$ and  two  physical critical points in $d=5$ (as well as single physical critical points).

 In addition, for fixed potential and $k=-1$ hyperbolic quartic black holes, while there are potential critical points
for $d=4,6$, these all have $\gamma^2<0$ (since $\alpha_4<0$) and so there are no physical critical points for these dimensions (and likewise none for $d=5$).  This situation is the same as for the fixed charge ensemble, confirming that both cubic and quartic coupling must exist to obtain physical critical points.

\section{Holographic hydrodynamics} \label{sec: holog}
\label{sec:holo_hydro}

One of applications of the AdS/CFT correspondence is the computation of the ratio of shear viscosity to entropy density $\eta/s$. We investigate in this section this ratio for  the quartic theory.

It is well-known that for the field theories dual to Einstein gravity, the shear viscosity to entropy density ratio is lower-bounded by $\ie$ $\eta/s = 1/(4\pi)$, and it has been suggested that  this   lower bound is universal, holding for any matter~\cite{Kovtun:2004de}. This conjecture thus states that $\eta/s \ge 1/ (4\pi)$, and is the so-called  KSS bound.

However  higher derivative order contributions can cause the violations of this bound~\cite{Brigante:2007nu}. We therefore evaluate $\eta/s$ for field theories dual to the quartic generalized quasi-topological theory for general $d$  to see whether the KSS bound is satisfied.

Here, we employ the   planar black hole solutions described by the following metric
\beq
ds^2 = \frac{r^2}{\ell^2} \left(-g(r) dt^2 + \sum_i dx_i^2 \right) + \frac{\ell^2 dr^2}{r^2 g(r)} \, .
\eeq
We define a new coordinate $z = 1 - r_+^2/r^2$, to compactify the  region outside the horizon. This gives
\beq
\label{eqn:zMetric}
ds^2 = \frac{r_+^2}{\ell^2 (1-z)} \left(-g(z) dt^2 + \sum_i dx_i^2  \right) + \frac{\ell^2}{4 g(z)} \frac{dz^2}{(1-z)^2},
\eeq
where $g(z)$ vanishes at $z=0$, and $g(1) = f_\infty$.

Expanding $g(z)$ near the horizon
\beq
g(z) = g_0^{(1)} z + g_0^{(2)} z^2 + g_0^{(3)} z^3 + \cdots,
\eeq
we solve  the field equations to  determine $g_0^{(i)}$ for $i \neq 2$. As we discussed previously, the second derivative of the metric function near the horizon, $g_0^{(2)}$, is not determined by the field equations. However, its value can be chosen such that the numerical solution approaches its associated asymptotic solution. It is easy to check that by the coordinate transformation, the parameters $g_0^{(i)}$ are written in terms of the parameters $a_i$ appearing in the near horizon expansion  \eqref{eqn:nh_ansatz}. We find
\begin{align}
g_0^{(1)} &= \frac{2 \pi T \ell^2}{r_+} \, , \quad g_0^{(2)} = - \frac{L^2}{4 r_+} \left(2 \pi T - r_+ a_2 \right),
	\nn\\
	g_0^{(3)} &= -\frac{L^2}{8 r_+} \left(2 \pi T - r_+ a_2 - r_+^2 a_3 \right)
\end{align}
where the explicit form of $a_3$ from section~\ref{sec:bhsolution} for the planar black holes is
\beqa
a_3&=&-\frac{1}{288\pi^3(3d-16)  \ell^2\lambda  T^3r_+^2}\bigg[3(d-1)(d-2)r_+^4-24\pi(d-3) \ell^2T r_+^3\nonumber\\
&&\left.+16 \pi^4(-5472+15d^3-341d^2+2434d)\ell^2\lambda\ T^4\right.\nonumber\\
&&\left.+2( 16\pi^3(d-5)(21d-160)\lambda T^3
-3r_+^3)\ell^2 r_+ a_2+
96\pi^2(3d-16) \lambda \ell^2 T^2r_+^2 a_2^2\bigg].\right.\nonumber\\
\eeqa

Using methods described  in~\cite{Paulos:2009yk}, we perform a shift  on the metric~\eqref{eqn:zMetric}
\beq
dx_i \to dx_i + \epsilon e^{-i\omega t} dx_j,
\eeq
with  perturbation parameter $\epsilon$.
Computing the Lagrangian for the perturbed metric, and performing a series expansion for small $\epsilon$, we obtain
\beqa
\sqrt{-g}{\cal L} &=&\frac{1}{16 \pi} \bigg[ \cdots-\frac{\epsilon^2\omega^2 r_+^
{d-3}}{4\ell^{d-4}z g_0^{(1)}}\bigg(1+\frac{16\hat{\lambda}}{5\ell^6}\Big((-1050+911d-261d^2+24d^3)(g_0^{(1)})^3\nonumber\\
&&\left.+4(842-414d+53d^2)(g_0^{(1)})^2 g_0^{(2)}+48(-22+5d)(g_0^{(1)})^2g_0^{(3)}\right.\nonumber\\
&&\left.+16 (15d-68)g_0^{(1)}(g_0^{(2)})^2\Big)
\bigg)+ {\rm Regular}
 \bigg],\right.
\eeqa
where $\hat{\lambda}$ is the coupling in the action and is related to the coupling appearing in the equation of motion, using \reef{rescall}.

The 'time formula' gives the shear viscosity
\beq
\eta = - 8 \pi T \lim_{\omega,\epsilon \to 0} \frac{{\rm Res}_{z=0} \sqrt{-g}{\cal L}}{\omega^2 \epsilon^2},
\eeq
whose explicit form for the case at hand is
\beqa
\eta &=&\frac{T r_+^
{d-3}}{8\ell^{d-4} g_0^{(1)}}\bigg(1+\frac{16\hat{\lambda}}{5\ell^6}\Big((-1050+911d-261d^2+24d^3)(g_0^{(1)})^3\nonumber\\
&&\left.+4(842-414d+53d^2)(g_0^{(1)})^2 g_0^{(2)}+48(-22+5d)(g_0^{(1)})^2g_0^{(3)}\right.\nonumber\\
&&\left.+16 (15d-68)g_0^{(1)}(g_0^{(2)})^2\Big)
\bigg).
 \right.
\eeqa
 where $\hat{\lambda}$ appears in the action \reef{action0} and is related to $\lambda$ via \reef{rescall}.

The entropy density for planar black holes is
\beq
s = \frac{S}{\ell^{d-2} {\rm Vol}\left(\mathbb{R}^{d-2}\right)} =  \frac{r_+^{d-2}}{4\ell^{d-2}}\left[1-16\pi^3(d-2)(3d-16)\frac{\lambda T^3}{3 r_+^3} \right],
\eeq
which follows from eq.\reef{ENT} with $k=0$.

 Taylor expanding  $\eta/s$ about $\lambda=0$ we obtain
\begin{align}
\frac{\eta}{s} &= \frac{1}{4\pi}\bigg[1-\frac{\lambda}{240 \ell^6(d-1)(3d-16)(d^5-14d^4+79d^3-224d^2+316d-170)}
	\nn\\
	&\times \Big(-(891d^8-19104d^7+172820d^6-868818d^5+2696601d^4
\nn\\
	& -5403214d^3+6944072d^2-5233280d+1740800)(d-1)^4
	\nn\\ &+96 \ell^8d(5d-22)(d-3)(3d^2-18d+19)\dot{a}_2(0)\Big)\bigg]+\cO(\lambda^2),
\label{etaos-taylor}
\end{align}
where $\dot{a}_2(0)$ denotes derivative of $a_2$ with respect to
$\lambda$ and then setting $\lambda=0$.  To compute the value of $\eta/s$ numerically , one needs to determine the parameter $a_2$ for the choice of the other parameters, where the value of temperature is known from expansion near the horizon. In the above computation we replaced the parameter $a_3$ in terms of $a_2$ using the second order expansion near $r=r_+$ of eq. \reef{Feq0}.

The Taylor expansion yields an $\dot{a}_2(0)$ term that is left undetermined at this level. Only under the condition that the expression at first order in $\lambda$ in the above series expansion  for higher dimensions with some positive coupling (but in   $d=4$   for certain negative coupling) can
 we  conclude that the KSS bound $\eta/s \ge 1/(4\pi)$ holds in all dimensions in the quartic generalized quasi-topological theories at small coupling.

\begin{figure*}[h]
\centering
\begin{tabular}{cc}
\includegraphics[width=0.45\textwidth]{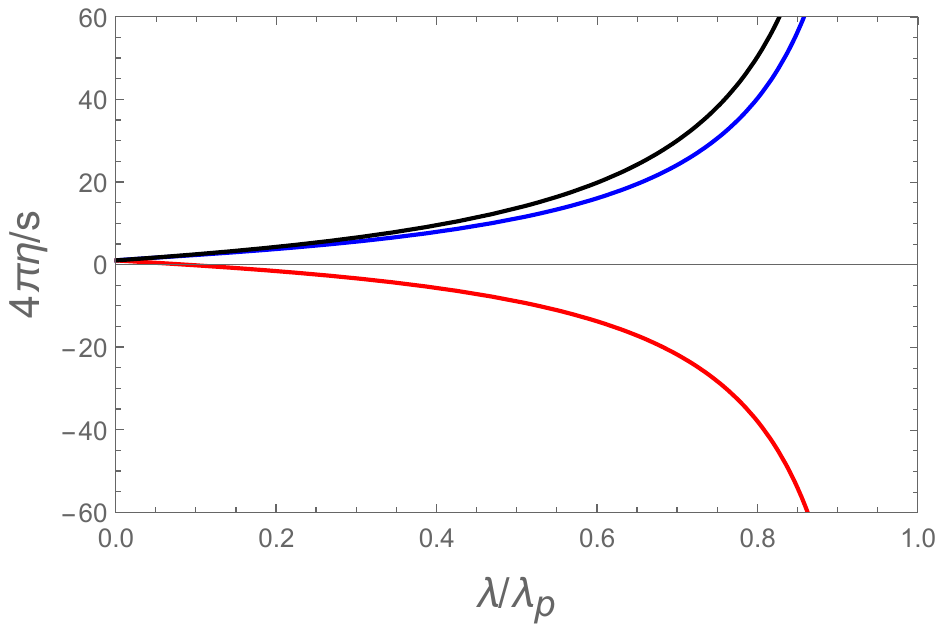}&\quad
\includegraphics[width=0.45\textwidth]{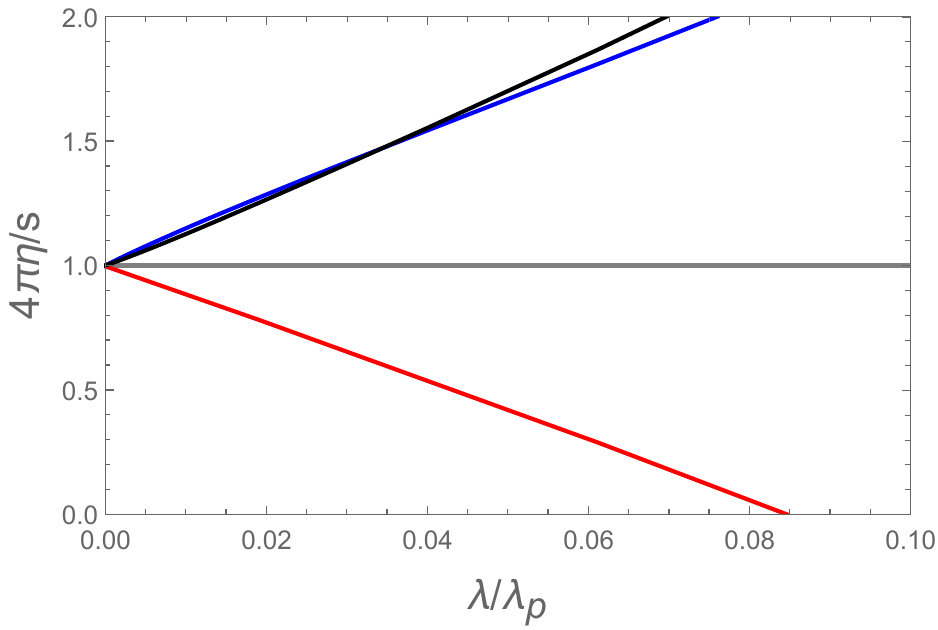}
\\
\end{tabular}
\caption{{\bf Ratio of shear viscosity to entropy density}: \textit{Left}: Plot of the ratio $\eta/s$ in four (blue) dimensions, five (red) dimensions and six (black) dimensions. \textit{Right}: A close up for small values of  the coupling.
The thin grey line shows the universal Einstein gravity value of $\eta/s  = 1/(4\pi)$. we used, a $[5|5]$ order Pad\'e approximant for $a_2$ (see the appendix) and we have set $\ell=1$ and $r_+=1$.
}
\label{fig:etaOvers}
\end{figure*}

The generalized quasi-topological term can cause the entropy density of black branes to change sign~\cite{Hennigar:2017umz}. Hence there is a certain $T_p$ for which the ratio $\eta/s$ exhibits a pole. Using the first order near-horizon expansion from \reef{Feq0}, the corresponding quartic coupling is
\beq
T_p = \frac{(d-2) r_+}{3 \pi \ell^2} \, , \quad  \lambda_p =  \frac{81\ell^6}{16 (d-2)^4(3d - 16)} \, ,
\eeq
where the subscript ``$p$'' stands for ``pole''.
 It is interesting to note that in four dimensions $\lambda_p$ is equal the value $\lambda_c = -81 \ell^6/1024$ that we encountered in the critical limit of the theory in section \ref{vacua}; however, this coincidence does not hold in higher dimensions.
However in  any dimension it leads to the Einstein branch of the theory.
 The leading order term in $(\lambda - \lambda_{p})$ that describes the behaviour of the entropy near zero is
\beq
s_{\lambda \to \lambda_{p}} = -\frac{r_+^{d-2}}{81 \ell^{d+4}} (d-2) (3d^5-37d^4+166d^3-348d^2+344d-128)  \left(\lambda - \lambda_{p} \right)+\cO\left((\lambda - \lambda_{p})^2\right)
\eeq
Furthermore the corresponding expansion for the shear viscosity reads as

\begin{align}
\eta &= \frac{r_+^{d-2}}{640\pi \ell^{d-2} \Big((3d-16)^2(d-2)^3(d^5-14d^4+79d^3-224d^2+316d-170)\Big) }
\nn\\
	&\times
\bigg[-8\Big(279d^8-7053d^7+72868d^6-394627d^5+1186685d^4-1893868d^3
+1248176d^2
\nn\\&
+193856d-435200\Big)
(d-2)^2
-18\ell^2d(d-2)(d-3)(3d^2-18d+19)(99d^3-1016d^2
\nn\\&
+3038d-2272)a_2(\lambda_p)
+135d\ell^4(d-3)(d-4)(3d-16)(3d^2-18d+19)a^2_2(\lambda_p) \bigg]
\nn\\&+
\frac{d(3d^3-27d^2+73d-57)r_+^{d-2}}{25920 \pi\ell^{d+4}(3d-16)^2(d-2)^3
(d^5-14d^4+79d^3-224d^2+316d-170)}
\nn\\
&\times
\bigg[
-48\ell^2(3d-16)(42d^4-363d^3+601d^2+1057d-1488)(d-2)^5a_2(\lambda_p)
\nn\\&
+90\ell^4(d+7)
(d-4)(3d-16)^2(d-2)^4a^2_2(\lambda_p)
\nn\\
	&
-729\ell^8(d-2)(99d^3-1016d^2+3038d-2272)\dot{a}_2(\lambda_p)
\nn\\&
-16(d-1)(3d-16)
(93d^4-1543d^3+8894d^2-20420d+14656)(d-2)^6
\nn\\&
+10935\ell^{10}(d-4)(3d-16)a_2(\lambda_p)\dot{a}_2(\lambda_p)\bigg]\left(\lambda-\lambda_p\right) 	 + \cdots\label{etaexp}
\end{align}
It is interesting to note that the entropy density vanishes linearly as $\lambda \to \lambda_p$. By contrast, the shear viscosity does not vanish in this limit as it contains a term
  consisting of some powers of $a_2$ evaluated at the pole. Explicit numerical evaluation shows that this  term does not vanish  as $\lambda \to \lambda_p$.
Consequently the ratio of shear viscosity to entropy density has a pole at $\lambda = \lambda_p$. Figure \ref{fig:etaOvers} shows that there is a smooth curve connecting $\eta/s = 1/(4\pi)$ (for $\lambda = 0$) and $\eta/s = \infty$ (for $\lambda = \lambda_p$).

Since including quartic quasi-topological or Lovelock terms into the action does not alter the black hole entropy from its Einstein gravity value~\cite{Hennigar:2017umz}, only quartic GQG contributions are responsible for the occurrence of this pole. A previous study of cubic GQG found that  similar behaviour  for $\eta/s$ took place \cite{mir:2018mmm}.

To determine how the ratio $\eta/s$ can be recast in terms of $\lambda$, we use the Pad\'e approximant method to evaluate the parameter $a_2$. For our consideration involving quartic term, the computations become cumbersome very  rapidly while going to higher orders, so we only present only a few corresponding curves in four, five and six dimensions in figure~\ref{fig:etaOvers}. From this figure it is obvious that $\eta/s$ starts from $1/(4\pi)$ for $\lambda=0$ and then it grows until it diverges as expected at  $\lambda = \lambda_p$.
 For four dimensions this figure illustrates that the KSS bound for $\eta/s$ holds.

From the relations \eqref{MT01} and \eqref{MT02} the explicit formulae for the temperature and mass are
\beqa
M_{d=5}= \frac{3 \left(\ell^6+16 \lambda \right)r_+^4}{16 \pi \ell^8}, \quad\quad T_{d=5}= \frac{r_+}{\pi \ell^2},
\label{mass-d5}
\eeqa
 where we see that $T$ is independent of $\lambda$ in five dimensions.

{In figure \ref{fig:etaOvers} we plot $\eta/s$ for the values of $\lambda$  such that physical constraints are satisfied.} To determine this,  one needs $f_{\infty}>0$ for an asymptotic AdS spacetime  and $P(f_{\infty})>0$ to satisfy the no-ghost criterion. The behaviour of $f_{\infty}$ in four dimensions is given in right diagram in figure \ref{finfplot}; similar behaviour is observed in $d=5$, but with slightly different contributions from the different solution branches. In four and five dimensions we find that $\lambda$ must be negative so that $\gamma^2>0$.   The expression for the mass in \eqref{mass-d5} sets a lower bound  on the coupling.
In conjunction with the condition  $\gamma^2>0$ in \reef{gamma2} required for a well-defined asymptotic region,
 we must have  $0 > \lambda >\lambda_p =-\ell^6/16$  for any choice of AdS length and black hole horizon, where the lower bound $\lambda_p$ corresponds to zero mass
and vanishing entropy; note that  $|\lambda_p|$ is smaller than $|\lambda_c|$ given in \reef{muclac} .  We note from
figure \reef{fig:etaOvers} that the KSS bound is violated for $\lambda/\lambda_p>0$ or in other words for
all physically acceptable values of $\lambda$ since $\lambda_p < 0$.

 In six dimensions solving equations \eqref{MT01} and \eqref{MT02} yields four different solutions for the mass and temperature.  However the only branch that satisfies the criteria for a physical solution is
 \beqa
M_{d=6}&=&\frac{8 r_+^5-3 \pi  \ell^2 r_+^4 \sqrt{Y}+3 \pi  \ell^2 r_+^4 \sqrt{\frac{3 r_+^3}{4 \pi ^3 \lambda \sqrt{Y}}-Y}}{2\pi  \ell^2},\nonumber\\
T_{d=6}&=&\frac{\sqrt{Y}}{2}-\frac{1}{2} \sqrt{\frac{3 r_+^3}{4 \pi ^3 \lambda \sqrt{Y}}-Y},
\eeqa
where
\beqa
Y= \frac{W}{4 \sqrt[3]{2} \pi ^2
\lambda \ell^2}+\frac{5 r_+^4}{2^{2/3} \pi ^2 W}, \qquad
W = \sqrt[3]{9 \lambda \ell^6 r_+^6+\sqrt{81 \lambda^2 \ell^{12} r_+^{12}-4000 \lambda^3 \ell^6 r_+^{12}}}
\quad
\eeqa
Here, the mass is positive for $0<\lambda<\lambda_p$ and
  vanishes (along with the entropy) at $\lambda=\lambda_p$; \ from eq. \reef{deq6d} we find
 \beqa
 \gamma^2&=&\frac{2 \left(8 f_{\infty}^3 \lambda+3 \ell^6\right)^3}{675 f_{\infty} \ell^{16} \lambda m^2},
\eeqa
which  is positive both for $\lambda > 0$ and for sufficiently negative $\lambda$. On the other hand, even though the overall structure of vacuum solutions looks like what is exhibited in the right graph in figure \ref{finfplot},  in this case the upper branch with negative $\lambda$ is associated with the existence of ghosts -- it is therefore  excluded from further consideration. The lower negative $\lambda$ branch yields negative $\gamma^2$. Hence only   positive values of $\lambda$ yield physical solutions as well as $\gamma^2>0$. However there is an upper bound for the positive coupling that is enforced by positive mass.

 From figure~\ref{fig:etaOvers} starting from the same value of the ratio at zero coupling,   for small absolute values of  $\lambda$,
 in four dimensions the ratio is initially larger than for $d=6$ but then declines to smaller values. Both curves blow up as the pole is approached.  In  these cases the KSS bound holds, and mass remains positive as $0<|\lambda|<|\lambda_p|$.

We anticipate similar behaviour in seven and higher dimensions, namely that positive coupling
is required for correct asymptotic behaviour and physical mass, and also satisfies the KSS bound.

It is known that there is an upper bound  on the coupling restricting the existence of acceptable CFT duals, so the whole range of $\lambda \in (0, |\lambda_p|)$ cannot  possess a holographic interpretation~\cite{Hofman:2008ar, Myers:2010jv}. We postpone how to address this issue to future investigations.

 If both cubic and quartic couplings are nonzero, there is still a point in parameter space for which the entropy vanishes.  This is a singular point. It cannot be removed since $a_2$, which  appears at zeroth order in the $\eta$ expansion (similar to Eq. \reef{etaexp}),  is computed for the values of the couplings at the pole.
Corresponding results for cubic gravity are discussed in \cite{mir:2018mmm}.

\section{Discussion}
\label{discuss}

 We have investigated charged static spherically symmetric AdS black holes for both spherical ($k=1$) and hyperbolic
 ($k=-1$) geometries   in generalized quasi-topological gravity (GQG). These theories are of notable interest since
 this class of solutions  has  a single metric function, analogous to  Lovelock and quasi-topological gravity at the same order.  We have considered both cubic and quartic GQG  to see how these additional terms  modify the results for Einstein gravity in four, five, and six dimensions.

 Although the metric function cannot be obtained analytically, it is feasible to find both the asymptotic and near horizon behaviour of the metric perturbatively.  We then apply the shooting method to verify that these solutions match in the intermediate region.  The near horizon expansion characterizes the mass and temperature of black holes, and therefore,
 the thermodynamics of the black hole can be completely understood, despite the lack of an exact solution. Furthermore our numerical considerations demonstrate that for either fixed cubic coupling or quartic, increasing the electric charge correspondingly decreases the horizon radius. On the other hand at fixed charge, enlarging the coupling has the effect of  inceasing the horizon radius. Investigating the solutions near the origin $r = 0$  in four dimensions we find that
the curvature scalar singularity is   softened.

Taking the cosmological constant and cubic and quartic couplings as thermodynamic variables, we examined the
 thermodynamic properties in the given spherically symmetric configuration in detail, including verification of the extended first law and Smarr relation.  However not all solutions obey standard physical requirements (positive mass,
 positive entropy, AdS asymptotics, and the no-ghost condition), so we constructed the physical constraints between the couplings and the charge that gives the domain for parameters to yield physical critical points.

In some regions of parameter space black hole entropy can be negative. The sign of the entropy depends on the spacetime dimension and is a function of temperature in terms of horizon radius.  For a fixed charge ensemble or for neutral black holes,  there are situations (for small black holes) in which the entropy becomes negative  as $r_+\to 0$. One can add the absolute value of this amount to the entropy to ensure that the vacuum state has zero entropy. Under these circumstances the associated free energies are shifted by the same amount, so the thermodynamic properties are not affected.

 Another way to shift entropy to a positive value is by adding an explicit Gauss-Bonnet contribution to the action \cite{Bueno:2016lrh,Castro:2013pqa} or by including the volume form of the induced metric on the horizon in the Lagrangian  \cite{Clunan:2004tb} where, similar to adding the absolute value of the negative entropy as $r_+\to 0$ as mentioned above,  one must ensure $S\rightarrow 0$ as $M\rightarrow 0$.

 More generally, in  even dimensional spacetimes adding  Euler densities to the action will shift the entropy by an arbitrary constant without changing the solution of the field equations (see Appendix A of
\cite{mir:2018mmm} for the case of Gauss-Bonnet gravity in four dimensions). This arbitrary constant must be chosen such that the entropy vanishes for AdS spacetime. The more general problem of how to deal with negative entropy in any dimension we leave  to future work.

Working in both the fixed charge ensemble and fixed chemical potential ensemble, we classified the phase structure and critical points for these black holes. In the fixed charge ensemble, and in four dimensions we found out that even for zero charge, there are still physical critical points for $k=1$ when at least one of couplings is non-zero  (in contrast to Einstein gravity) and also for $k=-1$ when both couplings are nonzero.  A  first order VdW phase transition between small and large black holes  was seen.

For the first time we have observed critical points  for a neutral hyperbolic black hole in any dimension provided both the cubic and quartic couplings are  nonvanishing. This emphasizes the importance of the non-linearity of GQG in inducing new phase behaviour.

In five dimensions we observed  the occurrence of two physical critical points when both cubic and quartic couplings are nonzero,  even for vanishing charge. These respectively correspond to the   end point of a standard
VdW transition and the  starting point of  a reverse VdW transition.  However for the reverse VdW transition the pressures are smaller than the second critical pressure, and so one cannot choose parameters such that these two end points merge
to obtain an isolated critical point --  the pressure in the second line of first-order phase transitions does not increase with temperature unlike the first coexistence line.

For hyperbolic black holes in five and six dimensions there are regions in parameter space that yield  negative mass.
In the $d=6,~k=+1$ case, we noted the existence of three and two critical points  under  the respective condition that three and two of the parameters (charge and two couplings) are non-zero.
 Since the coexistence lines are both  increasing functions of pressure with respect to  temperature, it is possible to
 find parameter choices for which the critical points merge into an  isolated critical point.

We obtained the critical temperature and volume in terms of the electric charge and couplings in various dimensions  up to first order in the quartic coupling $\lambda$.  For spherical black holes in quartic gravity the universal relationship in Einstein gravity given by the first term in \reef{pvtq} receives dimension-dependent corrections -- it is no longer universal.  These correction are negligibly
  small  in four and five dimensions;  however in higher dimensions they cause the critical ratio \eqref{pvtq} to increase.

 We also analyzed the existence of physical critical points in the grand canonical ensemble   in the quartic theory.  Obtaining the relevant thermodynamic quantities, we confirmed the presence of a first order phase transition in four dimensions that is absent in the corresponding situation for Schwarzschild black holes in both cases, for which {either} the chemical potential or the pressure are fixed.  We also investigated  phase transitions in
 higher dimensions.

Our study of black hole thermodynamics will, of course, be modified by the inclusion of
Lovelock terms in $d > 4$ and by standard quasitopological terms \cite{Oliva:2010eb,Oliva:2010zd,Myers:2010ru}.  While it is conceivable that such terms in combination with those in the quartic theory will yield new phase behaviour, their inclusion would considerably broaden the parameter space and so we shall leave this investigation for future work.

Our goal was to isolate the thermodynamic behaviour of black holes in the quartic theory.  The reason for omitting the Lovelock terms (which in principle could be present in more than 4 dimensions) was primarily for simplicity. Their inclusion broadens the parameter space, and would make our paper notably longer than it already is.  A study of how the Lovelock terms affect our results is possible, of course, but we prefer to leave this for future work.

 Finally, in the context of the AdS/CFT, we computed the ratio of shear viscosity to entropy density $\eta/s$ for field theories dual to the quartic generalized quasi-topological theory in all dimensions and concluded that in four dimensions the KSS bound held for choices of the quartic coupling yielding positive mass and temperature.
 In striking constrast, we found that this behaviour did not hold in five dimensions -- the range
of coupling required for a physical solution entailed a violation of the KSS bound. However, the bound remains valid in four and six dimensions, and we anticipate it is satisfied in higher dimensions.
It would be interesting to see if including yet higher order terms in the curvature can yield black holes
satisfying both the KSS bound and all physically reasonable requirements in five dimensions as well as other dimensions.
Further constraints could be imposed by   configuring other conditions in the corresponding CFT, such as  causality constraints and positivity of energy flux; the implications of this require further investigation.

\section*{Acknowledgements}

 We thank Robie Hennigar for discussion in early stages of this work.
M. M. appreciates the hospitality of the University of Waterloo where this work was initiated. This work was supported in part by the Natural Sciences and Engineering Research Council of Canada.

\appendix
\section{Explicit form of Lagrangian densities}
\label{app:lag_dens}

At the quartic level, there are five generalized quasi-topological terms in dimensions larger than four
and six in four dimensions
\cite{Ahmed:2017jod}.  However, in each case, the field equations are identical for the ansatz
\eqref{eqn:metricAnsatz} with $N(r)$ constant
and we can, without loss of generality, chose any representative of this class.  Here we have chosen the following expression,

\begin{align}
\mathcal{S}_{4, d} &= \frac{1}{3 (d - 3)^2 (d - 2) (d -1) d (11 - 6 d + d^2) (19 - 18 d + 3 d^2) (-22 + 26 d - 9 d^2 + d^3)} \times
\nn\\
&\times \big[
-4 (d - 2) (-718080 + 2405582 d - 3666144 d^2 + 3359133 d^3 - 2057938 d^4
\nn\\
&+ 887142 d^5 - 276120 d^6 + 62662 d^7 - 10296 d^8 + 1182 d^9 - 86 d^{10} + 3 d^{11}) R_{a}{}^{c} R^{ab} R_{b}{}^{d} R_{cd}
\nn\\
&- 4 (707880 - 2115012 d + 2700668 d^2 - 1809780 d^3 + 561468 d^4 + 61133 d^5 - 134394 d^6
\nn\\
&+ 60426 d^7 - 15005 d^8 + 2238 d^9 - 189 d^{10} + 7 d^{11}) R_{ab} R^{ab} R_{cd} R^{cd}
\nn\\
&+ 16 (d - 2) (d -1) (-8670 + 30262 d - 47247 d^2 + 43299 d^3 - 25747 d^4 + 10271 d^5
\nn\\
&- 2734 d^6 + 466 d^7 - 46 d^8 + 2 d^9) R_{a}{}^{c} R^{ab} R_{bc} R + 4 (198900 - 592178 d + 790224 d^2
\nn\\
&- 617415 d^3 + 313537 d^4 - 109500 d^5 + 27237 d^6 - 4900 d^7 + 624 d^8 - 51 d^9 + 2 d^{10}) R_{ab} R^{ab} R^2
\nn\\
&+ 48 (d - 2) (-68340 + 203532 d - 268574 d^2 + 203038 d^3 - 95967 d^4 + 29190 d^5
\nn\\
&- 5665 d^6 + 667 d^7 - 42 d^8 + d^9) R^{ab} R^{cd} R R_{acbd} - 6 (192100 - 603774 d + 820554 d^2
\nn\\
&- 605255 d^3 + 237492 d^4 - 22951 d^5 - 24843 d^6 + 14329 d^7 - 3890 d^8 + 609 d^9 - 53 d^{10}
\nn\\
&+ 2 d^{11}) R^2 R_{abcd} R^{abcd} - 48 (d - 2) (d -1) (-63580 + 183572 d - 244118 d^2
\nn\\
&+ 192444 d^3 - 97734 d^4 + 32893 d^5 - 7308 d^6 + 1032 d^7 - 84 d^8 + 3 d^9) R_{a}{}^{c} R^{ab} R^{de} R_{bdce}
\nn\\
& + 12 (d - 2) (d -1) (-29920 + 120000 d - 196892 d^2 + 175930 d^3 - 93864 d^4
\nn\\
&+ 30115 d^5 - 5212 d^6 + 193 d^7 + 99 d^8 - 18 d^9 + d^{10}) R^{ab} R^{cd} R_{ac}{}^{ef} R_{bdef}
\nn\\
&+ 4 (d - 3) (d - 2) (d -1) (2550 - 15414 d + 28633 d^2 - 26167 d^3 + 13715 d^4
\nn\\
&- 4351 d^5 + 830 d^6 - 88 d^7 + 4 d^8) R R_{ab}{}^{ef} R^{abcd} R_{cdef} + 6 (-60520 + 414664 d
\nn\\
&- 945458 d^2 + 1097752 d^3 - 719367 d^4 + 242784 d^5 - 3125 d^6 - 36155 d^7 + 17569 d^8
\nn\\
&- 4430 d^9 + 659 d^{10} - 55 d^{11} + 2 d^{12}) R_{ab} R^{ab} R_{cdef} R^{cdef}
\nn\\
& - 12 (d - 3) (d - 2) (d -1) (-22 + 26 d - 9 d^2 + d^3) (-340 + 494 d - 70 d^2 - 185 d^3
\nn\\
&+ 112 d^4 - 25 d^5 + 2 d^6) R^{ab} R_{a}{}^{cde} R_{bc}{}^{fh} R_{defh} \big]
\nn\\
&+  R_{ab}{}^{ef} R^{abcd} R_{ce}{}^{hj} R_{dfhj},
\end{align}
which corresponds to $\mathcal{S}_{d}^{(4)}$ in~\cite{Ahmed:2017jod}.

\section{Pad\'e approximants procedure to construct the shooting parameter}

To find an analytic approximation for the free parameter $a_2$ that appears in the first order expansion of equation of motion near the horizon, we use the method discussed in the appendix of~\cite{Hennigar:2018hza}. We apply the method to find the results of section~\ref{sec:holo_hydro}.

The near horizon expansion for the metric function is
\beq
f(r) = 4 \pi T (r-r_+) + a_2 (r-r_+)^2 + \sum_{i=3}^{\infty} a_i (r-r_+)^3.
\eeq
As discussed earlier the field equations at each  order determine the parameters in the expansion in terms of $a_2$, but $a_2 = f'(r_+)/2$ itself remains undetermined. One approach for fixing this free parameter is to use the shooting method such that the numerical solution for the metric function reaches the known asymptotic solution at large $r$.

Another method~\cite{Hennigar:2018hza} entails considering
\beq
a_2 = g(\lambda).
\eeq
So by knowing the formula for different coefficients of the near-horizon expansion in term of $g(\lambda)$, we demand that $a_i$ for $i\ge 2$ behave smoothly as $\lambda \to 0$. It means the coefficients of terms with inverse powers of $\lambda$ must vanish and this fixes the components of the series expansion of $g(\lambda)$ about $\lambda \to 0$. For example, in  $d=4$ we obtain
\begin{align}
a_3  &=-\frac{\ell^6 g(0) }{81  r_+\lambda}+\frac{-4 \ell^8 g'(0)+9  \ell^2 \left(79-16\ell^2 g(0) \right)g(0)-810}{324 \ell^2 r_+}  + \cdots \, ,
	\nn\\
a_4 &= \frac{\ell^{12} g(0) }{17496  r_+^2\lambda^2}+\frac{\ell^4 \left(2 \ell^8 g'(0)+9  \ell^2 \left(60 \ell^2 g(0) -169\right)g(0)+567\right)}{34992  r_+^2\lambda}  + {\rm finite} \, .
\end{align}
For first relation for $a_3$, we obtain $g(0) = 0$ that also removes the $1/\lambda^2$ pole in $a_4$. Furthermore the choice of $g'(0) = -567/(2\ell^8)$ eradicates $1/\lambda$ divergence in $a_4$ and it gives the value for the finite part as anticipated from the Einstein case, $\ie$ $a_3 = 1/(r_+ \ell^2) = f^{(3)}(r_+)/6$.
Continuing to higher orders, we find that $g^{(n)}(0)$ is specified at any arbitrary order provided $a_{n+3}$ does not have a singularity as $\lambda \to 0$, and $a_{n+2}$ receives the expected Einstein value at the same limitation.

Because the coefficients of the derivatives increase rapidly, the Taylor series does not have a non-zero radius of convergence and the analytic expression for $a_2$ is not valid. However inserting the $g^{(n)}(0)$ terms
(which implicitly contain derivatives of the temperature) into a
Pad\'e approximant yields a satisfactory result. The coefficients of the Taylor series diverge due to the existence of a pole at positive $\lambda$. This problem comes from the fact that in our calculations we consider the temperature as a function of the coupling, although it is not a real analytic function.

\begin{figure*}[h]
\centering
\begin{tabular}{cc}
\includegraphics[width=0.45\textwidth]{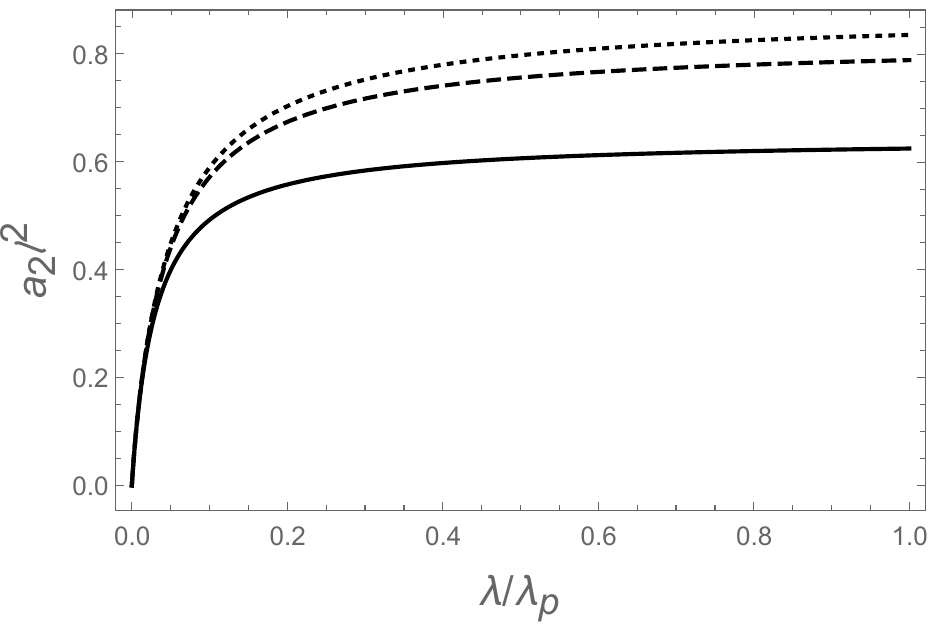}&\quad
\includegraphics[width=0.45\textwidth]{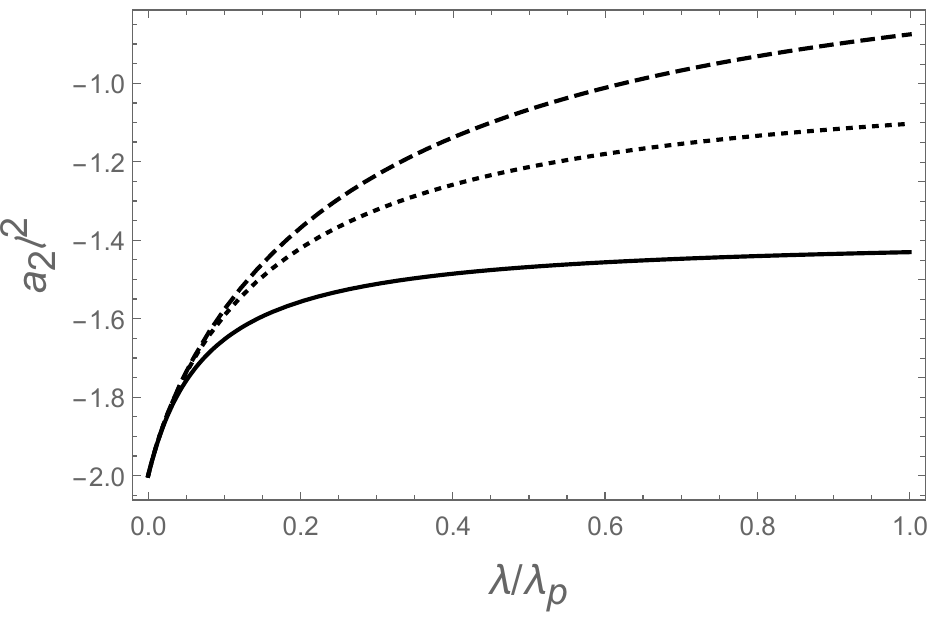}
\\
\end{tabular}
\caption{{\bf Approximation of shooting parameter in terms of coupling}: The shooting parameter $a_2$ for black branes in four (left) and five (right) dimensions. In these curves we demonstrate $[2/2]$, $[4/4]$ and $[5/5]$ order Pad\'e approximants (solid, dashed and dotted lines). }
\label{fig:shooting_parameters}
\end{figure*}

Computation of the Pad\'e approximants for higher orders is quiet tedious, although according figure~\ref{fig:shooting_parameters}, for small coupling one can make use of lower order Pad\'e approximant to obtain accepted outcome. For example, $[2|2]$ Pad\'e approximants for some dimensions in terms of $x = \lambda/\lambda_p$ are,
\begin{align}
a_2^{d = 4} \ell^2 &= \frac{104976 x  (690897408759 x +3684497152)}{\left(112691150931793887 x^2+4079675994292224 x +17247657525248\right) }  \, ,
	\nn\\
a_2^{d = 5} \ell^2 &=-\frac{2 \left(77197 x ^2+8623 x +10\right)}{\left(111329 x ^2+8663 x +10\right) }  \, ,
	\nn\\
a_2^{d = 6} \ell^2  &= -\frac{40 \left(1090352694831737484375 x ^2+109068070186354176000 x +789926926826340352\right)}{\left(16690216546970054015625 x ^2+940561493376835584000 x +6319415414610722816\right) }  \, .\nonumber
\end{align}

\bibliography{LBIB}
\bibliographystyle{JHEP}

\end{document}